%% file: btov.tex
\documentclass[a4paper,11pt,headings=big,DIV=12]{scrartcl}
\pdfoutput=1
\usepackage{amsmath,amssymb}
\usepackage{color,xcolor}
\definecolor{blue}{rgb}{0,0,0.5}
\usepackage[colorlinks,linkcolor=blue,urlcolor=blue,citecolor=blue]{hyperref}
\usepackage{graphicx}
\addtokomafont{disposition}{\rmfamily\boldmath}
\usepackage{enumerate}
\usepackage{slashed}
\usepackage{cite}
\usepackage{multirow}
\allowdisplaybreaks
\newcommand{\notebook}{\texttt{notebookBSZ.nb}}
\newcommand{\pol}{\iota}
\newcommand{\vev}[1]{\langle #1 \rangle} 
\newcommand{\state}[1]{|#1\rangle}

\newcommand{\matel}[3]{\langle #1|#2|#3\rangle}
\newcommand{\al}{\alpha}
\newcommand{\be}{\beta}
\newcommand{\ga}{\gamma}
\newcommand{\de}{\delta}
\newcommand{\la}{\lambda}

\newcommand{\GeV}{\,\mbox{GeV}}
\newcommand{\MeV}{\,\mbox{MeV}}

\newcommand{\fL}{f_V^\parallel}
\newcommand{\fT}{f^\perp_V}

\newcommand{\eom}{EOM }
\newcommand{\MSbar}{$\overline{\mathrm{MS}}$ }
\newcommand{\MSbarm}{\overline{\mathrm{MS}} }
\newcommand{\MSbard}{$\overline{\mathrm{MS}}$-}
\newcommand{\mi}{\!-\!}
\newcommand{\V}{{\cal V}}
\begin{document}

\begin{flushright}
\begin{tabular}{l}
TUM-HEP-957/14\\
CP3-Origins-2015-010 DNRF90 \\
DIAS-2015-10
\end{tabular}
\end{flushright}
\vskip1.5cm

\begin{center}
{\huge\bfseries \boldmath
$B\to V\ell^+\ell^-$ in the Standard Model\\[0.3 cm]
from Light-Cone Sum Rules
}\\[0.8 cm]
{\Large%
Aoife Bharucha$^{a,b}$,
David M. Straub$^c$,
and Roman Zwicky$^d$
\\[0.5 cm]
\small
$^a$ Physik Department T31, Technische Universit\"at M\"unchen, James-Franck-Str.~1, 85748~Garching, Germany
\\[0.2cm]
$^b$ CNRS, Aix Marseille U., U. de Toulon, CPT, UMR 7332, F-13288, Marseille,
France
\\[0.2cm]
$^c$ Excellence Cluster Universe, Technische Universit\"at M\"unchen, Boltzmannstr.~2, 85748~Garching, Germany
\\[0.2cm]
$^d$ Higgs Centre for Theoretical Physics, School of Physics and Astronomy,\\
University of Edinburgh, Edinburgh EH9 3JZ, Scotland
}
\\[0.5 cm]
\small
E-Mail:
\texttt{\href{mailto:aoife.bharucha@cpt.univ-mrs.fr}{aoife.bharucha@cpt.univ-mrs.fr}},
\texttt{\href{mailto:david.straub@tum.de}{david.straub@tum.de}},
\texttt{\href{mailto:roman.zwicky@ed.ac.uk}{roman.zwicky@ed.ac.uk}}.
\end{center}

\bigskip

\begin{abstract}\noindent
We present $B_q\to\rho$, $B_q\to\omega$, $B_q\to K^*$, 
$B_s\to K^*$ and $B_s\to \phi$ form factors from light-cone sum rules (LCSR) 
at ${\cal O}(\al_s)$ for twist-2 and 3 and ${\cal O}(\al_s^0)$ for twist-4 with updated hadronic input parameters.
Three asymptotic light-cone distribution amplitudes of twist-$4$ (and $5$) are determined, necessary  for the form factors to obey the equations of motion. It is argued that the latter 
constrain the uncertainty of tensor-to-vector form factor ratios thereby
  improving the prediction of zeros of helicity amplitudes 
 of major importance for  $B\to K^*\ell\ell$ angular observables.
We provide easy-to-use fits to the LCSR results, including the full error correlation matrix, in all modes at low $q^2$
as well as combined fits to LCSR and lattice results covering the entire kinematic range 
for $B_q\to K^*$, 
$B_s\to K^*$ and $B_s\to \phi$. 
The error correlation matrix  avoids  the problem of overestimating the uncertainty in phenomenological applications. 
Using the new form factors and recent computations of 
non-factorisable contributions we provide Standard Model predictions for  
 $B\to K^*\gamma$  as well as $B\to K^*\ell^+\ell^-$ and $B_s\to\phi\mu^+\mu^-$ at low dilepton invariant mass.  
Employing our $B \to (\rho,\omega)  $ form factor results 
we extract the CKM element $|V_\mathrm{ub}|$ from the semileptonic decays 
$B\to(\rho,\omega) \ell\nu$ and find good agreement with other exclusive determinations.
\end{abstract}

\newpage

\setcounter{tocdepth}{2}
\tableofcontents
\newpage

\section{Introduction}

Exclusive semi-leptonic $B$ decays are important tools to test the Standard Model (SM) and to look for new physics. 
Among these processes, the decays $B\to K^*(\to K \pi)\mu^+\mu^-$ and $B_s\to\phi(\to K^+K^-)\mu^+\mu^-$
%, that both involve a vector meson in the final state, 
are of particular relevance as their angular distributions give access to a host of observables that are  sensitive to new physics (e.g. \cite{Blake:2016olu} for arecent review). Predicting these observables, either within the SM or beyond, requires the knowledge of the form factors (FFs) -- in the case of $B\to V$ transitions, these are 7 functions of the dilepton invariant mass squared $q^2$. 
In the low $q^2$ region, where the vector meson is energetic, the FFs can be computed 
using the method of sum rules on the light cone (LCSR) whereas at high $q^2$ the FFs can be computed using lattice QCD.

In this work we present an update of the FF computation in \cite{Ball:2004rg}, for 
the modes  $B_q\to\rho$, $B_q\to\omega$, $B_q\to K^*$, 
$B_s\to K^*$ and $B_s\to \phi$ (with $q=u,d$),  
using current hadronic input and a concise discussion of the role of \emph{the equation of motion} (\eom\!\!) 
in correlating  vector and tensor FFs. The FFs are fitted to the 
$z$-expansion parameterisation in the helicity basis, \emph{retaining all correlations} among the expansion coefficients\footnote{Similar fits retaining correlations have recently been performed for the $B\to \pi$ FFs \cite{Imsong:2014oqa}.}. 
This information is  made publicly available as ancillary files on the arXiv web pages in a form which is easy to use for phenomenology.

Crucially the correlation of the uncertainties avoids overestimating uncertainties in observables.  
A particularly important  example
%, where this approach turns out to beneficial,  
are the angular observables in $B\to K^*\mu^+\mu^-$-type decays  since 
they are sensitive to ratios of FFs and zeros of helicity amplitudes. For  the latter two, 
the uncertainty is considerably reduced when taking correlation into account.

We argue, extending the work in Ref.~\cite{Hambrock:2013zya}, that the use of the \eom enforces the correlation of the non-parametric, sum rule specific, input parameters. 
This can be seen as an application of the large energy limit (LEL) ideas \cite{Charles:1998dr} to the sum rules on the light-cone. 
It is in giving numerical predictions
and not relying on the heavy quark limit that the LCSR computations go beyond the LEL ideas \cite{Charles:1998dr} (this includes the case  factorisable hard  $\alpha_s$-corrections \cite{Beneke:2000wa}).
The LCSR therefore give corrections to the LEL  \cite{Charles:1998dr}  and soft-collinear effective theory (SCET) \cite{Beneke:2000wa} relations.  
Going beyond the SCET framework of two soft FFs and hard $\alpha_s$-correction 
in the heavy quark limit involves  using the numerical predictions 
from LCSR, e.g.~\cite{Descotes-Genon:2014uoa,Jager:2014rwa}.  Going beyond the SCET framework 
has become increasingly important since observables designed to minimise the impact of 
the soft FFs  \cite{Descotes-Genon:2013vna} are, of course, sensitive to $1/m_b$-corrections.

We perform \emph{combined fits} of the FFs to the LCSR at low $q^2$ 
and a recent lattice computation at high $q^2$ \cite{Horgan:2013hoa,Horgan:2015vla}. 
This serves to test the consistency of the two complementary methods and 
provides FF sets valid over the entire kinematical region.
We extract the CKM element $|V_\mathrm{ub}|$ from 
$B\to(\omega,\rho)\ell\nu$   BaBar- and Belle-data using 
the $B \to (\omega,\rho)$ FF predictions of this paper.
This can either be viewed as an  extraction of  $|V_\mathrm{ub}|$  or as a check 
of the normalisation of the FF when compared to global fits or $B \to \pi \ell \nu$ extractions of 
$|V_\mathrm{ub}|$.
In addition to the FFs, the calculation of $B\to V\ell^+\ell^-$ observables involves \emph{non-factorisable contributions} from the weak hadronic Hamiltonian. 
Some of these contributions have been recently computed within LCSR. Including all of these ingredients, we present SM predictions for the branching ratios and angular observables of $B\to K^*\mu^+\mu^-$ and $B_s\to\phi\mu^+\mu^-$. We also compare the prediction for the branching ratio of $B\to K^*\gamma$, that has been measured precisely at the $B$ factories, to the data.

The paper is organised as follows. In section \ref{sec:LCSR-form} we present and discuss 
the seven $B \to V$ FFs within the LCSR context, discussing the implication of the \eom, finite width effects, input parameters and the interpolating fits to the lattice data. In section \ref{sec:pheno} phenomenological aspects of 
$B \to K^*\mu\mu$, $B \to K^* \gamma$, $B_s \to \phi \mu\mu$, $B \to K^*\mu\mu$ versus 
$B_s \to \phi \mu\mu$ and the determination 
of $|V_\mathrm{ub}|$ from $B \to (\rho,\omega) \ell\nu$ are discussed: see subsections \ref{sec:lowKs}, \ref{sec:Ksga}, \ref{sec:lowphi}, \ref{sec:versus}  and \ref{sec:Vub}
respectively.
Conclusions figure in section  \ref{sec:conc}.
Appendix  \ref{app:LCSR} assembles aspects of the \eom\!\!, explicit tree level 
results, scheme dependence and remarks on the Borel parameters.  
The determination of the three light-cone distribution amplitudes  
$\mathbb{A}_\parallel$ (twist-4) and $\mathbb{G}_\perp^{v,a}$ (twist-5), using an alternative method,  
is discussed in appendix \ref{app:LCDA}.
A detailed discussion on the determination of the decay constants from 
experiment is given in appendix \ref{app:decay}. Conversion between 
bases, further plots  and fit coefficients are given in appendixes \ref{app:FF-hel}, \ref{app:z},  
and \ref{sec:coeff} respectively. 
The effect of the sizeable $B_s$-lifetime is worked out  in appendix \ref{app:ADeltaGamma}.

\section{\texorpdfstring{$B\to V$}{B->V} form factors from light-cone sum rules}
\label{sec:LCSR-form}

The short distance matrix elements, relevant for the dimension six effective Hamiltonian, are  
parameterised by seven FFs\footnote{Due to the composition 
of the wave functions of the  $\rho^0 \sim 1/\sqrt{2} (\bar u u - \bar d d  )  $ and $\omega \sim 1/\sqrt{2} (\bar u u - \bar d d  )$, extra factors $c_V^{b \to q}$ have to attached to the matrix elements on the left-hand side, cf.~\cite{Ball:2004rg}. They read: 
$c^{ u}_{\rho^0} = -c^{ d}_{\rho^0}  =c^{ u}_{\omega} = c^{ d}_{\omega}= 
  \sqrt{2}$ and $c_V=1$ in all other cases.}
 \begin{alignat}{2}
 &     \matel{K^*(p,\eta)}{\bar s \gamma^\mu(1 \mp \gamma_5) b}{\bar B(p_B)}  
 &\;=\;&   \;\;   P_1^\mu \, \V_1(q^2) \pm P_2^\mu \, \V_2(q^2) \pm P_3^\mu \,  \V_3(q^2)  \pm P_P^\mu \V_P(q^2) 
   \; ,\nonumber  \\[0.1cm]
  & \matel{K^*(p,\eta)}{\bar s iq_\nu \sigma^{\mu\nu} (1 \pm \gamma_5) b}{\bar B(p_B)} 
  &\;=\;& \;\;  P_1^\mu  T_1(q^2)   \pm  P_2^\mu  T_2(q^2) \pm  P_3^\mu  T_3(q^2) 
   \; , 
   \label{eq:ffbasis}
\end{alignat}
where the Lorentz structures $P_i^\mu$ are defined as in \cite{Lyon:2013gba}
\begin{alignat}{2}
\label{eq:Vprojectors}
& P_P^\mu = i (\eta^* \cdot q) q^\mu \; ,&P_1^\mu  =&  2 \epsilon^{\mu}_{\phantom{x} \alpha \beta \gamma} \eta^{*\alpha} p^{\beta}q^\gamma  \; , \\
& P_2^\mu = i \{(m_B^2\mi m_{K^*}^2) \eta^{*\mu} \mi 
(\eta^*\!\cdot\! q)(p+p_B)^\mu\} \; , \qquad
&P_3^\mu =&  i(\eta^*\!\cdot\! q)\{q^\mu \mi  \frac{q^2 }{m_B^2\mi m_{K^*}^2} (p+p_B)^\mu \}   \;,
\nonumber 
\end{alignat}
with the $\epsilon_{0123}=+1$    convention for the Levi-Civita tensor.
The relation $T_1(0)= T_2(0)$ holds algebraically. The parameterisation \eqref{eq:ffbasis} makes the 
correspondence between vector and tensor FFs explicit.  
The correspondence of the   ${\cal V}_{P,1,2,3}$  to the more traditional FFs $A_{0,1,2,3}$ and $V$ 
is as follows
 \begin{alignat}{1}
& \V_P(q^2) =  \frac{- 2 m_{K^*}}{q^2} A_0(q^2) \;,  \quad \V_1(q^2) =  \frac{-V(q^2)}{m_B+m_{K^*}} \;, \quad     \V_2(q^2) =    \frac{-A_1(q^2)}{m_B-m_{K^*}} 
\;, \nonumber  \\[0.1cm]
 &  \V_3(q^2) =  \big( \frac{m_B+m_{K^*}}{q^2}    A_1(q^2) -   \frac{m_B-m_{K^*}}{q^2}    A_2(q^2) \big) \equiv \frac{2 m_{K^*}}{q^2} A_3(q^2) \; .
 \label{eq:VAs}
\end{alignat}
The relation $A_3(0) = A_0(0)$ assures finite matrix elements at $q^2 =0$. The last relation in 
\eqref{eq:VAs} indicates that one FF out of $A_{1,2,3}$ is redundant.\footnote{From the viewpoint of the projections the traditional nomenclature is unfortunate. It would have been better not to have $A_2$ at all and  use the notation $A_1 \to A_2$.}  The  pseudoscalar matrix element is related to $A_0$ through an axial Ward Identity:
\begin{equation}
\label{eq:WI5}
\matel{K^*(p,\eta)}{\bar s  \gamma_5 b}{\bar B(p_B)}  =  \left( \frac{P_P \cdot q }{m_s+m_b} \right) {\cal V}_P(q^2) = \left( \frac{2 m_{K^*}(\eta^* \cdot q) }{i(m_s+m_b)} \right) A_0(q^2) \;.
\end{equation}
The projection on the helicity basis, using the Jacob-Wick polarisation convention, is given
in appendix \ref{app:FF-hel}.
In the next section we briefly discuss the use of the method of LCSR before investigating the implications of the  \eom  on certain sum rule 
specific parameters.

\subsection{Calculation of the form factors in light-cone sum rules}
\label{sec:LCSR}

Light-cone sum rules (similar to QCD sum rules~\cite{Shifman:1978bx,Shifman:1978by}) for the FFs are derived 
by considering the correlator of the time-ordered 
product of two quark currents evaluated between the final state on-shell meson (in this case $V$) and the vacuum~\cite{Balitsky:1989ry,Chernyak:1990ag}. 
On expanding this correlator about the light-cone, one obtains a series of 
perturbatively calculable hard scattering kernels convoluted with non-perturbative,
universal light-cone distribution amplitudes, ordered by increasing twist (dimension minus spin).
Reasonable convergence of the LC-expansion is formally and by experience 
limited up to $q^2 \simeq {\cal O}(m_b \Lambda_\mathrm{QCD}) \simeq 14 \GeV^2$.
In the hadronic picture the correlator is expressed as  the sum
over excited states, the dominant state being the $B$-meson, and this is followed by the continuum. 
Assuming quark-hadron duality above a certain continuum threshold $s_0$ \cite{Shifman:1978bx,Shifman:1978bw}, an approximation referred to as the semi-global quark-hadron duality assumption, one arrives at an expression for the lowest lying hadronic parameter 
in terms of an expression of partonic QCD and $s_0$. 
A  Borel transformation which ameliorates both the hadron and the parton evaluation 
leads to a numerical improvement of the procedure. 

Light-cone sum rules results, with light-meson distribution amplitudes (DAs), 
have been computed for the $B \to P$ transition up to twist-3 $\mathcal{O}(\alpha_s)$  in \cite{Ball:2001fp, Ball:2004ye,Duplancic:2008ix} and for the $B \to V$ transition 
up to twist-4 at tree level  and twist-2 $\mathcal{O}(\alpha_s)$   \cite{Ball:1998kk} as well as twist-3 $\mathcal{O}(\alpha_s)$ \cite{Ball:2004rg}. 
In this paper we make use of the results 
in \cite{Ball:2004rg}.\footnote{In \cite{Ball:2004rg} the size of the twist-3 $\mathcal{O}(\alpha_s)$ corrections were not explicitly given. 
At $q^2 =0$ the twist-3 $\mathcal{O}(\alpha_s)$ corrections lead to a raise of around 
$10\%$  of the  FF $T_1$.} Alternatively the FFs can be determined using 
$V$-meson DA and an interpolating current for the $B$-meson.
 This program has been pursued in \cite{Khodjamirian:2006st}  at tree level in QCD  and in SCET~\cite{DeFazio:2007hw}. 
In this work we improve on the previous LCSR work \cite{Ball:2004rg} by 
\begin{itemize}
\item computation of the full twist-4 (and partial twist-5) 2-particle DAs contribution to 
the FF (appendix \ref{app:LCDA} -- available a downloadable Mathematica notebook),
\item determination of the DAs $\mathbb{G}_\perp^{v,a}$ (twist-5),  in the asymptotic limit, filling a gap in the literature (appendix \ref{app:LCDA}),
\item discussing the impact of the \eom on uncertainty correlations (section \ref{sec:eom}),
including the aspect of scheme-dependence (appendix \ref{app:scheme}),
\item explicit verification of  the \eom  at tree level (appendices \ref{app:eom-tree},\ref{app:T-results}) , for the asymptotic 2-particle DAs including ${\cal O}(m_s)$-corrections. 
\item verification of the compatibility of the composite operator renormalisation with the \eom
(appendix \ref{sec:RCOMP}),
\item discussion of  non-resonant background for vector meson final states  (section \ref{sec:finite}),
\item determination and usage of  updated hadronic parameters (section \ref{sec:choice}), specifically the decay constants (appendix \ref{app:decay}),
\item fits with full error correlation matrix for the $z$-expansion coefficients (section \ref{sec:fits}),
as well as an interpolation to the most recent lattice computation (section \ref{sec:lattice}).
\end{itemize}

\subsection{Equation of motion and form factors}
\label{sec:eom}

In this section we reiterate the use of the \eom \cite{Hambrock:2013zya}. 
As discussed in \cite{Hambrock:2013zya} this is of importance in reducing the uncertainty between certain FFs. Below we  give more details and strengthen the argument.
The following \eom 
\begin{alignat}{2} 
 &  i \partial^\nu (\bar s i \sigma_{\mu \nu} (\ga_5)b)   &=&
   - (m_s \pm m_b) \bar s \gamma_\mu (\ga_5) b +i \partial_\mu (\bar s(\ga_5)  b) 
   - 2 \bar s i \!\stackrel{\leftarrow}{D}_{\mu} (\ga_5) b \, ,
    \label{eq:eom_op}
\end{alignat}
are valid on physical states.  
Equations of the form \eqref{eq:eom_op} are sometimes also referred to as Ward identities. 
In particular, evaluated on $\matel{V}{\ldots}{B}$, Eq.~\eqref{eq:eom_op} 
yields
\begin{alignat}{1}
\label{eq:eom1}
&T_1(q^2)  + (m_b+m_s)\V_1(q^2) +  {\cal D}_1(q^2)     =  0 \;,    \\
\label{eq:eom2}
    &T_2(q^2) +  (m_b- m_s) \V_2(q^2) + {\cal D}_2(q^2)    = 0 \;,    \\
\label{eq:eom3}
&T_3(q^2)  + (m_b- m_s) \V_3(q^2) +  {\cal D}_3(q^2)   =   0 \;,  \\
\label{eq:eom4}
  &   (m_b-m_s)  \V_P(q^2)  +  \left(   {\cal D}_P(q^2) -  \frac{ q^2}{m_b+m_s} \V_P(q^2)     \right)   =  0 \;  .
\end{alignat}
One of the above four equations corresponds to each of the directions \eqref{eq:Vprojectors} \cite{Grinstein:2004vb,Hambrock:2013zya}, where  ${\cal D}_i$ are defined
\begin{equation}
\matel{K^*(p,\eta)}{\bar s  (2 i \!\stackrel{\leftarrow}{D})^{\mu}(1 \!\pm\!\gamma_5) b}{\bar B(p_B)}  
  =     P_1^\mu \,  {\cal D}_1(q^2) \pm P_2^\mu \, {\cal D}_2(q^2) \pm P_3^\mu \,  {\cal D}_3(q^2)  \pm 
  P_P^\mu {\cal D}_P(q^2) \;,
  \label{eq:DD}
\end{equation}
in complete analogy with \eqref{eq:ffbasis}.  Note that the $i\partial_\mu (\bar s(\ga_5)  b) $ operator 
only contributes to $P_P^\mu \sim q^\mu$, since the total derivative is
replaced by the momentum transfer $q^\mu$. 
In Eq.~\eqref{eq:eom4} we have included this contribution into round brackets
with the other derivative FF. 
Before discussing  Eqs.~\eqref{eq:eom1}--\eqref{eq:eom4} in various limits, we wish to stress that the equations are completely general and have to be obeyed by any FF determination.

\begin{figure}[tbp]
\begin{center}
\includegraphics[width=0.5\textwidth]{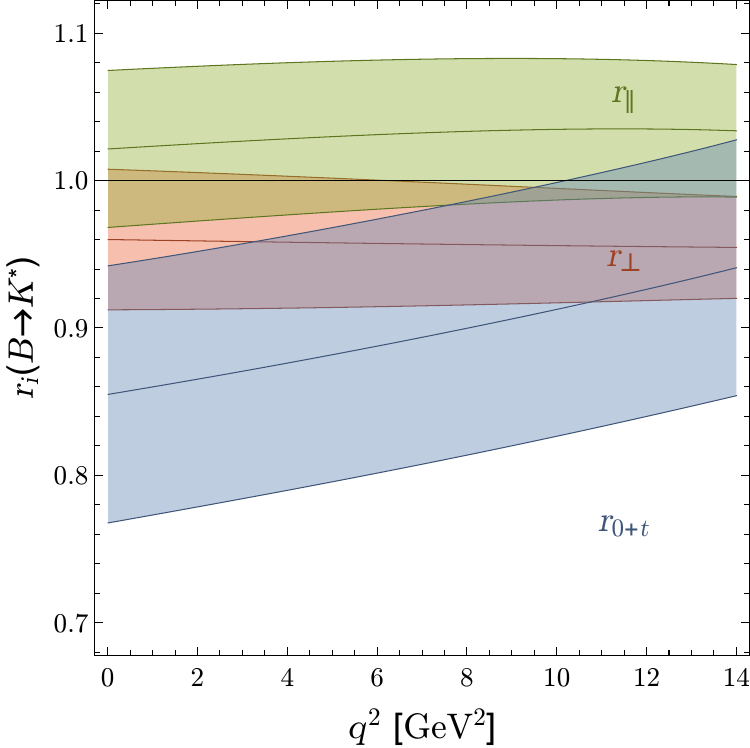}
\end{center}
\caption{\small Plots of $r_\perp$, $r_\parallel$ and $r_{0+t}$  Eq.~\eqref{eq:ri} 
(for the $B \to K^*$-transition) 
as a function of $q^2$. The deviation from unity  measures  the relative size of the derivative form 
factor with respect to the tensor and vector FFs.  The mass used in \eqref{eq:ri} is the pole mass, an issue 
discussed in appendix \ref{app:scheme}. For the explicit $m_b$ mass we use the central value and do not 
include an error since in the $B \to V \ell \ell$ helicity amplitudes 
the $m_b$-pole is not present.  
The fact that the quantities tend towards one for very high $q^2$ is expected from the viewpoint of the Isgur-Wise relations and proves a
certain robustness of the LCSR-results for high $q^2$. 
Similar ratios have been plotted in previous work in the context of Isgur-Wise relations 
\cite{Ball:1998kk} and LEL relations \cite{Altmannshofer:2008dz} but the observation that this might be useful 
for the correlating the continuum thresholds was not made. In particular the derivative form factor 
were not identified as a independent objects.}
\label{fig:IW-like-plots} 
\end{figure}

The Isgur-Wise relations \cite{Isgur:1990kf} follow from a clear physical picture. At low recoil 
the non-relativistic heavy quark effective theory applies and it can be shown that 
${\cal D}_i$ are suppressed by $\mathcal(\Lambda_\text{QCD}/m_b)$ with respect to the vector and tensor FFs  \cite{Grinstein:2004vb}.
This raises the question of whether there are combinations of  ${\cal D}_i$'s which are small at large recoil.
Eqs.~(\ref{eq:eom1},\ref{eq:eom2}) are direct candidates but Eqs.~(\ref{eq:eom3},\ref{eq:eom4}) require
some thought because of the common direction $q_\mu$. In fact in Eqs.~(\ref{eq:eom3},\ref{eq:eom4}) 
the poles in $q^2$ cancel between the FF $\V_{3},\V_{ P}$ and ${\cal D}_{3} , {\cal D}_{P}$ which implies that 
 ${\cal D}_{3} , {\cal D}_{P}$ are not individually small. 
Since the  $\matel{K^*}{ \bar s \ga_\mu \ga_5 b}{\bar B}$ matrix element is free from  singularities, adding 
Eqs.~(\ref{eq:eom3},\ref{eq:eom4})  yields a combination for which the derivative FFs are potentially small.  We define the following ratios 
\begin{alignat}{2}
\label{eq:ri}
& r_\perp(q^2) &\;=\;& - \frac{(m_b+m_s)\V_1(q^2)}{T_1(q^2)}  
= \frac{m_b+m_s}{m_B+ m_{K^*}} \frac{V(q^2)}{T_1(q^2)}\;, \nonumber \\[0.1cm]
& r_\parallel(q^2) &\;=\;&-  \frac{(m_b-m_s)\V_2(q^2)}{T_2(q^2)}  
= \frac{m_b-m_s}{m_B- m_{K^*}} \frac{A_1(q^2)}{T_2(q^2)}
\;, \nonumber \\[0.1cm]
& r_{0+t}(q^2) &\;=\;& 
- \frac{(m_b-m_s)(\V_2(q^2) -c_{23}(q^2) (\V_3(q^2)+\V_P(q^2)) )}{T_2(q^2)  - c_{23}(q^2) T_3(q^2)}  \nonumber \\[0.1cm]
&   &\;=\;& - \frac{ (m_b-m_s) ( \V_0(q^2) -c_{23}(q^2) \V_P(q^2)) }{T_0(q^2)} \nonumber
\\[0.1cm]
& &\;=\;&   \frac{m_b-m_s}{m_B- m_{K^*}} \frac{A_1(q^2) + c_{23}(q^2)\frac{2 m_{K^*}(m_B- m_{K^*}) }{q^2} (A_3(q^2)-A_0(q^2))}{T_2(q^2)  - c_{23}(q^2) T_3(q^2)}
\;,
\end{alignat}
where $X_0 = X_2 - c_{23}X_3$ for $X= T,\V$  with  $c_{23}(q^2)$ being a kinematic function defined in\eqref{eq:c23la}.
The deviations of these quantities from one measure the relative size of the derivative FFs with respect to vector and tensor FFs, 
\begin{equation}
r_\perp = 1+ \frac{{\cal D}_\perp}{T_1} \;, \quad 
r_\parallel = 1+ \frac{{\cal D}_\parallel}{T_2}\;, \quad  
r_{0+t} = 1+ \frac{{\cal D}_{0+t}}{T_0 } \;,
\end{equation}
where ${{\cal D}_\perp = {\cal D}_1}$, ${\cal D}_{\parallel} = {\cal D}_2$  
and ${\cal D}_{0+t} = {\cal D}_2 - c_{23}( {\cal D}_3+{\cal D}_P)$.
In Fig.~\ref{fig:IW-like-plots} we show plots of these ratios from $0 < q^2 < 14 \GeV^2$. 
The quantities  $r_{\perp,\parallel}$ and, somewhat less, $r_{0+t}$ are found 
to be very close to unity over this range.  
The basic idea is that if the ${\cal D}_i$ are considered 
as regular FFs with controlled uncertainty\footnote{For the $B \to K^*$ channel at $q^2 = 1 \GeV^2$ the corrections
due to twist-4 and $\alpha_s$-correction for $\{ (T_1,{\cal D}_\perp),(T_0, {\cal D}_{0+t})\}$
 are $\{ (4,6),(7,28)\}\%$ and $\{ (12,27),(11,31)\}\%$ respectively  indicating 
 regularity of ${\cal D}_i$  with regard to the twist- and the $\alpha_s$-
 expansion.   } 
 then this implies a high degree of correlation between vector and 
tensor FFs of a given polarisation.  This is partly reflected in the controlled error bands.

The aspect of the correlation between the continuum thresholds is discussed in some more detail in  appendix \ref{app:corr2}.  Here we just summarise the main argument and result.
Based on the \eom we argue, conservatively, that 
the continuum thresholds for tensor and vector FF cannot differ by more than $1 \GeV^2$ in order 
for the continuum thresholds of the derivative FF not to take on absurdly low or high values. 
This argument is less compelling for the $(0+t)$-direction, as can be inferred from the plots, resulting 
in lower correlation and larger error bands. 
We stress that if we were to impose standard error bands, say $s_0 = 35(2) \GeV^2$
 for the sum rule of the ${\cal D}_i$ FFs then 
the error bands for $r_i$-ratios would shrink to the $1\%$-level.

The smallness of the derivative form factors (cf. also \cite{Hambrock:2013zya}  for further references 
and more physical discussion)  is related to the  
findings of Charles et al \cite{Charles:1998dr} within the LEL  
and its extension into  SCET 
\cite{Beneke:2000wa,Bauer:2000yr}.
 The similarity is the use of the \eom 
of motion but the difference is that in this work the \eom are directly implemented 
within QCD whereas in the prior work the \eom are used at the level of an effective theory 
in $1/m_b$. This results in differences at the level of $1/m_b$ corrections. 
For example in \cite{Beneke:2000wa} the ratio analogous to $r_\perp$ \eqref{eq:ri},
which we shall denote by $r_\perp^\mathrm{BF} = m_B/(m_B+m_{K^*}) \, (V/T_1) = 1 + {\cal O}(m_b^{-1},\al_s m_b^0)$, 
differs from ours by $m_B \to  (m_b+m_s)$ which is indeed  ${\cal O}(m_b^{-1})$. 
For completeness let us mention that the symmetry breaking corrections to $r_\perp^\mathrm{BF}$ were computed 
to  ${\cal O}(\al_s m_b^0)$ and ${\cal O}(\al_s^2 m_b^0)$ in \cite{Beneke:2000wa} and \cite{Beneke:2005gs,Bell:2010mg} respectively.  
Even though the twist expansion of LCSR is not a $1/m_b$-expansion,
as first  stated for $B \to P$ and $B \to V$ decays in  \cite{Chernyak:1990ag} and 
\cite{Ali:1993vd}, it is of interest 
to examine the various twist-quantities from the viewpoint of the standard 
heavy quark scaling prescriptions \cite{Chernyak:1990ag}.
At the level 
of all explicit calculations in the literature it is found that
$T_1(0) \sim V(0)|_{\text{twist-2,3} } \sim m_b^{-3/2}$ and  
 $T_1(0) \sim V(0)|_\text{twist-4} \sim m_b^{-5/2}$. The derivative FFs follow
 ${\cal D}_1(0)|_{\text{twist-2,3} } \sim m_b^{-5/2}$ \cite{Hambrock:2013zya}  which is in agreement 
 with the explicit computation in \cite{Charles:1998dr}.
For this work  we have explicitly checked that  ${\cal D}_1(0)|_\text{twist-4} \sim m_b^{-7/2}$ and that  ${\cal D}_1(0) \sim {\cal O}(\alpha_s) m_b^{-3/2}$,  in accord with the results of Beneke and Feldmann \cite{Beneke:2000wa}.  
In summary we may state that the parametric statements and the previous numeric statements give a consistent picture.

\subsection{Discussion of non-resonant background  effects}
\label{sec:finite}

The signal final state in a $B \to \rho(\to \pi \pi) \ell \nu$-type decay, serving  
as a template for any $B \to V \ell_1 \ell_2$ decay,  is 
$  \pi\pi \ell \nu$. Hence in principle the decay ought to be analysed  via a
$B \to \pi \pi$ type form 
factor.\footnote{Within the framework of LCSR this could be done by using the a two-pion DA \cite{Diehl:1998dk,Polyakov:1998ze,Kivel:1999sd,Diehl:2003ny}.
The technology for pursuing a $B \to \pi \pi$  FF computation
 on the lattice has been put forward recently in reference
\cite{Briceno:2014uqa}.} The analysis of $B \to \rho  \ell \nu$  therefore becomes 
a matter  of how background, finite width and $S$,$P$-wave effects are treated or discerned. This question arises in any theoretical computation as well as in any experimental measurement. It is therefore important that both theory and experiment treat these issues in a consistent way.

Let us contrast the  $\rho$ final state with the $\pi\pi$-state from a pragmatic viewpoint relevant for this paper.
The orbital angular momentum of the  $\pi\pi$-state is either  $l =0,1,..$ ($S,P,..$-wave).
If the $\pi\pi$-state originates from a $\rho$-meson then it is necessarily in a $P$-wave state and shows a distinct angular distribution. 
Hence this contribution can be separated through an angular analysis from other type of partial waves.\footnote{The importance of separating the $S$-wave, in the context of $B \to K^*\ell \ell$-type decays, was emphasised not long ago in  \cite{Becirevic:2012dp}. 
Thereafter the $S$-wave FFs $B \to (K\pi)_\mathrm{S-wave}$ were computed in LCSR  in the tree level approximation \cite{Meissner:2013hya}.}
We therefore conclude that the $S$-wave contribution is not to be included 
in a $B \to \rho$ FF  and we therefore do not need to attribute any additional uncertainty to it.

We turn to the question of the treatment of the $P$-wave.  
For the sake of concreteness we  discuss the $B \to J/\Psi \pi^+\pi^-$ measurement of the LHC collaboration \cite{Aaij:2014siy}.
%From the viewpoint of a $\rho$-meson there are the issues of the 
%finite width and the non-resonant $\pi\pi$ $P$-wave background.
In a certain window around $m_{\pi\pi}^2 \simeq m_\rho^2$  
the $\pi\pi$-spectrum, in the $P$-wave, is fitted through an ansatz  
of a resonant $\rho$ and the two excited 
 $\rho(1450)$ and $\rho(1700)$ states. 
The main fit parameters are the amplitudes (complex residues) of the Breit-Wigner 
ansatz whose values determine the interference pattern. A non-resonant $P$-wave background is usually not fitted for since it is  assumed that the $S$-wave
is dominant in the non-resonant background.\footnote{In the cases where a background has been searched for in $B \to \rho \ell \nu$, 
it has been found to be consistent with zero  \cite{Behrens:1999vv,Adam:2007pv}. 
%Sibidanov:2013rkk} 
Whether or not future experiments can discern the background is difficult for us to judge but 
we  argue that from a pragmatic point of view this might not be necessary.} 
%Our main point of this section is that the very same is the case for 
%the $B \to \rho$ FFs from LCSR and that the treatment is therefore consistent in practice.  
This raises the question of how a theoretical framework like LCSR can accommodate 
this complex procedure. The answer is surprisingly pragramatic. As long as the 
experimental input into the LCSR is treated in the same way there is no systematic effect.

Let us argue this point in some more detail. 
In current LCSR determinations of the $B \to V$ FFs the $V$-meson 
is described by  vector meson DAs. 
The latter are mainly characterised by the longitudinal and transverse 
decay constants $ f_{V}^\parallel$ and $f_{V}^\perp$  respectively and it is therefore 
important to know how they are obtained.
The method of choice for determining $f^\parallel_{V}$  is experiment: 
$e^+ e^- \to V^0(\to PP)$ (for $V^0 = \rho^0,\phi,\omega$) 
and  $\tau^+ \to V^+(\to PP)  \nu$ (for $V^+ = K^{*,+}, \rho^{+}$) cf.~appendix  \ref{app:decay}.  
As long as the experimental treatment of the $\rho$-meson versus 
$\pi\pi$-signal event is the same as for the semileptonic decay $B \to (\rho \to \pi\pi) \ell \nu$
the decay constant encodes the same definition of the $\rho$-meson as used 
in $B \to \rho \ell \nu$.  The other quantities associated with the $\rho$-DA are not 
directly accessible in experiment. For example 
the transverse decay constant $f_{V}^\perp$  is obtained  through ratios of $f^\parallel_V /f_V^\perp$ from lattice QCD and sum rules where one would 
expect effects of the treatment of the $\rho$-meson  to cancel to a large extent or to be 
taken care of by the respective uncertainties.

We may discuss the same reasoning from the viewpoint
 of a computation using 
a two-pion DA instead of a $\rho$ DA. 
Let us consider for example the contribution of the DA that couples to the vector current. The latter is described 
at lowest order in the conformal spin expansion by $f_\rho^\parallel$ times the  
asymptotic DA which follows from conformal symmetry. From the formulae in \cite{Polyakov:1998ze} it is seen that the analogue
of the two-pion DA is given by the pion FF $f_+^{(\pi)}(q^2)$ times the asymptotic DA. Somewhat symbolically the transitions in terms of $\rho$ and 2-pion DAs 
are  given by 
\begin{alignat}{3}
\label{eq:rho}
& \mathrm{vector }\;\rho\mathrm{-DA:}  \qquad  & &  f_\rho^\parallel \cdot \mathrm{BW}( m_{\pi\pi}^2)  
&\;\to\;& \frac{f_\rho^\parallel m_\rho  g_{\rho \pi\pi}}{m_{\pi\pi}^2 - m_\rho^2 - i m_\rho \Gamma_\rho} 
+ ... \;,  \\
\label{eq:pipi}
& \mathrm{vector }\;\pi\pi\mathrm{-DA:} & &  f_+^{(\pi)}( m_{\pi\pi}^2 ) &\;\to\;&    \frac{f_\rho^\parallel m_\rho g_{\rho \pi\pi}}{m_{\pi\pi}^2  - m_\rho^2 - i m_\rho \Gamma_\rho} +  ... \;,
\end{alignat}
for $ m_{\pi\pi}^2 \simeq m_\rho^2$ where $g_{\rho \pi \pi}$ is the $\rho \to \pi \pi$ decay constant,  BW stands for some type of Breit-Wigner ansatz and the dots stand for 
all  contributions other than the $\rho$-resonance from the  $\pi\pi$ $P$-wave. Our argument is that unless one is specifically interested in the local $m_{\pi\pi}^2$-behaviour this contribution can and is effectively absorbed into $f_\rho^\parallel$ upon integration over the $\rho$ mass window 
in the experimental analysis. For higher order conformal spin corrections, i.e. 
higher Gegenbauer moments, and other decay constants the same reasoning applies.  
The strong rescattering phases in the $\pi\pi$-channel are universal in each partial wave 
and do not distort the result.

In summary, from a pragmatic viewpoint as long as the treatment of the $\rho(\to \pi\pi)$-meson is the same that is used for the extraction of $f_\rho^\parallel$, the LCSR should 
not suffer from sizeable additional  uncertainties.
It therefore seems that in practice the uncertainty is a small fraction of the 
the $P$-wave background  which itself is around $5\%$.\footnote{Despite this aspect it is of interest to estimate the non-$\rho$ background. One can get an idea by analysing the
 pion FF $f^{(\pi)}_+(q^2) (p_1-p_2)_\mu = \matel{\pi(p_1) \pi(p_2)}{j_\mu^\mathrm{em}}{0}$.
A measure of the non-resonant background around the $\rho$-meson peak 
is given by the difference of the model-independent determination of the pion FF 
using data on $\pi\pi$-scattering phase shifts and the Omn\`{e}s-dispersion relation versus
a fitted $\rho$-meson Breit-Wigner ansatz.
Around the $\rho$-meson mass window the difference is found to be  $~5\%$\cite{GilbertoPeter}. Similar conclusions can  be reached when considering 
the figures in
\cite{Jegerlehner:2011ti} with and without the $\rho(1450)$  and $\rho(1700)$ contributions.
We note that $5\%$ is of the same order as the $S$-wave background found in  
$B \to \pi\pi \ell \nu$ \cite{Meissner:2013pba}.} In view of all other sizeable uncertainties we refrain from adding any further error due to this effect 
and reemphasise the importance of comparing 
our result only with  the $P$-wave contribution of the corresponding $\pi\pi$-pair.
Whereas the analysis in this section questions the practical impact of using a two-pion DA around the 
 the $\rho$-meson mass, it is of course interesting to look at the $B_{\ell_4}$ decay $ B \to \pi\pi \ell \nu$ in other regions of phase space. 
For recent theoretical developments of $B_{\ell_4}$ we refer the reader to \cite{Faller:2013dwa,Kang:2013jaa}  which are though not yet at the level of maturity of  $K_{\ell_4}$
\cite{Colangelo:2015kha}.

\subsection{Input parameters and uncertainties}
\label{sec:choice}

\begin{table}
\addtolength{\arraycolsep}{3pt}
\renewcommand{\arraystretch}{1.5}
$$
\begin{array}{c l l l l l l l }
\hline
 & f^\parallel [\GeV] & f^\perp [\GeV] & a_2^\parallel  & a_2^\perp & a_1^\parallel & a_1^\perp  & \zeta_3^\parallel   \mbox{\cite{Ball:2007rt}} \\
 \hline
\rho & 0.213(5) & 0.160(7) & 0.17(7) & 0.14(6)  & - & -  &  0.030(10) \\
\omega & 0.197(8) & 0.148(13) &  0.15(12) &  0.14(12) & - & -  & \mathrm{idem} \\  
K^*  & 0.204(7)  & 0.159(6) &  0.16(9) & 0.10(8) & 0.06(4) & 0.04(3)  & 0.023(8) \\
\phi & 0.233(4) & 0.191(4) & 0.23(8) &  0.14(7) & - & - & 0.024(8) \\
\hline
\end{array}
$$
%\vspace*{-20pt}
\caption[]{\small 
The determination of $f^\parallel$ is discussed in some detail in appendix \ref{app:decay}.
The fine structure constant $\al$, relevant for the extraction of $\fL$, is evaluated at $\mu = 1 \GeV \sim m_V$.
Scale dependent quantities, e.g.~$f^\perp$, $a_{1,2}^{\parallel,\perp}$ and $\zeta^\parallel_3$, are evaluated  at  $\mu_F = 1 \GeV$.
The parameters $a_{1,2}^{\parallel,\perp}$  are taken to be the same as in \cite{Dimou:2012un} 
which include computations from \cite{Arthur:2010xf,Ball:2005vx,Ball:1996tb,Ball:2007rt,Ball:2006nr}. The $f^\perp$ decay constants are obtained from 
$f^\parallel$ through  ratios $r[X] = f^\perp_X(2 \GeV)/f^\parallel_X$ with 
$r[\rho] = 0.687(27)$, $r[K^*] = 0.712(12)$ and $r[\phi] = 0.750(8)$ from lattice QCD \cite{Allton:2008pn}.  For the $\omega$-meson we adopt $r[\omega] \simeq r[\rho]$ in view of a lack of a lattice QCD determination of this quantity.
Twist-3  DA parameters are taken from  the values for 
$\omega_3^\perp, \omega_3^\parallel$ and $\tilde{\omega}_3^\parallel$ \cite{Ball:2007rt} 
which include $\zeta_3^\parallel$ (quoted in the table), $\omega^\parallel_3$, $\omega_3^\perp$. 
The twist-4 3-particle DA parameters are neglected since they are at the sub per mill domain. Again, 
for the $\omega$-meson we adapt the same values as for the $\rho$-meson 
since a specific determination is lacking.
}
\label{tab:input}
\end{table}

The uncertainty of the LCSR results for the FFs
is determined from the uncertainties on the input parameters, 
the factorisation scale $\mu_F$ and the Borel parameter $M^2$ as well as the effective continuum threshold $s_0$.
The values of input parameters 
used in our calculation, along with the errors assigned can be found in Tab.~\ref{tab:input}.
We draw the reader's attention to the fact that
it is the quantity $[ F(q^2) \cdot f_B]$, where $F$ stands for any of the seven form 
factors, that is determined  from the correlation function. Therefore one needs to divide 
by $f_B$ in order to obtain the FF $F$. 
It is well known and appreciated that the uncertainty in $\al_s$ is considerably reduced 
when sum rule in  $f_B$ is taken to same order as for the quantity $[ F \cdot f_B]$.
 For example $f_B$ increases by ~$\sim 9\%$ at $\mathcal{O}(\alpha_s^2\beta_0)$ whereas
the combination $(f_+^{B \to \pi} f_B)_\text{LCSR}/ (f_B)_\text{SR}$ only increases by $2\%$ 
\cite{Bharucha:2012wy}.
Therefore we make use of the QCD sum rules result at $\mathcal{O}(\alpha_s)$~\cite{Aliev:1983ra,Bagan:1991sg} for $f_B$.

%\subsubsection{Borel parameter and continuum threshold}
%\label{sec:borels0}

The two sum rule specific parameters are the Borel parameter $M^2$ and the effective 
continuum threshold $s_0$. For reasons of consistency the Borel parameter is to be 
chosen at an extremum (cf.~appendix \ref{app:borel}) which serves as a quality control parameter. 
The continuum threshold is more problematic and the final result does depend on the choice.
Hence our recipe for the error analysis is to assume a sizeable uncertainty for 
the continuum threshold.
The new ingredient of our analysis is that we have argued  that the \eom results in correlations between  continuum thresholds of certain FFs; 
(cf.~ appendix \ref{app:corr2} for an elaborate discussion). 
The correlations used are summarised in and in between   Eqs.~\eqref{eq:corr1} and \eqref{eq:corr2}. 
The correlation of the continuum thresholds  
are such that the relative uncertainty is $1 \GeV^2$  which has to be compared to the individual uncertainty of 
$2\GeV^2$ or the uncertainty of their sum which is close to $4  \GeV^2$. 
%The small deviations from these constraints are covered by the variation of $s_0^{\perp,0}$  themselves.  
The influence of the Borel parameter on the light-cone sum rule is negligible as compared 
to the continuum threshold of the light-cone sum rule and we therefore do not vary them separately 
for each FF. The Borel parameter dependence of the $f_B$-sum rule is sizeable and is taken 
into account and contributes to the uncertainty of the normalisation of the FFs.
The intermediate states for the light-cone and the $f_B$ sum rule are the same since they couple to the same interpolating current $J_B$. 
It would therefore seem absurd, or contradictive to the method, if the corresponding 
continuum thresholds were far apart. We implement this reasoning by correlating 
 $s_0^{ f_B}$ and $s_0^\mathrm{LC}$
  at the $50\%$-level which implies
that the uncertainty on the difference is $2\GeV^2$; a factor of $\sqrt{2}$ lower than without correlations.

We turn to the choice of the actual central values of the continuum threshold and the Borel parameter.
It is useful to recall that if the sum rules were perfect then the LCSR FF would be independent of 
the Borel parameter. In reality a small Borel parameter is desirable from the viewpoint of suppressing 
any higher states in the spectrum whereas a large Borel parameter improves the convergence 
of the  light-cone operator product expansion (LC-OPE). In practice one therefore chooses a compromise value which is usually found 
as an extremum. The flatness of the FF around this extremum as a function of the Borel 
parameter is a measure of the quality of the sum rule. 
In appendix \ref{app:borel} it is shown that extremising in the Borel parameter is formally  equivalent 
to imposing a daughter sum rule for $m_B^2$.
From the viewpoint of the physics, the  effective continuum threshold  is expected to lie somewhere between
 $(m_B+ 2m_\pi) \simeq 30.9 \GeV^2 $ and $(m_B + m_\rho)^2  \simeq 36.6 \GeV^2 $ with the true value being 
 closer to the latter since the production of a $\rho$-meson is much more likely than the production of two 
 non-resonant pions. 
 The twist-4 contribution for $\pm$-helicity ($T_{1,2}$, $V$, and $A_1$) 
 is around $5\%$ whereas for $0$-helicity ($T_{23}$, $A_{12}$ and $A_0$) they are just below the $10\%$-range.
  Guided by 
the relative size of the twist-2 and twist-3 radiative versus tree contribution\footnote{We remind the reader that the actual impact of the radiative corrections 
on the FF result is considerably smaller since a large part is absorbed by 
the radiative corrections to $f_B$ (cf.~the beginning of this subsection).}
we estimate the uncertainty due to the missing ${\cal O}(\alpha_s)$ 
twist-4 contribution\footnote{More precisely no  ${\cal O}(m_V^2)$ are included at ${\cal O}(\al_s)$.
(cf.~table II in \cite{Ball:2004rg}).  We impose  a  $50\%$ uncertainty on the corresponding tree-level terms.} 
by associating a Gaussian error of $50\%$ to the latter.

In order  to limit contamination due to higher states we verify that the
continuum contribution does not exceed $30\%$. 
If one assumes that semi-global quark hadron-duality itself works at the 
$30\%$-level the additional suppression reduces this error to just below the $10\%$-level. 
The sum rule parameters, with some more details in the caption, are given in Tab.~\ref{tab:SRpara}.

\begin{table}
\addtolength{\arraycolsep}{3pt}
\renewcommand{\arraystretch}{1.6}
$$
\begin{array}{c  c  c  c  c  }
\hline
 B_q    &   M_{f_{B_q}}^2 & s_0^{f_{B_q}} &    M_\mathrm{LC}^2 & s_0^\mathrm{LC} \\ \hline
 B_d & 4.1(1)    & 34.2( 2)    &  c_c/\langle u  \rangle_{q^2} M_{f_{B_d}}^2     &  35(2) \\  
 B_s &  4.4(1)  & 35.4( 2)    &  c_c/\langle u \rangle_{q^2}  M_{f_{B_s}}^2     & 36 ( 2) \\  
 \hline
\end{array}
$$
%\vspace*{-20pt}
\caption[]{\small  Sum rule parameters for $B_d$ and $B_s$ sum rules. All numbers are in units 
of $\GeV^2$, $M_{f_B(\mathrm{LC})}^2$ and $s_0^{f_B(\mathrm{LC})}$ denote the Borel parameter and continuum threshold of the $f_B$ sum rule and the LCSR of $f_B F(q^2)$ (where $F$ stands for a FF) respectively. The difference between the $B_d$ and $B_s$ continuum thresholds follows  
$(m_{B_d} + \Delta)^2 = s_0|_{B_d}$ and $(m_{B_s} + \Delta)^2 = s_0|_{B_s}$. 
The average momentum fraction of the transition quark $\langle u  \rangle_{q^2}$ (cf.~\cite{Ball:2004ye} for the definition) 
varies smoothly from $0.86$ at $q^2 = 0\GeV^2$ to 
$0.77$ at $q^2 = 14\GeV^2$. 
Dividing the sum rule parameter by this quantity serves to take into account $q^2$-dependence  the 
Borel parameter under the extremisation procedure.  The value $c_c = 2.2$  is determined through the mentioned procedure of extremisation. 
The criteria in the text imply that the Borel parameter of the LCSR is considerably higher than that from 
the $f_B$-sum rule  \cite{Ball:2004ye}. 
}
\label{tab:SRpara}
\end{table}

\subsection{Series expansion fits to LCSR form factors}\label{sec:fits}

As mentioned in the introduction, for phenomenological analyses of rare decays, it is crucial to take into account the theoretical uncertainties of the $B\to V$ FFs and the correlations among them. In order to facilitate the use of the LCSR results, we perform fits of the full analytical result to a simplified series expansion (SSE), which is based on a rapidly converging series in the parameter
\begin{equation}
z(t) = \frac{\sqrt{t_+-t}-\sqrt{t_+-t_0}}{\sqrt{t_+-t}+\sqrt{t_+-t_0}} \;,
\end{equation}
where
$t_\pm \equiv (m_B\pm m_V)^2$ and $t_0\equiv t_+(1-\sqrt{1-t_-/t_+})$.
We write the FFs as
\begin{equation}
F_i(q^2) = P_i(q^2) \sum_k \al_k^i \,\left[z(q^2)-z(0)\right]^k\,,
\label{eq:SSE}
\end{equation}
where $P_i(q^2)=(1-q^2/m_{R,i}^2)^{-1}$ is a simple pole corresponding to the first resonance 
in the spectrum. 
The appropriate resonance masses are given in Tab.~\ref{tab:Res}.
\begin{table}
\addtolength{\arraycolsep}{3pt}
\renewcommand{\arraystretch}{1.4}
$$
\begin{array}{ l c  c  c }
\hline
F_i & J^P &  m_{R,i}^{b \to d}/\GeV & m_{R,i}^{b \to s}/\GeV  \\ \hline
A_0 & 0^-  & 5.279 & 5.366 \\
T_1, V & 1^- & 5.325 & 5.415 \\
T_{2}, T_{23}, A_{1} , A_{12} & 1^+ & 5.724 & 5.829\\
\hline
\end{array}
$$ 
%\vspace*{-20pt}
\caption{
Masses of resonances of quantum numbers $J^P$ as indicated necessary for the parameterisation of FF $F_i$ for $b\to d$ and $b\to s$ transitions.
}
\label{tab:Res}
\end{table}
We consider fits that are truncated after the quadratic term in $z$, i.e.\ we will have three fit parameters $\al_{0,1,2}$ for each of the seven FFs. We will see in section \ref{sec:lattice} that a three-parameter fit is sufficient for a combined fit to lattice and LCSR results in the entire kinematic range relevant for $B\to V\ell^+\ell^-$ decays.

Note that the parameterisation \eqref{eq:SSE} differs from that used in \cite{Horgan:2013hoa,Horgan:2015vla}. It has the advantage that the value of the FF at $q^2=0$ is among the fit parameters,
$F_i(0)=\alpha_0^i$.
We prefer this parameterisation as it allows to impose the exact kinematical relations
$A_0(0) =  (8 m_B m_V)/(m_B^2 - m_V^2) \,A_{12}(0)$ (which is equivalent to $A_0(0)=A_3(0)$)
and
$T_1(0)=T_2(0)$
at the level of the SSE coefficients as 
\begin{align}
\alpha^{A_0}_0 &= \frac{8 m_B m_V}{m_B^2 - m_V^2} \,\alpha^{A_{12}}_0
\,,
&
\alpha^{T_1}_0&=\alpha^{T_2}_0
\,.
\label{eq:coeffrel}
\end{align}
The results for the FFs at $q^2=0$ are provided in Tab.~\ref{tab:ff0}.
\begin{table}[tp]
\renewcommand{\arraystretch}{1.3}
\centering
\begin{tabular}{lccccc}
\hline
\input{FF-at-zero}
\end{tabular}
\caption{\small Values of the FFs at $q^2=0$ and their uncertainties.  
The tensor FFs are renormalised at the pole mass of the $b$-quark 
$\mu_\mathrm{UV} = 4.8 \GeV$.
For a more detailed error breakdown we refer the reader to the table 7
of the previous LCSR FF work \cite{Ball:2004rg}.}
\label{tab:ff0}
\end{table}
To determine the fit coefficients $\al_i$, the uncertainties, and the correlations between them, we first generate an ensemble of $N=500$ input parameter sets where the values of the input parameters are randomly distributed according to a multivariate normal distribution with the location given by the central values and the covariance given by the uncertainties and correlations of the input parameters discussed above. We then compute all FFs at integer values of $q^2$ between 0 and 14~GeV$^2$.
Finally, we fit the $z$ expansion to all seven FFs for the $N$ ensembles of FF values and extract the mean, variance, and correlation of the $z$ expansion coefficients $\al_{0,1,2}$.
Since we impose the exact conditions \eqref{eq:coeffrel} throughout, the number of independent fit parameters is 19.

The resulting mean and variance are shown in Tab.~\ref{tab:fit}.
We do not reproduce the full $21\times21$ correlation matrices in the paper but rather provide them as downloadable ancillary files which are available on the arXiv preprint page (see appendix~\ref{sec:coeff} for details). Here we merely note that these correlations are sizeable and it is crucial to include them when using the FFs in phenomenological analyses.

With these results at hand, the uncertainty of an  observable $\Psi$  (e.g.~angular observable) can be computed via
\begin{equation}
\sigma^2(\Psi) =
\sum_{k,l,i,j}
\frac{\partial \Psi( F_i) }{\partial \al_k^i}
\,
\text{cov}(\al_k^i,\al_l^j)
\,
\frac{\partial \Psi( F_i) }{\partial \al_l^j}
\,.
\end{equation}
where $i,j=1\ldots7$ denotes the FF index and $k,l=0\ldots2$ parameterises the expansion coefficients 
of the $z$-series. The covariance matrix is defined as
\begin{equation}
\text{cov}(\al_k^i,\al_l^j) = \text{corr}(\al_k^i,\al_l^j) \, \sigma(\al_k^i) \,\sigma(\al_l^j) \qquad\text{(no sums)}
\end{equation}
in terms of the correlation matrix and the variances.  

As an example let us write the formula relevant to the ratio ${\cal R}_1(q^2) = (m_B + m_V)/m_B 
T_1(q^2)/V(q^2)$ whose difference from $1$ marks difference from the large energy 
limit \cite{Beneke:2000wa}. At $q^2 =0$ the error of the FF ratio is given by
\begin{equation}
\sigma\left(\frac{T_1(0)}{V(0)}\right)^2 =
\left(\frac{\alpha_0^{T_1}}{\alpha_0^{V}}\right)^2
\left[
\left(\frac{\sigma(\alpha_0^{T_1})}{\alpha_0^{T_1}}\right)^2
+
\left(\frac{\sigma(\alpha_0^{V})}{\alpha_0^{V}}\right)^2
-
2\frac{\sigma(\alpha_0^{T_1})\sigma(\alpha_0^{V})}{\alpha_0^{T_1}\alpha_0^{V}} \text{corr}(\alpha_0^{T_1},\alpha_0^{V})
\right]
\end{equation}
from which we obtain ${\cal R}_1(0)^{B \to K^*} = 0.97 \pm 0.04 $.

\subsection{Interpolating between lattice and LCSR form factors}
\label{sec:lattice}

The LCSR and lattice FF calculations are complementary since the former is valid at low $q^2$ and the latter at high $q^2$. Performing a combined fit of the SSE parameterisation to both lattice and LCSR results is useful for two reasons. First, whether a good fit to two completely independent methods in two different kinematical regions is possible at all is a powerful consistency check of those methods. Second, in phenomenological analyses constraining physics beyond the SM combining both observables at low and at high $q^2$, one needs a consistent set of FFs for the full $q^2$ range.

To obtain this combined fit, we first generate pseudo-data points with correlated theoretical uncertainties at three $q^2$ values both at low and at high $q^2$. For LCSR at low $q^2$, we proceed as in the previous subsection. For the lattice FFs at high $q^2$, we make use of the parameterisation of lattice FFs provided in \cite{Horgan:2015vla}. We generate an ensemble of series expansion coefficient sets randomly distributed according to a multivariate normal distribution, using the fitted central values and covariance given in \cite{Horgan:2015vla}. For each of the sets, we then evaluate the FFs at the three $q^2$ values and  extract the uncertainties and correlation of these pseudo-data points.

We then construct a $\chi^2$ function
\begin{multline}
\chi^2(\al_0^1,\ldots,\al_2^7)
=\\
+
\sum_{ijkl}
\left[F_\text{LCSR}^i(q^2_k)-F_\text{fit}^i(q^2_k;\al_m^i)\right]
(C_\text{LCSR}^{ijkl})^{-1}
\left[F_\text{LCSR}^j(q^2_l)-F_\text{fit}^j(q^2_l;\al_n^j)\right]
\\
+
\sum_{ijkl}
\left[F_\text{latt}^i(q^2_k)-F_\text{fit}^i(q^2_k;\al_m^i)\right]
(C_\text{latt}^{ijkl})^{-1}
\left[F_\text{latt}^j(q^2_l)-F_\text{fit}^j(q^2_l;\al_n^j)\right]
\end{multline}
where $F_X^i$ are the central values of the pseudo data points of FF $i$ and $C_X^{ijkl}$ the corresponding covariance matrices (taking into account both the correlation between different FFs and different $q^2$ values).
We then sample a likelihood $L=e^{-\chi^2/2}$ using a Markov Chain Monte Carlo (MCMC) approach with flat priors for the series expansion coefficients. From the stationary distribution of the MCMC, we extract the central values and covariance of the coefficients.

\begin{figure}
\includegraphics[width=\textwidth]{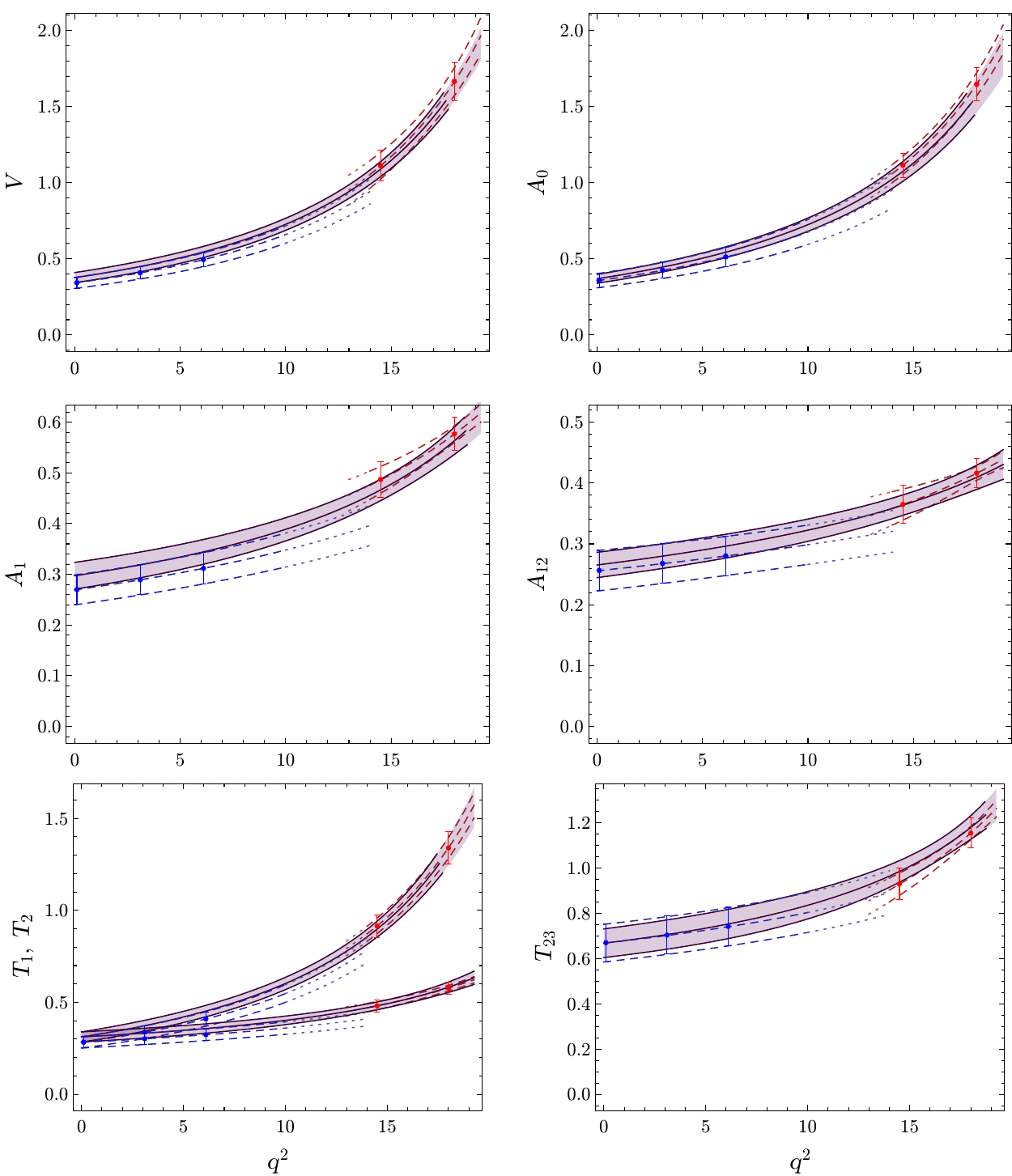}
\caption{\small Combined LCSR  and lattice fit to $B\to K^*$ FFs, where lattice data points are indicated in red, LCSR points in blue, the gray solid band shows the combined 3-parameter fit and the red dashed band the 2-parameter lattice fit from Ref.~\cite{Horgan:2015vla}. In the lower left plot $T_1 > T_2$ for $q^2 > 0\GeV^2$ above .}
\label{fig:lcsr-latt-bks-q2}
\end{figure}

Fig.~\ref{fig:lcsr-latt-bks-q2} shows the fit result for the $B\to K^*$ FFs in the variable $q^2$. 
The  FF plots,  in the $z(q^2)$-variable, for the modes 
$B \to K^*$, $B_s \to \phi$ and $B_s \to \bar K^*$ are shown in Figs.~\ref{fig:lcsr-latt-bks},\ref{fig:lcsr-latt-bsphi},\ref{fig:lcsr-latt-bsks} of appendix    \ref{app:z}.
The LCSR and lattice pseudo-data points are shown in blue and red. The light red dashed band shows the 2-parameter fit from \cite{Horgan:2015vla}. 
The solid gray band is our combined 3-parameter fit result. 
The numerical fit coefficients, of both fits, are reproduced in Tab.~\ref{tab:fit-combined} 
in appendix \ref{sec:coeff}. As for the pure  LCSR fit, the central values, the uncertainties and all correlations are provided as ancillary files on the arXiv preprint page (see appendix~\ref{sec:coeff}).

We would like to add that the fits are valid under the premise that the LCSR and lattice QCD data points and their 
uncertainties, including correlations, are valid as well as the $z$-expansion being a reasonable model function. There is no evidence against the latter, as we have found that adding higher orders in the $z$ expansion and using different parameterisations does not change matters. This is an aspect that could change in the future with more precise  
FF determination from LCSR or lattice QCD.
Overall the agreement is good.
For the central values, we find a $\chi^2$ of
$7.0$ for $B \to K^*$,
$10.2$ for $B_s \to \phi$,
and $19.5$ for $B_s \to K^*$,
for 16 degrees of freedom.
The fits in the $z(q^2)$-variable, shown in Figs.~\ref{fig:lcsr-latt-bks},\ref{fig:lcsr-latt-bsphi},\ref{fig:lcsr-latt-bsks},
are particularly useful in judging the quality of the fits.
In these figures the FFs times  $P=(q^2- m_R^2)$ is plotted.
The latter serves 
to cancel  the first physical pole, at the resonance $R$, in the $q^2$-spectrum. 
The remaining slope therefore 
is a measure of the behaviour of the higher poles or cuts in the $q^2$-spectrum.  

An  interesting qualitative feature is the  behaviour of the $B \to K^*$ versus $B_s \to \phi$ lattice FFs $P \cdot T_{23}(q^2)$ (and to some extent also $P \cdot A_{12}$). 
From  Figs.~\ref{fig:lcsr-latt-bks},\ref{fig:lcsr-latt-bsphi} is seen
that the slopes are  opposite in direction for the
two cases. 
In a LCSR computation, valid at low $q^2$, such a qualitative difference cannot arise since 
the main difference between the FFs for $B \to K^*$ versus $B_s \to \phi$ is due to hadronic input data (which is numerically similar). It is possible that by going closer to 
the hadronic spectrum, at high $q^2$,   a more distinct pattern arises in accordance with  the lattice QCD computation. 
It  will be interesting to see whether this qualitative feature which is not yet statistically significant is confirmed 
in future lattice predictions with higher statistics and a more complete treatment of 
the vector mesons (e.g.~physical quark masses).

To this end we would like to add that differences in normalisation of $\parallel,\perp$ 
($V$, $A_{1}$ and $T_{1,2}$) versus $0$-helicity ($A_0$ $A_{12}$ and $T_{23}$) in the LCSR computation are 
highly sensitive to $\fL$ and $\fT$ decay constants. 
For instance the $0$-helicity FFs depend to $\sim 75\%$ on the normalisation of $\fL$ with the situation being just the opposite for the $\parallel,\perp$-helicity FFs.

\section{Phenomenological applications}
\label{sec:pheno}

We make use of the updated FFs and their error correlations in predicting experimentally accessible observables.
More specifically we consider the $b \to s$ flavour-changing neutral current transitions (FCNC)
$B\to K^*\mu^+\mu^-$, $B\to K^*\gamma$, $B_s\to \phi\mu^+\mu^-$  sensitive to physics beyond the SM 
and the branching fractions of the tree-level decays  $B\to (\rho,\omega) \ell\nu$. 
The latter are of interest to extract the CKM matrix element $|V_\mathrm{ub}|$ and conversely serve as a test of the FF normalisation (and shape)  when 
$|V_\mathrm{ub}|$ is taken as an  input from other channels and global fits.

\subsection{\texorpdfstring{$B\to K^*\mu^+\mu^-$}{B->K*mu+mu-} at low \texorpdfstring{$q^2$}{q2}}
\label{sec:lowKs}

The decay $B\to K^*\mu^+\mu^-$, being one of the golden channels of LHCb,  requires no introduction. 
It has received a great deal of attention, particularly in the last decade. 
By making use of  the large energy relation \cite{Charles:1998dr}, 
observables have been identified which have reduced uncertainties with respect to FFs (e.g~\cite{Descotes-Genon:2013vna}).  
Recent measurements and analyses of several of these observables by LHCb~\cite{Aaij:2013iag,Aaij:2013qta,Aaij:2014pli,Aaij:2015oid}, CMS~\cite{Chatrchyan:2013cda}, ATLAS~\cite{ATLAS:2013ola} and Belle 
\cite{Abdesselam:2016llu} have revealed a number of potential tensions with the SM predictions. 
Whether or not this is due to new physics or hadronic effects 
  is a subject of vital debate \cite{Descotes-Genon:2013wba,Altmannshofer:2013foa,Beaujean:2013soa,Lyon:2014hpa,Altmannshofer:2014rta,Descotes-Genon:2014uoa,Jager:2014rwa}.
This motivates reinvestigation into predictions of hadronic quantities such as the FFs undertaken in this work.

In the SM, neglecting the muon mass, the differential decay distribution of $B\to K^*\mu^+\mu^-$ can be written in terms of  
  of six
   helicity amplitudes 
\begin{align}
\label{eq:ampli}
H^V_\pol  &=  N \sqrt{q^2} \,\left(
C_9^\text{eff}(q^2) \,\V^{(\pol)}(q^2)
+     \frac{2 m_b}{q^2} \,C_7^\text{eff}(q^2)\, T^{(\pol)}(q^2)
+i\,\sqrt{\lambda}\,\frac{2 m_b}{q^2}\,\Delta_\pol(q^2) \right)
\,,\\
H^A_\pol &= N \sqrt{q^2} \, C_{10} \,\V^{(\pol)}(q^2)
\,,
\end{align}
where $\pol=+,-,0$ denotes the polarisation of the $K^*$-meson. 
The helicity FFs $T^{(\pol)},\V^{(\pol)}$  are defined as in appendix~\ref{app:FF-hel} and  $\Delta_\pol$ stands for various 
corrections to be discussed further below.  The quantity 
\begin{equation}
N \equiv V_\mathrm{tb}V_\mathrm{ts}^* \left[\frac{G_F^2 \alpha^2}{3\cdot 2^{10}\pi^5 m_B^3}
\lambda^{1/2} \right]^{1/2},
\end{equation}
%N \equiv \frac{2G_F }{\sqrt{2}}\frac{e^2}{16\pi^2}V_\mathrm{tb}V_\mathrm{ts^*} 
is a normalisation factor where $G_F$ stand for  Fermi's constant, $e$ for the electric charge and $V_\mathrm{tb(s)}$ are  CKM matrix elements. The differential branching ratio is then given by
\begin{equation}
\frac{d\text{BR}(B_q\to K^*\mu^+\mu^-)}{dq^2} = \tau_{B_q}
\frac{1}{2}\sum_{\pol=\pm,0}\sum_{X=V,A}|H^X_\pol|^2
\,.
\end{equation}

Factorisable quark loop contributions are absorbed into  Wilson coefficients $C_7^\text{eff}$ and $C_9^\text{eff}$  which therefore become $q^2$-dependent (e.g.~\cite{Altmannshofer:2008dz} for the definition). 
The  quantities $\Delta_\pol$ contain the  NNLL corrections to the matrix elements of the current-current operators \cite{Asatryan:2001zw} as well as various ``non-factorisable'' contributions.  The latter entail  the effect of weak annihilation,   the chromomagnetic operator contribution both computed in  LCSR  \cite{Lyon:2013gba,Dimou:2012un} as well as hard spectator scattering 
taken from QCD factorisation \cite{Beneke:2001at,Beneke:2004dp}. 

An important contribution arises due to the final state leptons emerging from charm quarks; so called charm loops.  
At high $q^2$ the effect of the broad charmonium resonances measured by the LHCb-collaboration \cite{Aaij:2013pta} has turned out to be substantially more sizeable than anticipated  \cite{Lyon:2014hpa}. 
More precisely, for $\text{BR}(B \to K \mu \mu)$ the resonance residues are found to be $\sim 2.5$ larger with opposite sign from naive 
factorisation indicating sizeable duality violations  \cite{Lyon:2014hpa}.
It therefore seems well-motivated to discuss the various contributions in some detail.
At low $q^2 < 6 \GeV^2$ such effects are thought to be captured in the partonic language of charm quarks and gluons.
The ${\cal O}(\al_s)$ hard vertex corrections  \cite{Asatryan:2001zw} factorise in the heavy quark limit into a $q^2$-dependent function times the FF. The part which does not contain gluon exchanges between the hadron transition and the charm loop factorises non-perturbatively 
by definition and the $q^2$-dependent function is given by the vacuum polarisation. The latter can be extracted in a model-independent way from $e^+e^- \to \mathrm{hadrons} $-data.\footnote{Cf.~\cite{Lyon:2014hpa} for a recent determination.}
 These contributions, as mentioned above, are included in the central values of the predictions of our work.
 In addition there is soft gluon emission from the charm loop into the $B$-meson as well as the $K^*$-meson. 
 Both effects have been assessed in LCSR, the former with a $B$-meson DA  \cite{Khodjamirian:2010vf} and the latter with 
 $K^*$-meson DA (for $B \to K^* \gamma $ only) \cite{Ball:2006eu,Muheim:2008vu}. 
The combination of the two results is not completely free from model-dependence.\footnote{The problem is that the two effects are computed 
in two, slightly, different frameworks. It would be best to compute, in either of the two frameworks, the radiative corrections which would 
then include both effects as well as the ${\cal O}(\al_s)$ vertex corrections. 
This could be a rather challenging as it would seem to require analytic results in order to be verify the dispersion relation.}
At $q^2$ approaching the charmonium resonance region, the contribution is predicted to be significantly enhanced, rendering the partonic theory prediction unreliable above about 6~GeV$^2$. 
These two effects, the  soft gluon emission  and the charmonium effect, 
can be captured in the region $q^2 \in [0,6 \GeV^2]$ by a linear parameterisation
\begin{equation}
\Delta_\pol^{c\bar c}(q^2) =
 \frac{i}{\sqrt{\lambda}}\,C_7^\text{eff}\,T^{(\pol)} \left[ a_\pol e^{i\phi_{a_\pol}} + b_\pol e^{i\phi_{b_\pol}} \left(\frac{q^2}{6\,\text{GeV}^2}\right)\right]
\,,
\label{eq:charm}
\end{equation}
where $a,b$ are positive numbers and $\phi_{a,b}$ are strong phases whose parameter
ranges we discuss further below.
Note that  \eqref{eq:charm} and the replacement of 
\begin{equation}
\label{eq:C7effc}
 C_7^\mathrm{eff}  \to  C_7^\mathrm{eff} \left[1 + a_\pol e^{i\phi_{a_\pol}} + b_\pol e^{i\phi_{b_\pol}} \left(\frac{q^2}{6\,\text{GeV}^2}\right) \right]
\end{equation}
 in \eqref{eq:ampli} are equivalent to each other.
%The factor in front of the brackets is the leading contribution that we just use as a normalisation factor.  
The parameterisation \eqref{eq:charm} is convenient for low $q^2$ since it 
incorporates the helicity hierarchy $\Delta_+\ll\Delta_-$\footnote{We refer the reader 
to the  appendix of  \cite{Dimou:2012un} and \cite{Jager:2012uw} for recent theoretical  discussions of this topic.}
through the FF parameterisation.  This results in  $a_+ \simeq a_-$ and 
$b_+ \simeq b_-$.  We find\footnote{Compared to the parameterisation used 
in \cite{Altmannshofer:2014rta} the value of $ b_0$ is considerably reduced. 
For the observables presented in this paper this effect has a negligible influence on the values of the uncertainty.}
\begin{alignat}{4}
\label{eq:range}
& & a_{\pm} & \in [0,0.05]  \;,\quad  &  b_{\pm} & \in [0,0.2]  \;,  \nonumber \\
& & a_0 & \in [0,0.2] \;,\quad  &  b_0 &\in [0,0.05] \;.
\end{alignat}
where $a_\pol$ is mainly fixed at low  $q^2$ by the  soft gluon emission \cite{Khodjamirian:2010vf,Ball:2006eu,Muheim:2008vu}  and $b_\pol$ is then determined to cover the 
$J/\Psi$ uncertainty.
We vary the phase of the $J/\Psi$-residue in the 
dispersion representation  in the full range motivated by the findings in   \cite{Lyon:2014hpa}. Note that the absolute value of the residues are known from the polarisation specific branching fractions $B \to J/\Psi K^*$. The asymmetry between the parameter values 
of $a_0,b_0$ and $a_\pm,b_\pm$ is due to the  $\pm$ directions being sensitive to the photon pole 
(contrary to the $0$-helicity direction). 
At intermediate $q^2$ this hierarchy disappears which can for example be seen 
from the polarisation fractions of the $B \to J/\Psi K^*$ amplitudes or 
the general result that the  helicity amplitudes are degenerate at the kinematic endpoint \cite{Hiller:2013cza}. In summary the uncertainty due to soft gluon emission and nearby resonances 
is covered by the parameterisation \eqref{eq:C7effc} with parameter ranges as given in 
\eqref{eq:range} and varying the phases $\phi_{a,b}$ in the full range.

%To estimate the uncertainty induced by the charm loop, we vary the phases between 0 and $2\pi$, $a$ between 0 and $0.05$, and $b$ between 0 and $0.2$. In this way, the central value of the parametrization provided in Ref.~\cite{Khodjamirian:2010vf} is within our uncertainty band.

Numerical predictions in different $q^2$ bins
for $B^0\to K^{*0}\mu^+\mu^-$ observables (see e.g. Refs.~\cite{Altmannshofer:2008dz,DescotesGenon:2012zf,Gratrex:2015hna} for definitions of the angular observables) and the $B^+\to K^{*+}\mu^+\mu^-$ branching ratio are given in Tabs.~\ref{tab:b0ks0mm-sm}, \ref{tab:b0ks0mm-sm-p} and \ref{tab:bpkspmm-sm} respectively.
Crucially uncertainties are split into parametric\footnote{The parametric uncertainties, with values adopted from the PDG \cite{Agashe:2014kda}, include $ |V_\mathrm{tb}V_\mathrm{ts}^*|=(4.01\pm 0.10 )\cdot 10^{-2}$, the scale variation $\mu = 4.8 \pm 0.8 \GeV$, the $b$-quark \MSbar mass  $m_b(m_b) = 4.18\pm 0.03 \GeV$  and the pole mass of the charm quark 
$m_c =1.4 \pm 0.2 \GeV$.}, FF 
and non-factorisable charm uncertainties as parameterised in Eq.~(\ref{eq:charm}). 
It is observed that the dominant uncertainty of the branching fraction is due to FFs 
 and amounts to about 20\% relative to the central value.
  In the case of the angular observables the error is considerably  reduced by the inclusion of the correlations.
Comparing the angular observables $S_{4,5}$ with the related observables $P_{4',5'}$ 
it is noted that the FF uncertainties are comparable.  
This improvement for $S_{4,5}$ observables is due to the inclusion of correlated uncertainties in the FFs.  
The error due to the $\Delta$-corrections  results in comparable uncertainties in both bases.
The advantage of using fully correlated errors in explicit computation over general parameterisation can be seen by 
comparing the uncertainties in our work versus those of Ref. \cite{Jager:2014rwa}.

For  comparison of the $B\to K^*\mu^+\mu^-$ observables to existing experimental measurements of 3 $\mathrm{fb}^{-1}$ LHCb data and the implications for new physics, we refer the reader to \cite{Altmannshofer:2014rta}, where the FF results of this work were used for a global analysis of $b\to s$ transitions.

Values of important parameters used for all the SM predictions are given in Tab.~\ref{tab:par}.

\begin{table}
\centering
\begin{tabular}{ccc}
\hline
Parameter & Value & Ref. \\
\hline
$\tau_{B^0}$ & $1.520(4)~\text{ps}$ & \cite{Amhis:2014hma} \\
$\tau_{B^\pm}$ & $1.638(4)~\text{ps}$ & \cite{Amhis:2014hma} \\
$\tau_{B_s}$ & $1.604(10)~\text{ps}$& \cite{Amhis:2014hma}\\
$\Delta\Gamma_s/\Gamma_s\equiv2y_s$ & $0.124(9)$& \cite{Amhis:2014hma}\\
$|V_{cb}|$ & $4.221(78)\times10^{-2}$  & \cite{Agashe:2014kda}\\
$|V_{tb}V^{*}_{ts}/V_{cb}|$ & $0.980(2)$ & \\
\hline
\end{tabular}
\caption{
Numerical inputs for the SM predictions.
\label{tab:par}
}
\end{table}

\begin{table}[tbp]
\begin{center}
\renewcommand{\arraystretch}{1.2}
\begin{tabular}{ccr}
\hline
\multicolumn{3}{c}{$B^0\to K^{*0}\mu^+\mu^-$}\\
\hline
Observable & $q^2$ bin & \multicolumn{1}{c}{SM prediction}\\
\hline
\input{Obs-BKs}
\hline
\end{tabular}
\end{center}
\caption{\small Standard model predictions for binned $B^0\to K^{*0}\mu^+\mu^-$ observables, where the uncertainties are split into parametric uncertainties, FF uncertainties, and our estimate of the uncertainties due to missing hadronic effects.
\label{tab:b0ks0mm-sm}}
\end{table}

\begin{table}
\begin{center}
\begin{tabular}{ccr}
\hline
\multicolumn{3}{c}{$B^0\to K^{*0}\mu^+\mu^-$}\\
\hline
Observable & $q^2$ bin & \multicolumn{1}{c}{SM prediction}\\
\hline
\input{Obs-PBKs}
\hline
\end{tabular}
\end{center}
\caption{\small Standard model predictions for binned angular $B^0\to K^{*0}\mu^+\mu^-$ observables, where the uncertainties are split into parametric uncertainties, FF uncertainties, and our estimate of the uncertainties due to missing hadronic effects.}
\label{tab:b0ks0mm-sm-p}
\end{table}

\begin{table}[tbp]
\begin{center}
\renewcommand{\arraystretch}{1.2}
\begin{tabular}{ccr}
\hline
\multicolumn{3}{c}{$B^+\to K^{*+}\mu^+\mu^-$}\\
\hline
Observable & $q^2$ bin & \multicolumn{1}{c}{SM prediction}\\
\hline
\input{Obs-BRBpKs}
\hline
\end{tabular}
\end{center}
\caption{\small Standard model predictions for the differential branching ratio of $B^+\to K^{*+}\mu^+\mu^-$, where the uncertainties are split into parametric uncertainties, FF uncertainties, and our estimate of the uncertainties due to missing hadronic effects.}
\label{tab:bpkspmm-sm}
\end{table}

\subsection{\texorpdfstring{$B\to K^*\gamma$}{B->K*gamma}}
\label{sec:Ksga}

The precise experimental determination of the branching ratio for $B\to K^*\gamma$ provides a good opportunity to
compare our results for the FFs to experiment.
The branching ratio of $ B\to  K^*\gamma$ is given by
\begin{equation}
\text{BR}(B_q\to  K^*\gamma) =  \tau_{B_q}  48\pi^2  \left(|H_+^q|^2+|H_-^q|^2 \right)
\,,
\end{equation}
where $q=u,d$.  
We have introduced the superscript $q$ in addition to the previous section because 
we give separate predictions for charged and neutral modes.
The amplitudes are related to the limit of the vector helicity amplitudes of $B\to K^*\ell^+\ell^-$ at vanishing dilepton invariant mass,
\begin{align}
H^q_\pm (  B\to  K^*\gamma ) = \lim_{q^2\to0} \frac{q^2}{e}H^{V,q}_{\pm}( B\to  K^*\ell^+\ell^-) \;.
\,
\end{align}

They can be written as
\begin{align}
H_\pm^q &=  \frac{N}{e}\, \sqrt{\la_0} \left(2 m_b \,C_7^\text{eff} \,(T_{\pm}(0)   +i\, 2\, m_b \, \Delta_\pm^q(0)\right) 
\,,
\end{align}
where and $\la_0 = \la|_{q^2=0} = (m_B^2-m_{K^*}^2)^2$ is the K\"all\'en-function for the photon final state and 
$T_{\pm}(0) =  T^{(\pm)}(q^2)/\sqrt{\la(q^2)}|_{q^2=0}$ which results in $T_+(0)=      2 T_1(0)$ and $T_-(0) = 0$.    
The quantity $ T^{(\pm)}$ is defined in appendix~\ref{app:FF-hel}. 
 For $\Delta_\pm(0)$,  the following contributions are included,

 %\begin{align}
% \Delta_-(0) &=  \underset{\text{vertex}}{(0.06 + 0.05 i)} +  \underset{\text{hard scattering}}{(0.03+0.05 i)}
% + \ldots \\
% \Delta_+(0) &=\ldots
% \end{align}
% %
% where \ldots

\begin{itemize}
\item Corrections to the matrix elements of current-current operators \cite{Asatryan:2001zw};
\item Hard scattering contributions computed in QCD factorisation \cite{Beneke:2001at,Beneke:2004dp};
\item Non-factorisable contributions of the chromomagnetic penguin operator $O_8$ computed in LCSR \cite{Dimou:2012un};
\item Weak annihilation computed in LCSR  \cite{Lyon:2013gba}.
\end{itemize}
The first of these corrections is by far the dominant one, leading to a $+60\%$ shift in the branching ratios. 
The three remaining ones contribute to the isospin asymmetry (e.g. \cite{Lyon:2013gba}) of which WA is the one which is most sizeable. 

Our predictions for the branching ratios are listed in Tab.~\ref{tab:BKsgamma} along with the experimental world averages and are  consistent  with the experimental results
at around $1\sigma$. 
We would like to emphasise that the $B \to K^*\gamma$ is a FCNC and that 
the consistency cannot be taken to be one to one with a FF measurement. 
The $B \to (\rho,\omega) \ell \nu$ decays, discussed in section \ref{sec:Vub}, are more
favourable in this respect.
\begin{table}[tbp]
\renewcommand{\arraystretch}{1.2}
\centering
\begin{tabular}{lcc}
\hline
& Theory & Experiment \\
\hline
$10^5\times\text{BR}(B^0\to K^{*0}\gamma)$ & $3.39 \pm 0.14\pm 0.70\pm 0.28$ & $4.33 \pm 0.18$ \\
$10^5\times\text{BR}(B^+\to K^{*+}\gamma)$ & $3.33 \pm 0.13\pm 0.72\pm 0.29$ & $4.21 \pm 0.15$ \\
\hline
\end{tabular}
\caption{\small SM predictions and experimental world averages for the branching ratios of $B^0\to K^{*0}\gamma$ and $B^+\to K^{*+}\gamma$. The theory uncertainty is split into parametric, FF, and non-factorisable power correction uncertainties.}
\label{tab:BKsgamma}
\end{table}

Another cross-check is the branching ratio of the decay $B\to K^*e^+e^-$ at very low $q^2$ that is dominated by the photon pole and that has been measured recently by LHCb \cite{Aaij:2013hha},
\begin{equation}
\text{BR}(B\to K^*e^+e^-)^{\text{30--1000\,MeV}}_\mathrm{exp}
=
(3.1{}^{+0.9}_{-0.8}{}^{+0.2}_{-0.3}\pm0.2)\times 10^{-7},
\end{equation}
where the superscript refers to $\sqrt{q^2}$. An interesting observable is the ratio of this branching ratio to the $B\to K^*\gamma$ branching ratio, since theoretical uncertainties, factorisable or non-factorisable, cancel to a high degree. In the SM, we predict
\begin{equation}
 R_{ee,\gamma} \equiv \frac{\text{BR}(B^0\to K^{*0}e^+e^-)^{\text{30--1000\,MeV}}}{\text{BR}(B^0\to K^{*0}\gamma)}=(6.3\pm0.2)\times 10^{-3} \,,
\end{equation}
where the residual error is dominated by FF uncertainties.
Combining experimental errors in quadrature, from the LHCb measurement and the world average of $\text{BR}(B^0\to K^{*0}\gamma)$, we obtain
\begin{equation}
 R_{ee,\gamma}^\text{exp} = (7.2\pm2.1)\times10^{-3} \,,
\end{equation}
which is consistent with  the prediction, albeit with sizeable uncertainties.
Finally, for the angular observable $F_L$, that has been measured recently in $B^0\to K^{*}e^+e^-$ at low $q^2$ \cite{Aaij:2015dea}, we predict
\begin{equation}
F_L(B\to K^*e^+e^-)^{\text{45--1058\,MeV}}
=
0.203\pm0.003\pm0.058\pm0.017 \,.
\end{equation}
This is in very good agreement with the experimental value,
\begin{equation}
F_L(B\to K^*e^+e^-)^{\text{45--1058\,MeV}}_\mathrm{exp}
=
0.16\pm0.06\pm0.03 \,.
\end{equation}

\subsection{\texorpdfstring{$B_s\to \phi\mu^+\mu^-$}{Bs->phi mu+mu-} at low \texorpdfstring{$q^2$}{q2}}
\label{sec:lowphi}

The decay channel $B_s\to \phi\mu^+\mu^-$ is proceeds via the same quark level transition 
as  $B\to K^*\mu^+\mu^-$ and may serve to compare possible deviations.
An important difference between the two channels is that the $\phi$-meson decays to $K^+K^-$, implying that the decay is not self-tagging in contrast to $B_d\to K^*\mu^+\mu^-$, where the flavour of the initial $B$-meson can be inferred from the charge of the $K\pi$ decay products of the $K^*$. As a consequence, among the observables discussed for $B\to K^*\mu^+\mu^-$, $A_\text{FB}$ and $S_5$ cannot be measured at a hadron collider. 

\begin{table}[tbp]
\begin{center}
\renewcommand{\arraystretch}{1.2}
\begin{tabular}{ccr}
\hline
\multicolumn{3}{c}{$B_s\to \phi\mu^+\mu^-$}\\
\hline
Observable & $q^2$ bin & \multicolumn{1}{c}{SM prediction}\\
\hline
\input{Obs-Bsphi}
\hline
\end{tabular}
\end{center}
\caption{\small Standard model predictions for binned, time-integrated $B_s\to\phi\mu^+\mu^-$ observables, where the uncertainties are split into parametric uncertainties, FF uncertainties, and our estimate of the uncertainties due to missing hadronic effects.}
\label{tab:bsphimm-sm}
\end{table}

Other than CP asymmetries, the most interesting observables are then the differential branching ratio, $F_L$, and $S_4$, in the SM and beyond. 
For these observables, the three possible sources of difference between the results for $B_s\to \phi\mu^+\mu^-$ and those for $B\to K^*\mu^+\mu^-$ are as follows,
\begin{itemize}
\item the FFs are different;
\item differences induced by spectator effects, e.g.~weak annihilation;
\item effects due to the sizeable $B_s$-$\bar B_s$ lifetime difference.
\end{itemize}
The FFs have already been discussed in sections \ref{sec:fits} and \ref{sec:lattice}. The spectator effects turn out to be very small in the SM and are not relevant compared to the FF uncertainties. For a discussion of effects beyond the SM we refer the reader to 
the appendix of Ref.~\cite{Lyon:2013gba}.
The lifetime effects are due to the $B_s$ and $\bar B_s$ lifetime difference of roughly $6\%$ absent for $B_d$-mesons.
This leads to a difference between the observables defined in the absence of neutral meson oscillations, as used in the case of $B_d\to K^*\mu^+\mu^-$, and time-integrated observables, as  measured experimentally  \cite{Descotes-Genon:2015hea}.
This difference has to be taken into account when comparing theory predictions to experimental data.
Details are discussed in appendix~\ref{app:ADeltaGamma}.

In Tab.~\ref{tab:bsphimm-sm}, we list our numerical predictions for the differential branching ratio and angular observables accessible from an untagged measurement $B_s\to \phi\mu^+\mu^-$. The uncertainties are treated in the same 
way as for $B \to K^* \mu^+\mu^-$.

\subsection{\texorpdfstring{$R_{K^* \phi}$: $B \to K^* \mu^+\mu^-$}{RK*phi: B->K*mumu} versus \texorpdfstring{$B_s \to \phi  \mu^+\mu^-$}{Bs->phimumu}}
\label{sec:versus}

The similarity of the $B \to K^* \mu^+\mu^-$ and $B_s \to \phi  \mu^+\mu^-$ channels 
implies that the uncertainties of ratios of these observables should be strongly reduced.\footnote{In this work we have not performed an error analysis on the ratios themselves. 
The latter would greatly reduce the error and could be undertaken if the experimental central values persist with smaller uncertainties.}
Theory predicts $B_s \to \phi  \mu^+\mu^-$  to have a higher transition than $B \to K^* \mu^+\mu^-$ which essentially comes from the decay constants (cf.~Tab.~\ref{tab:input}) showing this hierarchy. 
At low $q^2$ and for $\phi(K^*)\gamma$ final state (i.e. $q^2 =0$) the central values of 
the LHCb results show the opposite effect.

First, we  recapitulate the prediction for the branching ratio of $B_s\to\phi\gamma$ (see appendix A of Ref.~\cite{Lyon:2013gba} for more details) versus 
$B\to K^*\gamma$. The effect is driven by   $T_1^{B\to K^*}(0)/T_1^{B_s\to \phi}(0)=0.89\pm0.10$\footnote{The central value of this work, which is a more complete update, is $T_1^{B\to K^*}(0)/T_1^{B_s\to \phi}(0) = 0.91$.},  resulting  from the above 
mentioned decay constants, leads to
\begin{equation}
R^{(\ga)}_{ K^*\phi} = \frac{\text{BR}(B^0\to K^{*0}\gamma)}{\text{BR}(B_s\to\phi\gamma)} = 0.78 \pm 0.18
 \,,
\end{equation}
which is roughly $1.5$ standard deviations below the LHCb measurement for this ratio, $1.23\pm0.32$~\cite{LHCb:2012ab}.
Such a deviation can, of course, not be  regarded as statistically significant.

\begin{figure}
\centering
\includegraphics[width=.94\textwidth]{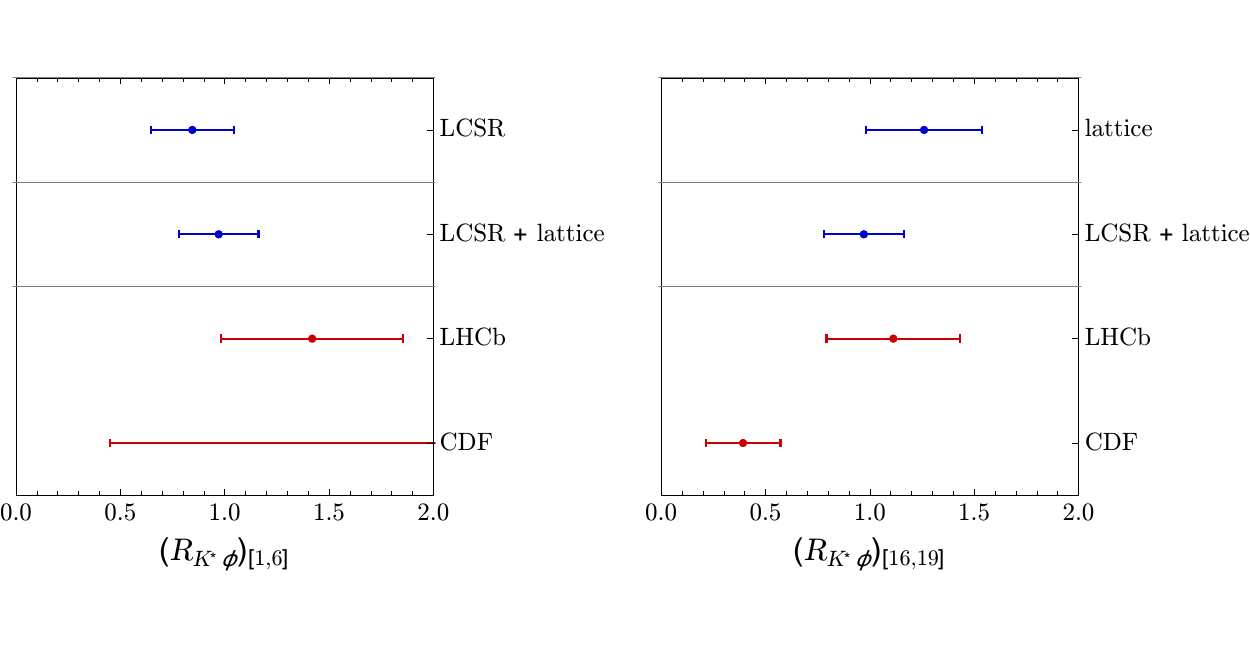}
\caption{\small Our predictions for $R_{\phi K^*}$ at low and high $q^2$ using LCSR, Lattice and a combination of the two, compared to experimental results from LHCb with
and integrated luminosity of 1~fb${}^{-1}$~\cite{Aaij:2013iag,Aaij:2013aln} and CDF~\cite{CDFupdate}.}
\label{fig:RKsphi}
\end{figure}

A similar ratio can also be considered for the decay to leptons,
\begin{equation}
 R_{ K^*\phi}[q_1,q_2]\equiv\frac{d\text{BR}(B^0\to K^{*0}\ell^+\ell^-)/dq^2|_{[q_1,q_2]}}{d\text{BR}(B_s\to\phi\ell^+\ell^-)/dq^2|_{[q_1,q_2]}}, 
\end{equation}
by considering ratios of the differential branching ratios integrated over specified ranges in $q^2$.
We show a graphical comparison of our predictions using LCSR, lattice and combinations of the two for the ratio $R_{\phi K^*}$ to the results of LHCb~\cite{Aaij:2013iag,Aaij:2013aln} and CDF~\cite{CDFupdate} at both low and high $q^2$ in Fig.~\ref{fig:RKsphi}.  
Again, the results per se are not statistically significant. 
On the qualitative level it is though interesting that the theoretical and the experimental ratio are 
below and above unity respectively.  It is hard to see how the theoretical value can move 
above one, through redetermination of parameters, without uncovering a new physical effect.
We stress once more that we have not undertaken an analysis with correlated errors 
for this quantity. One could easily expect the theory error to reduce down by a factor of two
 which would result
in $R_{K^*\phi}|_{[1,6]} < 1$ within uncertainties. We are looking forward to the  $3 \mathrm{fb}^{-1}$ results to reexamine this issue. 

\subsection{\texorpdfstring{$|V_\mathrm{ub}|$}{|Vub|} from \texorpdfstring{$B\to(\rho,\omega)\ell\nu$}{B->rho,omega l nu}}
\label{sec:Vub}

FCNC decays such as $B\to K^*\mu^+\mu^-$ and $B_s\to\phi\mu^+\mu^-$ are potentially affected 
by new physics and therefore do not provide an unambiguous  environment to test FF predictions.  
The semi-leptonic decays $B\to( \omega,\rho)\ell^+\nu$ based on the  charged current $ b \to u$ transition occur at tree-level, and are therefore less likely to be affected by new physics and 
serve  to test FF predictions. In particular in view of the current discrepancies between  
the $B\to K^*\mu^+\mu^-$ and $B_s\to\phi\mu^+\mu^-$ branching fraction measurements the normalisation of the FFs per se has become an issue of considerable interest.

The differential branching ratios of these decays, for $m_\ell =0$,  are given by
\begin{align}
\label{eq:Brholnu}
\frac{d\text{BR}(B^{0}\to \rho^{-}\ell^+\nu)}{dq^2}
&=
\tau_{B^{0}}|V_\mathrm{ub}|^2
\frac{G_F^2}{192\pi^3 m_B^3}\sqrt{\lambda} \left(
|\V^{(+)}|^2+|\V^{(-)}|^2+|\V^{(0)}|^2
\right)
\,,\\
\frac{d\text{BR}(B^{+}\to \rho^{0}\ell^+\nu)}{dq^2}
&=
\frac{\tau_{B^+}}{2\tau_{B^{0}}}\frac{d\text{BR}(B^{0}\to \rho^{-}\ell^+\nu)}{dq^2}
\,,
\end{align}
%%
%\begin{align}
%H_\pm &= - (m_B+m_\rho) A_1 \pm \frac{\sqrt{\lambda}}{m_B+m_\rho} V\,,
%&
%H_0 &= -\frac{8  m_B m_\rho}{\sqrt{q^2}}\, A_{12} \,,
%\end{align}
where definitions of $\V^{(\pol)}$ for $\pol=+,-,0$ as well as $\la$, the K\"all\'en-function, can be found in appendix~\ref{app:FF-hel}, with the adaption $m_{K^*} \to m_\rho$. The one for $B^+ \to \omega \ell^+ \nu$ is analogous to $B^+ \to \rho^0 \ell^+ \nu$ with obvious replacements.

The most recent measurements of the branching ratios have been performed for  $B \to \rho \ell \nu$ by
 BaBar~\cite{delAmoSanchez:2010af} and Belle~\cite{Sibidanov:2013rkk} and for $B \to \omega \ell \nu$  by BaBar~\cite{Lees:2012mq,Lees:2012vv}  and Belle~\cite{Sibidanov:2013rkk} respectively.
We extract $|V_\mathrm{ub}|$ from the BaBar and Belle data by minimizing the $\chi^2$ function that reads in both cases
\begin{align}
\chi^2(|V_\mathrm{ub}|)
&=
\sum_{ij}
\left[B_\text{exp}^i-B_\text{th}^i(|V_\mathrm{ub}|)\right]
(C_\text{exp}^{ij}+C_\text{th}^{ij})^{-1}
\left[B_\text{exp}^j-B_\text{th}^j(|V_\mathrm{ub}|)\right]
\,,
\end{align}
where $B_\text{exp}^i$ and $B_\text{th}^i$ are the experimental and theoretical central values for the branching ratios in one $q^2$-bin and the sum runs over all bins for the charged and neutral mode. $C_\text{th}$ is the theoretical covariance matrix that includes in particular the correlated FF uncertainties\footnote{In Ref.\ \cite{Bernlochner:2014ova},
the importance of uncertainty correlations to extract $V_{ub}$ from $B\to\rho\ell\nu$ decays has been emphasized.}.
In the case of Belle, we use the data up to $q^2=8\,\text{GeV}^2$ or $12\,\text{GeV}^2$ and take the full covariance matrix provided in Ref.~\cite{Sibidanov:2013rkk}. The BaBar dataset consists of a single bin 
in the  low-$q^2$ region  from 0 to 8 GeV$^2$ and the correlation between the charged and neutral decay is not provided.
For $B \to \rho \ell \nu$ we obtain 
\begin{alignat}{2}
&  |V_\mathrm{ub}|^{B\to\rho\ell\nu}_{\text{Belle, }q^2<8 \text{GeV}^2} &= (3.36\pm 0.17 \pm 0.34)\times10^{-3} 
\,,\\
&  |V_\mathrm{ub}|^{B\to\rho\ell\nu}_{\text{Belle, }q^2<12 \text{GeV}^2} &= (3.25\pm 0.14 \pm 0.34)\times10^{-3}
\,,\\
&  |V_\mathrm{ub}|^{B\to\rho\ell\nu}_{\text{BaBar, }q^2<8 \text{GeV}^2} &= (2.52\pm 0.42 \pm 0.56)\times10^{-3}
\,,
\end{alignat}
and those from $B^+ \to \omega \ell^+ \nu$  we get
\begin{alignat}{2}
& |V_\mathrm{ub}|^{B\to\omega\ell\nu}_{\text{Belle, }q^2<7 \text{GeV}^2} &= (2.49 \pm 0.34 \pm 0.32)\times10^{-3} 
\,,\\
& |V_\mathrm{ub}|^{B\to\omega\ell\nu}_{\text{BaBar, }q^2<8 \text{GeV}^2} &= (3.25 \pm  0.36 \pm 0.53)\times10^{-3}
\,,\\
& |V_\mathrm{ub}|^{B\to\omega\ell\nu}_{\text{BaBar, }q^2<12 \text{GeV}^2} &= (3.25 \pm  0.29 \pm 0.46)\times10^{-3}
\,,
\end{alignat}%}
where the first error is experimental and the second theoretical.
%Note that the BaBar results do not include the complete data set, and a new analysis is forthcoming.
For the FF $B \to \rho$ and $B \to \omega$ we have taken into account that it is a $b \to u$ and not a $b \to d$ 
transition by scaling the FFs as in \eqref{eq:scale-F}.

\begin{figure}
\centering
\includegraphics[width=0.8\textwidth]{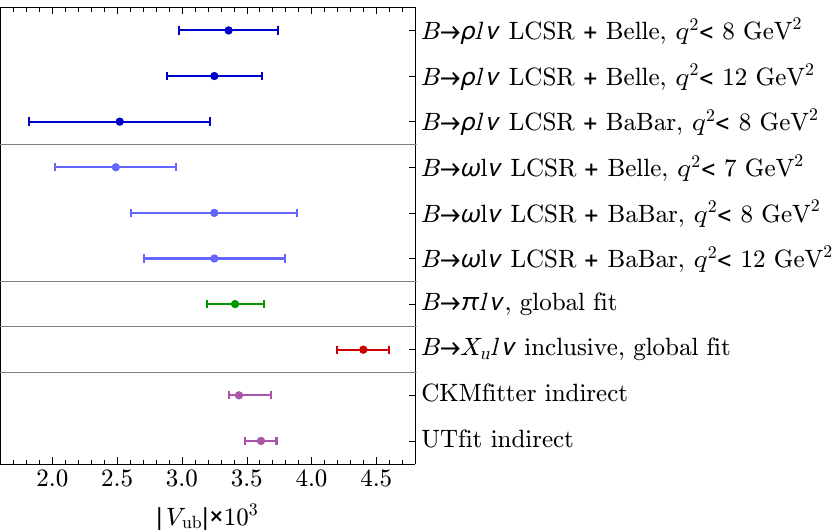}
\caption{\small Our predictions for $|V_\mathrm{ub}|$ from $B\to \rho\ell\nu$ and $B\to \omega\ell\nu$ (blue) compared to global fits to $V_\mathrm{ub}$ from exclusive\cite{Agashe:2014kda} and inclusive channels~\cite{Sibidanov:2013rkk} and indirect determinations from  fits of the unitarity triangle~\cite{Charles:2004jd,Bona:2006ah}.}
\label{fig:vub}
\end{figure}

Our results can be compared to the value extracted from $B\to \pi\ell\nu$ decays, obtained in Ref.~\cite{Sibidanov:2013rkk} from a global fit of BaBar and Belle data to lattice and LCSR computations,
\begin{equation}
|V_\mathrm{ub}|^{B\to\pi\ell\nu} = (3.41\pm 0.22)\times10^{-3} \,,
\end{equation}
or the average of the inclusive semi-leptonic $b\to u$ determinations \cite{Agashe:2014kda}
\begin{equation}
|V_\mathrm{ub}|^\text{incl.} = \left(4.41 \pm  15 ^{+15}_{-17}\right)  \times10^{-3} \,,
\end{equation}
where the first error is experimental and the second error is theoretical.
Finally we also compare our results to the values obtained indirectly from global fits of the CKM matrix \cite{Charles:2004jd,Bona:2006ah},
\begin{align}
|V_\mathrm{ub}|_\text{CKMfitter} &= \left(3.44^{+0.25}_{-0.08}\right)\times10^{-3}
\,,&
|V_\mathrm{ub}|_\text{UTfit} &= \left(3.61\pm 0.12\right)\times10^{-3}
\,.
\end{align}
  The various values for $|V_\mathrm{ub}|$ quoted in this section are summarised graphically  in Fig.~\ref{fig:vub}.  

The $B \to (\rho,\omega) $ FFs do not, and should not, incorporate an $S$-wave contribution
since the  $ (\rho,\omega) \to \pi \pi$ is necessarily in 
a $P$-wave (cf.~section \ref{sec:finite}). Hence the experimental branching ratios might be too large which in turn leads 
to a systematic upward shift  of  $|V_\mathrm{ub}|$ as extracted from these analyses. 
 In Ref.~\cite{Meissner:2013pba} (cf.~Fig.~9 of that reference) this effect has been analysed and it has been found that the integrated line-shapes of the $S$-wave over the interval $[m_\rho - \Gamma_\rho, m_\rho + \Gamma_\rho]$ is around $12\%$
of the corresponding $P$-wave contribution.  This means that if the $S$-wave is neglected 
altogether then we can expect an upward 
shift of $\sim 6\%$ of the $|V_\mathrm{ub}|$ values.  
In the BaBar and Belle analysis the $S$-wave has not been subtracted 
systematically. Hence if the precision below the $10\%$-level is to be reached then the experimental analyses have to perform an angular analysis\footnote{For the 
$\pi\pi$ final state one can make use of the isospin in order to deduce 
whether it is an $S$- or $P$-wave contribution \cite{Behrens:1999vv}.}
 in order to subtract the $S$-wave. 

%This assumption could be tested by doing an angular analysis in the future analyses. 

 Leaving aside the inclusive determinations 
 %and taking into account the issue of the $S$-wave for the BaBar analysis 
 we conclude that our $|V_\mathrm{ub}|$-values from the combined  Belle and BaBar analysis  
  are somewhat lower but surely 
  consistent with $B \to \pi \ell \nu$ determinations as well as the global CKM fits. 
  The values  of $|V_\mathrm{ub}|$ which are considerably lower than the average 
  come with large experimental uncertainties and are consistent at the level of one standard deviation. 
 The uncertainty is rather large and an updated analysis with the full BaBar data set will be more telling.

\section{Conclusions}
\label{sec:conc}

In this paper we present an update of the light-cone sum rules  FFs relevant for the $B \to V$ transitions $B_{d,s}\to K^*,\rho,\omega,\phi$ using new hadronic inputs such as decay constants (appendix \ref{app:decay}), 
the newly determined light-cone DAs   $\mathbb{G}_\perp^{v,a}$ (twist-5) (appendix \ref{app:LCDA}) with explicit results given in appendix \ref{app:T-results} as well as in form of a Mathematica notebook.
 To corroborate the robustness of our predictions, we have discussed in detail the role of the equations of motion in reducing the uncertainties of tensor-to-vector 
FF ratios and mass scheme dependence in section \ref{sec:eom} and appendix 
\ref{app:scheme} respectively. The impact of the $V$-meson being an unstable particle is analysed in section \ref{sec:finite}.

An important point of this work are the easy-to-use numerical expressions of the FFs, provided to the phenomenological 
community, that allow to retain all the uncertainty correlations among the FFs in phenomenological analyses.
This is of particular importance for predicting angular observables that involve ratios of FFs. 
A $z$-expansion fit, Eq.~\eqref{eq:SSE}, to the FFs is provided along  central values, uncertainties, and correlation matrices for the expansion coefficients; available in electronic form as ancillary files on the arXiv webages (see appendix~\ref{sec:coeff} for details 
and Tab.~\ref{tab:fit} for the central values). 
The parameterisation is chosen to transparently fulfil the two exact relations among the FFs at $q^2=0$.
In addition we  performed combined fits to LCSR and lattice computations of the FFs. This serves on  the one hand to obtain predictions for the FFs valid in the full kinematic range, on the other hand as a cross-check of the consistency between the two complementary approaches, as they have to coincide for intermediate $q^2$ values;
good agreement is observed between the two.
Likewise the $z$-expansion coefficients and the correlated uncertainties 
of the combined fits are  downloadable as ancillary files (and central values in Tab.~\ref{tab:fit-combined}).

A phenomenological analysis is performed 
using the updated predictions  
and a new treatment of theoretical uncertainties.
In sections \ref{sec:lowKs} and \ref{sec:Ksga}, we have given updated Standard Model 
predictions for $B\to K^*\mu^+\mu^-$ and $B\to K^*\gamma$ observables, taking into account 
LCSR calculations of several hadronic effects beyond FFs and an
estimate of the uncertainties due to missing hadronic effects, notably  
contributions from charm quarks.  
Potentially relevant long-distance effects which 
have not been fully computed are the complete  $ {\cal O}(\alpha_s)$ 
charm loops effects in one single framework as well as
weak annihilation at    $ {\cal O}(\alpha_s)$. The latter could be sizeable since 
the chiral suppression might be relieved for radiative corrections.
Likewise, in section~\ref{sec:lowphi} we have given predictions for $B_s\to\phi
\mu^+\mu^-$ observables, showing (in appendix~\ref{app:ADeltaGamma}) that the effect of 
the finite $B_s$-lifetime difference is negligible in the Standard Model.
Our predictions are particularly relevant in view of several apparent tensions between Standard Model expectations and experimental measurements observed recently. A crucial question to address in the near future will be whether these tensions are due to underestimated hadronic effects or physics beyond the Standard Model. Our improved FF predictions can play an important role in answering this question.

In section~\ref{sec:Vub},  the new FF predictions were used to extract the CKM element $|V_\mathrm{ub}|$ from BaBar and Belle measurements of $B\to(\rho,\omega)\ell\nu$ decays. Barring some disagreement among the experiments, we find good agreement of our predictions with other exclusive predictions, e.g.~$B\to\pi\ell\nu$ and global fits. 
Our results contribute to the enhancement of the ongoing tension between  
the exclusive and inclusive determination of $|V_\mathrm{ub}|$.  
From another viewpoint the encouraging agreement with other exclusive channels serve 
as a test of the FF normalisation which might become particularly important 
if the disagreement of $B \to K^* \ell^+ \ell^-$ versus $B_s \to \phi  \ell^+ \ell^-$ 
becomes more significant.
Our predictions for $|V_\mathrm{ub}|$ have a relative theory uncertainty at the level of 10\% from $B\to\rho$ and 12--15\% from $B\to\omega$, showing the potential of future, more precise measurements of these semi-leptonic decays to improve the precision on $|V_\mathrm{ub}|$.

We conclude by emphasising that our improved FF predictions 
are important for the tensions in both $b\to s$ channels and the determination of $|V_\mathrm{ub}|$.  
These questions can be further examined with future experimental data to which we look 
forward.

\subsection*{Acknowledgments}

We thank Wolfgang Altmannshofer, Frederik Beaujean, Martin Beneke, Christoph Bobeth, Peter Boyle, 
Vladimir Braun, Gerhard Buchalla, J\'er\^ome Charles, Gilberto Colangelo, Greig Cowan, Luigi Del Debbio, Jochen Dingfelder, Thorsten Feldmann, Gudrun Hiller, Lars Hofer, 
Fred Jegerlehner,  Alex Khodjamirian, Andreas Kronfeld, Vera Luth, Kim Maltman,
Stefan Meinel, Mikolai Misiak, Matthias Neubert, Nils Offen, Steve Playfer, Christoph Schwanda, Peter Stoffer, Javier Virto, Yuming Wang,
and Matthew Wingate for useful discussions. 
R.Z. is particularly grateful to Patricia Ball for collaboration on $B \to V$ form factor computations in the past.
The research of D.S. is supported by the DFG cluster of excellence ``Origin and Structure of the Universe''.

\subsection*{Addendum}

Significant changes from v1 to v2 of this paper are summarised here. 
In appendix \ref{app:decay} the extraction of the (longitudinal) decay constants $\fL$  has been redone, following earlier analyses in \cite{Ball:2006eu,Dimou:2012un}, using updated experimental results and more refined definitions. 
This led to changes in the decay constants, and therefore FFs, of around $2$--$3\%$ depending on 
the vector meson. 
The twist-4 DA $\mathbb{A}_\parallel$, differing from previous results in the literature, and  the twist-5 DAs $\mathbb{G}_\perp^{v,a}$ DAs have been determined and implemented into the calculation cf appendix \ref{app:DAnew}.
Using the latter the tree-level contribution to the FF have been recomputed. 
This led to downward changes of around $4\%$, $6\%$ and $10\%$ for the $\pm$-helicity, $A_{0,12}$ and $T_{23}$ FFs respectively.  The \eom in \eqref{eq:eom1}--\eqref{eq:eom4} have been explicitly checked 
at the  tree-level including in particular the new twist-4,5 contributions.
Explicit results are given in  appendix \ref{app:T-results} 
and in form of a  Mathematica \notebook.
The effect of the finite $B_s$-$\bar B_s$ lifetime difference discussed
in appendix~\ref{app:ADeltaGamma} is now included in all predictions of $B_s\to\phi\mu^+\mu^-$
observables. \\
There are only minor changes from v2 to v3. In particular the numerics remain unchanged. 
Changes include  clarifying remarks in the text and the adaption of the notation 
of the twist-4 DAs in accordance with the literature (cf. footnote \ref{foot:29}).

\appendix
\numberwithin{equation}{section}

\section{Aspects of the LCSR determination of the form factors }
\label{app:LCSR}

\subsection{Equation of motion and correlation functions}
\label{app:corr1}

 The LCSR FFs are computed from a correlation function of the type 
  \begin{equation}
  \label{eq:CGamma}
C[\Gamma]   = i \int d^4x e^{i p_B \! \cdot \! x} \matel{K^*(p)} { T \{  \bar s  \Gamma  b(0) J_B(x)  \} } {0} \;,
\end{equation}
where $\Gamma$ is the Dirac-structure and $J_B \equiv m_b \bar b i \ga_5 q$  an interpolating field for the $B$-meson. 
In fact, the projection on the $B$-meson through a dispersion relation and the Borel transformation can be seen as the substitute of the LSZ-formalism for the $B$-meson. It is well-known that at the level of correlation functions \eom \eqref{eq:eom_op} are corrected by contact terms. 
This results in  
\begin{alignat}{2}
\label{eq:eom_corr}
& q^{\nu} C[i \sigma_{\mu\nu} (\ga_5) ]  + \Delta^{(5)}_\mu &=&  -(m_s\pm m_b) C[\ga_{\mu} (\ga_5) ]  +  C[i (\stackrel{\leftarrow}{\partial} + \stackrel{\rightarrow}{\partial})_\mu (\ga_5) ] - 2 i C [\!\stackrel{\leftarrow}{D}_{\mu} (\ga_5)]  \;,
\end{alignat}
where  $\Delta^{(5)}_\mu $ denote the contact terms.
A heuristic derivation of the contact term follows from the time derivative 
acting on the time ordering of the operators which leads to a commutator expression 
\begin{equation}
\Delta^{(5)}_\mu = -  \int d^3x e^{- i \vec{p_B} \cdot  \vec{x}} \matel{K^*(p,\eta)} { [ \bar s  i \sigma_{\mu0} (\ga_5)  b(0) ,J_B(\vec{x},0)]  } {0} \;.
\end{equation}
Using  the canonical  equal-time commutation relation for the $b$-quarks, $\{ b^\dagger_\al(\vec{x},0),b_\be(0) \} = \delta^{(3)}(\vec{x}) \delta_{\al \be}$, leads to\footnote{In the computation we have assumed that the vector meson is at rest.  The result \eqref{eq:c-cov} is 
the covariantised version. Alternatively we could have derived the contact term directly from the path integral  through field transformations or the (covariant) $T^*$-product. Both of which should directly yield a covariant result.}
\begin{equation}
\label{eq:c-cov}
\Delta_\mu = 0  \text{       and      }    \Delta^{5}_\mu  = -i m_b f_V^\parallel \eta^*_\mu \;.
\end{equation}
The crucial point is  that the contact term is a local term which does not affect the extraction of the FFs at all since it does not enter the dispersion relation. Hence the FFs which are determined from 
the correlation function in LCSR obey the EOM.
More precisely the \eom impose constraints/correlations on Borel parameters and continuum thresholds
of sum rule parameters.

\subsection{Correlation of continuum thresholds and Borel parameters}
\label{app:corr2}

Each correlation function obeys a dispersion relation. 
Using the notation $C[\ga^\mu ] =  C_1[\ga^\mu ]  P_1^\mu + ..$, $C_1[\ga^\mu]$  can be written as
\begin{eqnarray}
\label{eq:disC}
 C_1[\ga^\mu ] = \int^\infty_\text{cut}  ds \frac{\rho_1[\ga^\mu ] (s)}{s-p_B^2-i0}  =
 %\left\{   \begin{matrix}   
   \frac{m_B^2 \V_1(q^2) f_B }{p_B^2 - m_B^2+i0} +  \int^\infty_{s_c}  ds \frac{\rho_1[\ga^\mu ](s)}{s-p_B^2-i0}  \;,
  %  \\  
  % \phantom{\frac{m_B^2 \V_1(q^2) f_B }{p_B^2 - m_B^2} +}   \int^\infty_\text{cut}  ds \frac{\rho^\text{LC-OPE}_1(s)}{s-p_B^2-i0}     \end{matrix}  \right.
\end{eqnarray}
where $s_c$ marks the continuum threshold and ``cut'' stands for the beginning of the discontinuity which is just below $p_B^2 = m_B^2$.  Since \eqref{eq:eom_corr} is valid for any $p_B^2$ it follows from 
the representation \eqref{eq:disC} that the \eom are valid for the densities $\rho_1(s)$ point by point, i.e. locally.  In particular subtracting the FF \eom \eqref{eq:eom1} from \eqref{eq:eom_corr} we obtain
\begin{equation}
q^{\nu} \int^\infty_{s_c}  ds \frac{\rho_1[i \sigma_{\mu\nu} (\ga_5) ] (s)}{s-p_B^2-i0} = - (m_s+m_b) \int^\infty_{s_c}  ds \frac{\rho_1[\ga^\mu ](s)}{s-p_B^2-i0} - 2 i \int^\infty_{s_c}  ds \frac{\rho_1[ \!\stackrel{\leftarrow}{D}_{\mu} ](s)}{s-p_B^2-i0} 
\end{equation} 
for the direction $P_1^\mu$ with  somewhat elaborate  notation.
This is of course true for the exact density as well as for the density $\rho_1^\text{LC-OPE}$ computed from the light-cone OPE.  The semi-global quark-hadron duality, or sum rule approximation, 
consists of replacing the integral on the right-hand side of \eqref{eq:disC} by:
\begin{equation}
\label{eq:sg}
\int^\infty_{s_c}  ds \frac{\rho_1(s)}{s-p_B^2-i0}  \simeq \int^\infty_{s^V_0}  ds \frac{\rho_1^\text{LC-OPE}(s)}{s-p_B^2-i0} \;, 
\end{equation}
where $s^V_0$ is some effective threshold parameter which is expected to lie somewhere between
 $(m_B+ 2m_\pi) \simeq 30.9 \GeV^2 $ and $(m_B + m_\rho)^2  \simeq 36.6 \GeV^2 $.  A simple way to achieve consistency with the \eom \eqref{eq:eom_corr} is 
 to impose $s_0^V = s_0^{T_1} = s_0^{{\cal D}_1}$.  From a physical perspective this is a natural choice since the currents are of identical quantum numbers and therefore couple to the same spectrum of states.
 Below, it is argued that the \eom strengthen this point  implying  a high degree of correlation of  the continuum thresholds.
 
Our main point is that  since ${\cal D}_1 \ll T_1,V$, which we infer from the closeness of $r_\perp$ \eqref{eq:ri} to unity (cf.~Fig.~\ref{fig:IW-like-plots}), 
  a relative difference between 
 $s_0^V$ and $ s_0^{T_1}$
  can only be compensated by a much larger change in $s_0^{{\cal D}_1}$.   
  The latter corresponds to a gross violation of 
  semi-global quark hadron duality which we exclude; partly on grounds of past experience with LCSR.
  
  Let us illustrate this more quantitatively by considering $r$-ratio which are not accidentally close to zero 
  cf.~Fig.~\ref{fig:IW-like-plots}. Let us choose $r_\perp(0) \simeq 0.957$,  
  for $s_0^{V} = 35 \GeV^2$ and $s_0^{T_1} = s_0^{V} \pm 0.5 \GeV^2 $, with fixed Borel parameters,
the \eom \eqref{eq:eom1} requires $s_0^{{\cal D}_1} =  \left( 35_{-4}^{+6}\right) \GeV^2$ which are considerable shifts.  This corresponds to a change in the FF ${\cal D}_1$ of roughly $50\%$ in both directions. 
The situation is similar for the other directions.\footnote{With respect to v1 of our paper the argument is even stronger as the twist-4 contributions do all satisfy the eom.}
\footnote{The main conclusions remain unchanged when other points are chosen. For example 
  $r_\parallel(10 \GeV^2) \simeq  1.022$  requires shifts of  
  $s_0^{{\cal D}_1} = 35 \left(_{+10}^{-5.5}\right) \GeV^2$ for $s_0^{T_2} = 35 \pm 0.5 \GeV^2$ 
  again corresponding to shift of around $50\%$ of ${\cal D}_2(10\GeV^2)$. In fact the $+10\GeV^2$ is only compatible within $10\%$ with the \eom\!\!. The \eom are only satisfied asymptotically for this case. 
  The situation for $r_{0+t}$ is less compelling and requires shift of $\pm 2 \GeV^2$ for an original deviation 
  of $0.5 \GeV^2$ which is still remarkable though.} 
  From this we infer that a difference of  $1\GeV^2$ on the two continuum thresholds $s_0^{T_1}-s_0^{V}$ is at the upper
boundary of what seems plausible.
For $s_0^{T_1,V} = 35(2)\GeV^2$ this can be imposed by correlating the two FFs by 
$7/8$  (i.e. $87.5\%$). The same line of reasoning applies to $r_\parallel$ and $r_{0+t}$ \eqref{eq:ri}.
Yet for $r_{0+t}$ the numerics are less compelling (cf footnote above) and we restrict the correlation between to $50\%$.
There are two further correlations at $q^2=0$, namely $T_1(0)=T_2(0)$ which is of the algebraic type and 
$A_0(0) = A_3(0)$ which is required to avoid an unphysical pole at $q^2 =0$. This leads to 
$s_0^{T_1} = s_0^{T_2}$ and $s_0^{\V_0} = s_0^{A_0}$. Strictly speaking the latter two are only 
exact at $q^2=0$ but since we refrain from assigning a $q^2$-dependence to $s_0$ the relation is assumed throughout. 

In summary the following correlations are imposed,
\begin{alignat}{2}
\label{eq:corr1}
\text{corr}(s_0^{T_1},s_0^V) = 7/8 \;, \quad \text{corr}(s_0^{T_2},s_0^{A_1}) = 7/8 \;, \quad \text{corr}(s_0^{A_{12}},s_0^{T_{23}}) = 1/2 \;,
\end{alignat}
and the full correlations $\text{corr}(s_0^{T_1},s_0^{T_2}) = 1$ and $\text{corr}(s_0^{A_0},s_0^{A_{3}}) = 1$ together with \eqref{eq:corr1} imply 
\begin{alignat}{2}
\label{eq:corr2}
\text{corr}(s_0^{T_1},s_0^{A_1}) = 7/8 \;, \quad \text{corr}(s_0^{T_2},s_0^{V}) = 7/8 \;, \quad \text{corr}(s_0^{A_{12}},s_0^{A_0}) = 1/2 \;.
\end{alignat}
 The reader is reminded  that we have argued in section \ref{sec:choice} for a correlation of the type $\text{corr}(s_0^{F},s_0^{f_B}) = 1/2$ where $F$ stands 
for any FF and $s_0^{f_B}$ is the continuum threshold for the $f_B$ sum rule.

So far we have not discussed the role of the Borel parameter.  In principle one could argue that the Borel parameter 
and the continuum threshold could conspire to satisfy the \eom\!\!\!. Whereas this does not seem to be very viable from the point of 
view of physics it is in addition not credible on grounds of the actual numerics.
For example doubling the Borel parameter of the light-cone sum rule $M^2_\mathrm{LC}$, 
keeping the Borel parameter $M^2_{f_B}$ of the $f_B$ sum rule 
fixed, leads to a change in the FFs $T_1$ and $V$ of just one percent. 
Doubling the sum rule parameter is outside the validity range since it enhances the continuum 
contributions relative to the $B$-pole contribution.  For example for $T_1(0)$ the continuum contribution becomes 
$42\%$ by doubling $M^2_\mathrm{LC}$.
Hence the Borel parameter cannot balance 
a change in the continuum threshold of $s_0$ of $1~\text{GeV}^2$.
Hence it is legitimate not to enter the $M_\mathrm{LC}^2$ in the discussion.
The sensitivity of the $f_B$ sum rule to the Borel parameter $M^2_{f_B}$ is slightly higher 
presumably because the local condensates are more vulnerable to (semi-global) 
quark hadron duality violations. 
This uncertainty is  important for the FF prediction per se but only enters the \eom 
by a global factor and is therefore not relevant for the discussion of this section.
Hence, we fully correlate the uncertainties of the Borel parameters  $\text{corr}(M^2_{f_B},M^2_{F}) = 1$ which is 
justified since the variation in $s_0$ are responsible for the bulk part of the 
uncertainty. It should be added that it is the variation 
of  the parameters $M^2$ and $s_0$ that addresses the validity of the semi-global quark hadron duality.

\subsection{Remarks on the explicit verification of the \eom at tree level}
\label{app:eom-tree}

In view of the importance of the \eom for the determination of the ratio of 
tensor-to-vector FFs we discuss in some more detail  the DAs that enter  the \eom 
and the consistent handling of the projection of the correlation functions on the FFs.
We have explicitly verified the \eom at the tree level  for all  five structures 
appearing in \eqref{eq:eom_corr}  including the derivative FFs as well 
as the strange quark mass terms at twist-$3$. The strange quark mass terms cancel non-trivially 
between the explicit term in \eqref{eq:eom_op} and ${\cal O}(m_s)$-correction in the DAs.  
More detail on the latter can be found in section \ref{app:LCDA}.
The mechanism that guarantees this interplay of light and heavy quark mass are the \eom of the light DAs. 
%The relevant \eom, which can be found in \cite{Ball:1998sk}, is 
%\begin{equation}
%\label{eq:WW}
%\int_0^u dv \left[\phi_\parallel(v) - g^{(v)}_\perp(v)\right] = \bar u\left(g^{(v)}_\perp(u) - \frac{g^{(a)'}_\perp(u)}{4}\right) - \frac{g_\perp^{(a)}(u)}{4}\;.
%\end{equation}
Explicit results are given in the next subsection \ref{app:T-results} and in form of 
a Mathematica notebook. In addition we have verified that the renormalisation of 
the composite operators  (cf.  section \ref{sec:RCOMP}) are compatible with the \eom as expected from first principles. 
We remind the reader that the completely tractable issue of contact terms has been discussed and resolved in 
section \ref{app:corr1}.
Let us add that the covariant derivative between the strange and beauty quark can be exchanged by using the following 
algebraic identity 
\begin{equation}
C[i (\stackrel{\leftarrow}{\partial} + \stackrel{\rightarrow}{\partial})_\mu] - 2 i C [\!\stackrel{\leftarrow}{D}_{\mu}]  = - C[i (\stackrel{\leftarrow}{\partial} + \stackrel{\rightarrow}{\partial})_\mu] + 2 i C [\!\stackrel{\rightarrow}{D}_{\mu}]  \;.
\end{equation}

To this end we would like to discuss 
the consistent handling of the projection onto the structures $P_i^\mu$ \eqref{eq:Vprojectors}.
First, we note that for the \eom to be satisfied the projection on the 
Lorentz structures ought to be handled consistently for  all structures appearing in \eqref{eq:eom_corr}.
For $P_{2,3}^\mu$ extra care is in order, see e.g.~discussion in \cite{Dimou:2012un}, since  $q \! \cdot \! (p+p_B) = p_B^2 - m_V^2 $ equals 
$m_B^2 - m_V^2$ only if $p_B^2$ is on-shell. 
Hence in the computation the following projectors ought to be used
\begin{alignat}{1}
\label{eq:Vprojectors2}
& p_2^\mu =     i \{q \! \cdot \! (p+p_B)  \eta^{*\mu} \mi 
(\eta^*\!\cdot\! q)(p+p_B)^\mu\}  \;, \qquad \nonumber \\
&p_3^\mu =  i(\eta^*\!\cdot\! q)\{q^\mu \mi  \frac{q^2 }{q \! \cdot \! (p+p_B) } (p+p_B)^\mu \}  \;,
\end{alignat}
which we denote by a lower case $p$.  
The important point is that these projectors are transverse $q \cdot p_i = 0$ even in the case 
where $p_B^2 \neq m_B^2$ (off-shell). 
We should add that the actual effect  on the standard tensor and vector FFs  due to the difference of using 
either $p_{2,3}^\mu$ or $P_{2,3}^\mu$ is rather small (numerically around $2\%$)
since the  the sum rule aims at imposing $p_B^2 \simeq m_B^2$ by construction. The latter might be taken 
as a measure of the quality of the sum rule. See also the discussion on the optimisation of the Borel 
parameter given in appendix \ref{app:borel}.

\subsection{Explicit tree level results}
\label{app:T-results}
 
The FF densities $R_{T_{1,2,3},\V_{1,2,3,P}}$, given in appendix \ref{app:T-exresults}  relate to the FFs as follows
\begin{equation}
\label{eq:F}
F(q^2) = c_F  \frac{m_b\,e^{ m_B^2/M^2}}{m_B^2 f_B}\hat{B} \left[ \int_{0}^1     \left(  R_F(u,q^2) \right)  du \right]
\end{equation}
with 
\begin{equation}
c_{T_{1,2,3}} = 1 \;, \quad c_{V,A_1} = - (m_B \pm m_V) \;, \quad  c_{A_{3,0}} = \pm \frac{q^2}{m_V} \;,
\end{equation}
being a matching factor, with a slight abuse of notation, to translate from  $\{ V,A_1, A_3,A_0 \} \leftrightarrow \{\V_1,\V_2,\V_3,\V_P   \}$, cf.~\eqref{eq:VAs}. The symbol $\hat{B}$ denotes the subtracted Borel transformation  explained in the next subsection.
As an example we write 
\begin{eqnarray}
V(q^2) &=& -(m_B+m_V)\frac{m_b\,e^{ m_B^2/M^2}}{m_B^2 f_B}  \hat{B} \left[    \int_{0}^1  du \frac{- f_V^\perp}{2 \Delta} \phi_\bot(u)   + {\cal O}(m_V^2) + ...\right]   \nonumber \\
&=&     (m_B+m_V)\frac{m_b\,}{m_B^2 f_B} \int_{u_0}^1 \frac{du}{u} e^{(m_B^2-h(u,q^2))/M^2} \, 
 f_V^\perp  \phi_\bot(u) + {\cal O}(m_V^2) + ... \;,
\end{eqnarray}
where the dots stand for contributions from other DAs and $h(u,q^2)$ and $u_0$ 
are defined in  Eqs.~\eqref{eq:I1h} and \eqref{eq:u0} respectively.
%with vector meson DA as defined in Eqs \eqref{eq:DAvec}.
Simpler and more definite expression for the DAs $\phi_{\perp,\parallel}$, $g_{v,a}^\perp$ can be found 
in v1 of this paper which fully agree with the current version.
Those  results agree with the expressions given 
in \cite{Altmannshofer:2008dz} for $T_1(0)$, $V(q^2)$ and $A_0(q^2)$ but differs slightly 
 for $A_1(q^2)$  due to the previously discussed  handling of the projections (cf.~section \ref{app:eom-tree}). 
 In practice the results  for each FF are numerically small but for this work is of importance since we are interested in a precise determination of the ratio of tensor-to-vector 
FF. For example $A_1$ differs by a prefactor
$(m_B^2-m_V^2)/(q \cdot (p+p_B)) = (m_B^2-m_V^2)/(p_B^2-m_V^2)  =  (m_b^2 - \bar u q^2 )/(u m_B^2) + {\cal O}(m_V^2)$.  

\subsubsection{The subtracted  Borel transformation}

The correlation functions of the explicit tree-level results are given by integrals of the type
\begin{equation}
I_n = \int_0^1 du \frac{F(u)}{\Delta^n} \;, \quad \Delta \equiv m_b^2 - u p_B^2 - \bar u q^2 \;.
\end{equation}
The subtracted Borel transformation $\hat B$ is defined as the two-fold operation of taking the Borel transformation $B$ followed by the so-called continuum subtraction.
From $\hat{B} [I_1]$ the higher $\hat{B} [I_{n \ge 1} ]$ can be found by
\begin{equation}
\label{eq:rec}
\hat{B} [I_n]  =  \frac{1}{\Gamma[n]} \left( - \frac{d}{d m_b^2}\right)^{(n-1)}  \hat{B} [I_1] \;.
\end{equation}
Hence, if   $\hat{B} [I_1]$ is found explicitly then the problem is solved.
To do so one can write 
\begin{equation}
\label{eq:I1h}
I_1 = \int_0^1 du \frac{F(u)}{\Delta}  =  \int_0^1 \frac{du}{u}  \frac{F(u)}{h(u)-  p_B^2 }  \;, \quad h(u,q^2)= \frac{1} 
{u} ( m_b^2  - \bar{u} q^2) \;,
\end{equation}
and apply the standard Borel transformation 
$B[ 1/(m^2 - p_B^2)] = e^{-m^2/M^2}$ and the subtraction (corresponding to cutting 
of the dispersion integral at $s=s_0$)  is implemented by imposing $u \geq u_0$
\begin{equation}
\label{eq:u0}
\hat{B} [I_1] = \int_{u_0}^1  \frac{ F(u) e^{-\frac{h(u,q^2)}{M^2}} }{u}  du     \;, \quad  u_0 \equiv \frac{m_b^2-q^2}{s_0-q^2} \;.
\end{equation}
 Using formula \eqref{eq:rec}, taking into account the $m_b$-dependence of $u_0$ and $h$, results in 
\begin{eqnarray}
\hat{B} [I_2] &=&  \left( \int_{u_0}^1
 \frac{F(u)
   e^{-\frac{h(u,q^2)}{M^2}}}{M^
   2 u^2} \, du +   \frac{F(u_0) e^{-\frac{s_0}{M^2}}}{u_0
   \left(s_0-q^2\right)}  \right)  \;, \nonumber \\[0.3cm]
 \hat{B} [I_3] &=& \frac{1}{2} \left(\int_{u_0}^1 \frac{F(u)
   e^{-\frac{h(u,q^2)}{M^2}}}{M^
   4 u^3} \, du+
   \frac{
   e^{-\frac{s_0}{M^2}}}{u_0^2
   \left(s_0-q^2\right){}^2}\left(  F(u_0)    \left( 1+ x_{s_0}  \right)  -  u_0 F'(u_0)   \right)
  \right)  \;, \nonumber \\[0.3cm]
  \hat{B} [I_4] &=& \frac{1}{6} \Big( \int_{u_0}^1 \frac{F(u)
   e^{-\frac{h(u,q^2)}{M^2}}}{M^
   6 u^4} \, du+
 \quad   \frac{
   e^{-\frac{s_0}{M^2}}}{u_0^3
   \left(s_0-q^2\right){}^3} \times \\
  & &   \left(   F(u_0) (2+ 2 x_{s_0} +   x_{s_0}^2) - u_0 F'(u_0)  (2 + x_{s_0}) + u_0^2 F''(u_0)   \right) 
 \Big) \;, \nonumber
\end{eqnarray}
where $x_{s_0} \equiv (s_0-q^2)/M^2$. 

To this end we wish to comment on the different techniques 
of Borel transformation. 
Method i): substitute explicit DAs and then integrate over the DA-parameters to obtain analytic functions in 
$p_B^2$ (and $q^2)$, take the discontinuity, obtain the dispersion relation and then perform the
Borel transform on the dispersion relation which corresponds to the standard Borel transformation.
This method has been  pursued for example in  \cite{Ball:2004rg} 
 for the radiative corrections. Method ii): using partial integration rewrite the integrals over the DA-parameters 
 such that they take the form of a dispersion relation and then apply the standard Borel transformation. 
 This method has been applied in \cite{Altmannshofer:2008dz} and the v1 of this paper to 
 present tree-level results of  a few DAs. Method iii): the one described above and used here.   
Methods ii) and iii) have an advantage in that one can substitute other DAs after performing the Borel transformation. It should be added though that with method ii) that for performing the Borel transformation 
assumptions on the endpoint behaviour of the DAs might have been made.

\subsubsection{Explicit tree-level correlation functions}
\label{app:T-exresults}

The relation of the correlation functions $R_F$ (cf.~also Mathematica notebook \notebook) given below to the FFs is given in 
\eqref{eq:F}  and $ \Delta \equiv m_b^2 - u p_B^2 - \bar u q^2$.
The definition of the DAs and the actual form and values chosen are described 
in section \ref{app:LCDA}. We find (with   $R_{T_P} = 0$ by definition) 
\begin{eqnarray*}
\label{eq:RT}
R_{T_1} \!\!\!\!&=&  \frac{m_b f_V^\perp  \left(u
   \bar{u} m_V^2+\Delta \right)\phi _{\bot } }{2 \Delta
   ^2}-\frac{u m_b m_V^2
    f_V^\perp h_{\perp,3}^{\text{(1)}}{}}{2 \Delta
   ^2}+\frac{f_V^\parallel m_V \phi_\parallel^{(1)}}{2 \Delta
   }+  \nonumber \\[0.1cm] & &\frac{f_V^\parallel m_V \left(m_b^2+\Delta
   +q^2\right) \tilde{g}_a^{\bot }}{8 \Delta
   ^2}- 
   \frac{m_b^3 m_V^2 f_V^\perp A_{\bot
   }}{4 \Delta ^3}-\frac{m_b m_V^2 f_V^\perp
   h_\parallel^{\text{(t,2)}}}{\Delta
   ^2}+\frac{u f_V^\parallel m_V g_v^{\bot }}{2
   \Delta }   \;, \nonumber \\[0.3cm]
 R_{T_2} \!\!\!\!  &= & -\frac{u m_b m_V^2 
   f_V^\perp \left(q^2 \left(q^2 \bar{u}-\Delta
    (u-2)\right)+q^2 u
   m_b^2+m_b^4\right)h_{\perp,3}^{\text{(1)}}}{2 \Delta ^2
   \left(q^2
   \bar{u}+m_b^2\right){}^2}+\frac{f_V^\parallel m_V
   \left(m_b^2-q^2\right) \phi_\parallel^{(1)}}{2 \Delta 
   \left(q^2
   \bar{u}+m_b^2\right)}+ \nonumber \\[0.1cm] & & 
    \frac{ \left(q^2
   \left(q^4 \bar{u}^2+\Delta  q^2 u
   \bar{u}+\Delta ^2 u\right)+q^2 m_b^2
   \left(\Delta  u-q^2 \left(u^2-4
   u+3\right)\right)+q^2 (3-2 u)
   m_b^4-m_b^6\right)m_b^3
   m_V^2 f_V^\perp A_{\bot }}{4 \Delta ^3
   \left(q^2
   \bar{u}+m_b^2\right){}^3}  + \nonumber \\[0.1cm] & & 
     \frac{f_V^\parallel m_V
   \left(m_b^2-q^2\right){}^2  \left(q^2 \bar{u}+m_b^2+\Delta
   \right)g_a^{\bot
   }}{8 \Delta ^2 \left(q^2
   \bar{u}+m_b^2\right){}^2}  
   + \frac{u f_V
   m_V \left(m_b^2+q^2\right) g_v^{\bot
   }}{2 \Delta  \left(q^2
   \bar{u}+m_b^2\right)}-
    \nonumber \\[0.1cm] & &
   \frac{m_b m_V^2
   f_V^\perp \left(q^2
   (u-2) m_b^2+m_b^4+q^2 \left(q^2
   (-u)+q^2-\Delta  u\right)\right) h_\parallel^{\text{(t,2)}}}{\Delta
   ^2 \left(q^2
   \bar{u}+m_b^2\right){}^2}+  \nonumber \\[0.1cm] & &
    \frac{m_b
   f_V^\perp \left(q^2 (u-2)
   m_b^2 \left(u \bar{u} m_V^2+\Delta
   \right)+m_b^4 \left(u \bar{u}
   m_V^2+\Delta \right)-q^2 \left(q^2
   \bar{u} \left(u \bar{u} m_V^2+\Delta
   \right)+\Delta  u^3
   m_V^2\right)\right) \phi _{\bot }}{2 \Delta ^2
   \left(q^2 \bar{u}+m_b^2\right){}^2}   \;,  \nonumber \\[0.3cm]
   R_{T_3} \!\!\!\!&=&  -\frac{ \left(q^2 \left(q^2 \bar{u}-\Delta
    (u-2)\right)+q^2 u
   m_b^2+m_b^4\right) u m_b m_V^2  \fT h_3^{\text{(1)}}
    }{2 \Delta ^2
   \left(q^2
   \bar{u}+m_b^2\right){}^2}+ 
   \frac{
   \left(m_b^2-q^2\right) f_V^\parallel m_V \text{$\Phi$g}_v^{\text{(1)}}}{2 \Delta 
   \left(q^2  \bar{u}+m_b^2\right)}+  \nonumber \\[0.1cm] 
 & &   \frac{
   \left(m_b^2-q^2\right){}^2 \left(q^2 \bar{u}+m_b^2+\Delta
   \right)f_V^\parallel m_V g_a^{\bot
   }  }{8 \Delta ^2 \left(q^2
   \bar{u}+m_b^2\right){}^2} +  m_b^3
   m_V^2 \fT A_{\bot } \times  \nonumber \\[0.1cm] 
   & & \frac{ \left(q^2
   \left(q^4 \bar{u}^2+\Delta  q^2 u
   \bar{u}+\Delta ^2 u\right)+q^2 m_b^2
   \left(\Delta  u-q^2 \left(u^2-4
   u+3\right)\right)+q^2 (3-2 u)
   m_b^4-m_b^6\right)}{4 \Delta ^3
   \left(q^2
   \bar{u}+m_b^2\right){}^3}+   \nonumber \\[0.1cm] 
 & &   \frac{ \left(m_b^2+q^2\right)u f_V
   m_V g_v^{\bot
   }}{2 \Delta  \left(q^2
   \bar{u}+m_b^2\right)}-\frac{ \left(q^2
   (u-2) m_b^2+m_b^4+q^2 \left(q^2
   (-u)+q^2-\Delta  u\right)\right)m_b m_V^2
   \fT h_L^{\text{(t,2)}}}{\Delta
   ^2 \left(q^2
   \bar{u}+m_b^2\right){}^2}+  \nonumber \\[0.1cm] 
& &   \frac{ \left(q^2 (u-2)
   m_b^2 \left(u \bar{u} m_V^2+\Delta
   \right)+m_b^4 \left(u \bar{u}
   m_V^2+\Delta \right)-q^2 \left(q^2
   \bar{u} \left(u \bar{u} m_V^2+\Delta
   \right)+\Delta  u^3
   m_V^2\right)\right)m_b
   \fT \phi _{\bot }}{2 \Delta ^2
   \left(q^2 \bar{u}+m_b^2\right){}^2}  \;,
      \end{eqnarray*}
   \begin{eqnarray*}
   \label{eq:RV}
 R_{\V_1} \!\!\!\!  &= & -\frac{f_V^\perp  \left(u
   \bar{u} m_V^2+\Delta \right)\phi _{\bot }}{2 \Delta
   ^2}-\frac{m_b f_V^\parallel m_V \tilde{g}_a^{\bot }}{4
   \Delta ^2}+\frac{m_V^2 \left(2
   m_b^2+\Delta \right) f_V^\perp A_{\bot
   }}{8 \Delta ^3} \;, \nonumber \\[0.3cm]
 R_{\V_2} \!\!\!\!  &= &
\frac{u m_V^2  f_V^\perp
   \left(m_b^2 \left(2 q^2 \bar{u}+3 \Delta
   \right)+\Delta  q^2 \bar{u}+2
   m_b^4\right)h_{\perp,3}^{\text{(1)}}}{2 \Delta ^2 \left(q^2
   \bar{u}+m_b^2\right){}^2}-
   \frac{u m_b
   f_V^\parallel m_V g_v^{\bot }}{\Delta 
   \left(q^2
   \bar{u}+m_b^2\right)}+  \nonumber \\[0.1cm] & & 
   \frac{m_V^2 f_V^\perp
   \left(q^2 (u-2)
   m_b^2+m_b^4+q^2 \left(q^2
   (-u)+q^2-\Delta  u\right)\right) h_\parallel^{\text{(t,2)}}}{\Delta
   ^2 \left(q^2
   \bar{u}+m_b^2\right){}^2}-\frac{u m_V^2
   f_V^\perp \tilde{h}_{\parallel }^{(s)}}{2
   \Delta  \left(q^2
   \bar{u}+m_b^2\right)}- \nonumber \\[0.1cm] & & 
    \left(m_b^2 \left(q^2 (u-2)
   (u \bar{u} m_V^2+\Delta \right)+2
   \Delta  u^2 m_V^2\right)+ 
   m_b^4 \left(u
   \bar{u} m_V^2+\Delta \right)-q^2
   (q^2 \bar{u} \left(u \bar{u}
   m_V^2+\Delta \right)   -  \nonumber \\[0.1cm] & & \Delta  (u-2) u^2
   m_V^2)) \frac{f_V^\perp \phi
   _{\bot } }{2 \Delta ^2
   \left(q^2 \bar{u}+m_b^2\right){}^2}   +  
      (-q^2 m_b^2
   \left(2 q^4 \bar{u}^2+\Delta  q^2
   \left(u^2+2 u-3\right)+3 \Delta ^2
   u\right)- \nonumber \\[0.1cm] & & 
   \Delta  q^4 \bar{u} \left(q^2
   \bar{u}+\Delta  u\right)+q^2 m_b^4
   \left(2 q^2 \left(u^2-4 u+3\right)-3
   \Delta \right)+m_b^6 \left(\Delta +q^2
   (4 u-6)\right)+2 m_b^8) \times \nonumber \\[0.1cm] & &
   \frac{m_V^2
   f_V^\perp A_{\bot }} {8 \Delta
   ^3 \left(q^2
   \bar{u}+m_b^2\right){}^3} \;,   \nonumber \\[0.3cm]
 R_{\V_3} \!\!\!\!  &= &  \frac{ \left(q^2
   \bar{u}+m_b^2-\Delta
   \right)3 m_b^3 f_V^\parallel m_V^3
   A_{\parallel
   }^{\text{(1)}}}{2 \Delta ^4 q^2
   u}+ \frac{
   \left(q^2
   \bar{u}+m_b^2\right)
   m_V^2 \fT h_{\parallel
   }^s{}_{\text{}}}{2
   \Delta ^2 q^2}+  \nonumber \\[0.1cm] & &
   \frac{
   \left(-m_b^2 \left(2 u
   \bar{u} m_V^2+\Delta
   \right)+q^2
   \left(-\bar{u}\right)
   \left(2 u \bar{u}
   m_V^2+\Delta \right)+\Delta
    \left(\Delta +u (2 u-1)
   m_V^2\right)\right)m_b f_V^\parallel m_V
   \text{$\Phi
   $g}_v^{\text{(1)}}}{\Delta
   ^3 q^2 u}+  \nonumber \\[0.1cm] & &
   \frac{
   \left(-2 q^2 \bar{u}-2
   m_b^2+\Delta
   \right)m_b f_V
   m_V^3 g_3^{\text{(2)}}}{\Delta ^3
   q^2}+\frac{
   \left(u \bar{u}
   m_V^2+\Delta \right)
   m_b f_V^\parallel m_V g_{\bot }^v}{\Delta ^2
   q^2} +   \nonumber \\[0.1cm] & &
   \frac{ \left(2
   q^4 \bar{u}+2 q^2 u m_b^2+2
   m_b^4-2 \Delta ^2+\Delta 
   q^2 (3 u-4)\right)m_V^2
   \fT h_{\parallel
   }^{\text{(t,2)}}}{\Delta
   ^3 q^2 u}+  \nonumber \\[0.1cm] & &
   \frac{ \left(2 \Delta 
   u m_V^2-q^2 \left(u \bar{u}
   m_V^2+\Delta
   \right)\right)\fT \phi
   _{\bot }}{2 \Delta ^2
   q^2}-\frac{
   \left(m_b^2+2 \Delta +q^2
   (-u)+q^2\right)m_V^2
   h_3^{\text{(1)}} \fT}{2 \Delta
   ^2 q^2}+ \nonumber \\[0.1cm] & &
   \frac{m_V^2 \left(2
   m_b^2+\Delta \right) \fT
   A_{\bot }}{8 \Delta
   ^3}-\frac{m_b^3 f_V^\parallel m_V^3
   \mathbb{G}_{\bot }^v}{2
   \Delta ^3 q^2}  \;,   \nonumber \\[0.3cm]
 R_{\V_P} \!\!\!\!  &= &
   \frac{m_b f_V^\parallel m_V  \left(m_b^2
   \left(2 u \bar{u} m_V^2+\Delta
   \right)+q^2 \left(-\left(2 u \bar{u}
   m_V^2+\Delta \right)\right)-\Delta 
   \left(\Delta +u (2 u-1)
   m_V^2\right)\right)\phi_\parallel^{(1)}}{\Delta ^3 q^2
   u}-  \nonumber \\[0.1cm] & & 
   \frac{m_b f_V^\parallel m_V \left(u \bar{u}
   m_V^2+\Delta \right) g_v^{\bot
   }}{\Delta ^2 q^2}+\frac{3 m_b^3 f_V
   m_V^3 \left(-m_b^2+\Delta +q^2\right)
   A_{\parallel }^{\text{(1)}}}{2 \Delta
   ^4 q^2 u}+  \nonumber \\[0.1cm] & & \frac{m_V^2
   h_{\perp,3}^{\text{(1)}} f_V^\perp \left(m_b^2+2
   \Delta -q^2\right)}{2 \Delta ^2
   q^2}+   
   \frac{m_b f_V^\parallel m_V^3
   \left(2 m_b^2-\Delta
   +2 q^2\right) g_{\parallel,3}^{\text{(2)}}}{\Delta ^g_{\parallel,3}3 q^2}- \nonumber \\[0.1cm] & & \frac{2
   m_V^2 f_V^\perp \left(-2 q^2
   m_b^2+m_b^4-\Delta ^2+q^4\right)
   h_\parallel^{\text{(t,2)}}}{\Delta ^3 q^2
   u}-\frac{m_V^2 \left(m_b^2+q^2\right)
   f_V^\perp \tilde{h}_{\parallel }^{(s)}}{2
   \Delta ^2 q^2}+ \nonumber \\[0.1cm] & & \frac{m_b^3 f_V^\parallel m_V^3
   \mathbb{G}^v_{\bot }}{2 \Delta ^3
   q^2}-\frac{u m_V^2 f_V^\perp \phi _{\bot
   }}{\Delta  q^2}\;,
   \end{eqnarray*}

\begin{eqnarray*}
R_{{\cal D}_1} \!\!\!\!  &= & -\frac{f_V^\parallel m_V \tilde{g}_a^{\bot }}{4 \Delta }  \;, \nonumber  \\[0.3cm]
R_{{\cal D}_2} \!\!\!\!  &= & \frac{f_V^\parallel m_V \left(q^2-m_b^2\right) 
   \Phi_\parallel^{(1)}}{\Delta  \left(q^2
   \bar{u}+m_b^2\right)}+\frac{u m_b m_V^2
   f_V^\perp h_{\parallel
   }^s{}_{\text{}}}{\Delta  \left(q^2
   \bar{u}+m_b^2\right)}  \;, \nonumber  \\[0.3cm]
   R_{{\cal D}_3} \!\!\!\!  &= &
   \frac{
   \left(m_b^2 \left(6 q^4
   \bar{u}+2 \Delta  q^2
   \bar{u}+\Delta
   ^2\right)-\Delta  \left(\Delta
   ^2+\Delta  q^2 (5-3
   u) -4
   q^4 \bar{u} \right)+6 m_b^4
   \left(\Delta -q^2
   (u-2)\right)-6
   m_b^6\right)}{4 \Delta ^4
   q^2 u} \times   \nonumber \\[0.1cm] & & 
  f_V^\parallel m_V^3 A_{\parallel
   }^{\text{(1)}}  -\frac{ \left(m_b^2 \left(u
   \bar{u} m_V^2+\Delta
   \right)+q^2 \bar{u} \left(u
   \bar{u} m_V^2+\Delta
   \right)-\Delta  u^2
   m_V^2\right)f_V^\parallel m_V g_{\bot
   }^v}{\Delta ^2
   q^2}     +  \nonumber \\[0.1cm] & &
  \frac{f_V^\parallel m_V^3
   \left(m_b^2 \left(4 q^2
   \bar{u}+\Delta \right)+3
   \Delta  q^2 \bar{u}+4
   m_b^4\right) g_3^{\text{(2)}}}{\Delta ^3
   q^2}-\frac{m_b m_V^2 \fT
   \left(q^2
   \bar{u}+m_b^2\right)
   h_{\parallel
   }^s{}_{\text{}}}{\Delta
   ^2 q^2}+  \nonumber \\[0.1cm] & & 
   \frac{
   \left(2 q^2 \bar{u}
   m_b^2-\Delta  \left(\Delta
   -2 q^2 \bar{u}\right)+2
   m_b^4\right)f_V^\parallel m_V^3
   \mathbb{G}_{\bot }^v}{4 \Delta ^3
   q^2}  + \Big(m_b^2 (q^2 (u-2)
   \left(2 u \bar{u}
   m_V^2+\Delta \right) -  \nonumber \\[0.1cm] & &    
   \Delta  \left(u \bar{u}
   m_V^2+\Delta
   \right))+m_b^4
   \left(2 u \bar{u}
   m_V^2+\Delta \right)-\bar u  q^4
   \left(2 u \bar{u}
   m_V^2+\Delta \right) + \Delta ^2 u^2
   m_V^2 +   \nonumber \\[0.1cm] & &   \Delta
    q^2 \left(\Delta +u (2
   u-1) m_V^2-\Delta 
   u\right) \Big) \frac{f_V^\parallel m_V
   \text{$\Phi$g}_v^{\text{(1)}}}{\Delta ^3 q^2 u} 
    \;,  \nonumber \\[0.3cm]
R_{{\cal D}_{P}}  \!\!\!\!  &= & 
\frac{ \left(m_b^2
   \left(u \bar{u}
   m_V^2+\Delta \right)+q^2
   \left(-\left(u \bar{u}
   m_V^2+\Delta
   \right)\right)-\Delta  u^2
   m_V^2\right)f^\parallel_Vm_V
   g_{\bot }^v}{\Delta ^2
   q^2}
   -  \nonumber \\[0.1cm]   & &  
   (-m_b^2 \left(2 q^2
   \left(2 u \bar{u}
   m_V^2+\Delta \right)+\Delta
    \left(u \bar{u}
   m_V^2+\Delta
   \right)\right)+m_b^4
   \left(2 u \bar{u}
   m_V^2+\Delta \right)+q^4
   \left(2 u \bar{u}
   m_V^2   
   +\Delta \right) +
    \nonumber \\[0.1cm]   & &
   \Delta
    q^2 \left(\Delta +u (5
   u-1) m_V^2\right)+\Delta ^2
   u^2 m_V^2 )\frac{f^\parallel_Vm_V \text{$\Phi
   $g}_v^{\text{(1)}}}{\Delta ^3
   q^2 u}+
    \nonumber \\[0.1cm]   & &
   \frac{
   \left(-6 m_b^4 \left(\Delta
   +2 q^2\right)+m_b^2
   \left(-\Delta ^2+6 q^4+10
   \Delta  q^2\right)+6
   m_b^6+\Delta  \left(\Delta
   ^2-4 q^4-3 \Delta 
   q^2\right)\right)
  f^\parallel_Vm_V^3  A_{\parallel
   }^{\text{(1)}}}{4 \Delta
   ^4 q^2 u}   +  
     \nonumber \\[0.1cm]   & &
   \frac{
   \left(m_b^2+2 \Delta
   -q^2\right)m_V^2 f^\perp_V
   h_3^{\text{(1)}}  }{\Delta ^2
   m_b}-\frac{
   \left(\Delta  m_b^2+4
   m_b^4+q^2 \left(5 \Delta -4
   q^2\right)\right)f^\parallel_Vm_V^3
   g_3^{\text{(2)}}}{\Delta
   ^3 q^2}+ \nonumber \\[0.1cm]   & & \frac{
   \left(m_b^4-q^4\right)
 m_V^2  f^\perp_V h_{\parallel
   }^s{}_{\text{}}}{\Delta
   ^2 q^2 m_b}-\frac{4  \left(-2 q^2
   m_b^2+m_b^4-\Delta
   ^2+q^4\right) m_V^2
   f^\perp_V h_{\parallel
   }^{\text{(t,2)}}}{\Delta
   ^3 u m_b} -  \nonumber \\[0.1cm]   & & 
   \frac{
   \left(-2 q^2 m_b^2+2
   m_b^4+\Delta  \left(2
   q^2-\Delta \right)\right)
  f^\parallel_Vm_V^3  \mathbb{G}_{\bot }^v}{4
   \Delta ^3 q^2}-\frac{2 u
   m_V^2 f^\perp_V \phi _{\bot
   }}{\Delta  m_b} \;.
\end{eqnarray*}

\subsection{Scheme dependence of the form factors}
\label{app:scheme}

In many determinations of $B \to V,P$ FF calculation in LCSR 
the pole mass scheme is assumed to be the appropriate scheme for the $b$-quark mass.  
For $B \to \pi$ FFs it has been found that a conversion 
to the \MSbard scheme leads to minor changes only 
\cite{Duplancic:2008ix,Bharucha:2012wy}. 
The explicit appearance of  $m_b$ in the \eom 
\eqref{eq:eom1}-\eqref{eq:eom4} deserves a reinvestigation of the  
issue  of  scheme dependence. 
 
In LCSR calculations one distinguishes between a factorisation scale 
$\mu_F^2 \simeq m_B^2 -m_b^2 \simeq {\cal O}(m_b \Lambda_\mathrm{QCD})$ and a renormalisation scale 
$\mu_\mathrm{UV} = m_b$. 
The former is the separation scale of the LC-OPE and the latter is the scale of the composite operators e.g.~the tensor or vector bilinear quark currents. 
For the analysis in this appendix, and throughout the paper, we adopt the strategy 
to lower $\mu_\mathrm{UV}$ to $\mu_{F}$ in the actual computation and then use 
renormalisation group running to scale the tensor FFs from $T_i(q^2)|_{\mu_\mathrm{UV}=
\mu_F}$ to $T_i(q^2)|_{\mu_\mathrm{UV}= m_b}$. This makes it clear how  the 
\eom are obeyed at any step of the computation.  More details on the renormalisation of the composite operators are given in the next section.

One can switch back and forth between the pole and \MSbard scheme by replacing 
$m_b^\mathrm{pole} =  \bar m_b(\mu_m) Z_m^{\text{\MSbar}}/Z_m^\mathrm{pole} =   \bar m_b(\mu_m) 
(1 + \frac{\alpha_s({\color{blue} \mu})}{4 \pi} C_F (4 - 3 \ln (m_b^2 / \mu_m^2))   + {\cal O}
(\al_s^2)) $ (with $C_F= 4/3$ in QCD) in the tree-level computation and 
expanding to first order in $\al_s$. 
The additional scale $\mu_m$ is introduced (through the \MSbar-scheme)  
 for the same reasons as  $\mu_F$-scale mentioned above.  
In Tab.~\ref{tab:scheme}  examples of  FF determinations in both schemes are given.
We infer that the impact of changing from the pole to the \MSbard scheme for the FFs 
is around $4\%$ which is sizeable but controlled. Yet the ratio of FFs changes by only $1\%$ which is rather small and therefore substantiates the robustness of the tensor-to-vector FF ratio which is one
of the main points of this paper.  
The $\mu$-dependence entering through the $\mu$-dependent \MSbard mass is reflected in 
the pole scheme through a larger uncertainty in the pole mass itself; $m^\mathrm{pole}_b = 4.8(1) \GeV$ as compared to 
$\bar{m}_b(\bar{m}_b) = 4.18(3) \GeV$ \cite{Agashe:2014kda}.

\begin{table}
\addtolength{\arraycolsep}{3pt}
\renewcommand{\arraystretch}{1.3}
$$
\begin{array}{c  c  c  c  c   }
\hline
  B \to K^*     &  \mu^2 [\GeV^2] &  T_1(0) & V(0) & T_1(0)/V(0)  \\ \hline
 \mathrm{pole} &  4.8  & 0.282 & 0.341 & 0.828 \\
 \MSbarm &  4.8  & 0.271 & 0.330 & 0.821 \\
 \MSbarm &  8  & 0.293 & 0.349 & 0.840 \\
 \hline
 \end{array}
$$
\caption{\small As mentioned in the text the tensor 
FFs are understood to be evaluated at the scale $\mu_\mathrm{UV}=m_b$ by one-loop
renormalisation group running. Note $\mu^2 = m_B^2 - m_b^2 \simeq 4.8 \GeV^2$ is the
standard factorisation scale of the LC-OPE used throughout. 
The values are for central values of the input parameters and differ slightly from 
that obtained from the Markov Chain Monte Carlo.}
%\vspace*{-20pt}
\label{tab:scheme}
\end{table}

\subsubsection{Renormalisation of composite operators  and compatibility with \eom}
\label{sec:RCOMP}

The aim of this section is to clarify the renormalisation of the composite operators 
entering the \eom \eqref{eq:eom_op} with particular focus on the $m_b$ quark mass.
The following shorthand notations for the operators
\begin{alignat}{2}
\label{eq:Os}
& O_1 = O_D =  2 \bar s i \!\stackrel{\leftarrow}{D}_{\mu}  b\;, \quad  & & O_2 = O_{\partial T} =   i \partial^\nu (\bar s i \sigma_{\mu \nu} b)  \;,  \nonumber  \\
& O_3 =  O_{mV} =   (m_s+m_b) \bar s \gamma_\mu b   \;, \quad  & & O_4 = O_{\partial S} =  i \partial_\mu (\bar s b) \;,
\end{alignat}
is introduced. 
The mixing matrix is defined by  
\begin{equation}
O_i^{(0)}   = Z_q Z_{ij} O_j \;,
\end{equation}
where $Z_q$ is the external leg or wavefunction renormalisation. 
Through an explicit computation it is found that
\begin{equation}
\label{eq:Z}
Z_{ij} = \delta_{ij} + C_F \frac{\al_s}{4 \pi} \frac{1}{4-d} \left(  \begin{matrix} 2 & 2 & 6 & 6 \\ 0 & 0 & 0 & 0 \\ 0 & 0 & (2-6) & 0 \\ 0 & 0 & 0 & 8    \end{matrix} \right)  \;.
\end{equation}
It is noteworthy that the renormalisation of the operator $O_D$ requires 
the additional diagrams where a gluon originates from the vertex   
through the covariant derivative. The operators $O_{mV,\partial T,\partial S}$ 
do renormalise multiplicatively since they are of lowest dimension (effectively three) 
and differ in  quantum numbers when the contraction of the total derivative is undone.
The operator $O_D$ is  of dimension four and the dimension three operators can  and do mix with $O_D$.  In the notation \eqref{eq:Os}, the operator identity \eqref{eq:eom_op} reads
\begin{equation}
\label{eq:eomO}
O_D + O_{mV} + O_{\partial T} - O_{\partial S} = 0 \;.
\end{equation}
It is readily verified that the renormalisation \eqref{eq:Z} is compatible with the \eom \eqref{eq:eomO}.
As an additional check let us mention that from the diagonal elements 
$Z_q \cdot \mathrm{diag}(Z)  \equiv  (Z_D , Z_{\partial T}, Z_m Z_V,Z_{\partial S})$ 
one infers $Z_{S} \equiv Z_{\partial_S} = 1 + 6 \Delta$, $Z_V = 1$ and $Z_T \equiv Z_{\partial T} = 
1-2 \Delta$ with $\Delta \equiv C_F \frac{\al_s}{4 \pi} \frac{1}{\epsilon} $, $Z_m = 1-6 \Delta$. 
From the latter  the well-known anomalous dimensions $\gamma_S^{(0)} = - 6 C_F $, $ \gamma_V^{(0)} = 0$ and $\gamma_T^{(0)} =2  C_F $ of these operators
follow (notation: $\gamma_X = \ga_X^{(0)}\frac{\al_s}{4 \pi} + {\cal O}(\al_s^2)$).

At last we turn to the issue of the impact of the mass renormalisation on the composite operators. 
From the mixing of operators in \eqref{eq:Z} 
it is clear that the renormalisation of $O_D$ is affected by a mass scheme change. 
This can be seen by writing somewhat symbolically $Z_{13} = Z_{D(mV)} = Z_m Z_{DV}$.
So in summary going to the pole scheme enforces a finite renormalisation of the 
operator $O_D$ since  changing from \MSbar to the pole scheme corresponds 
to a finite shift in the ratio of the $Z_m$-factors.
Most importantly the renormalisation of the composite operators $O_V$, $O_T$ and $O_{ S}$, on the other hand, is not affected by the mass scheme. 
Hence it is legitimate to use the  \MSbard scheme to renormalise them. This is fortunate since the 
 Wilson coefficients are evaluated in the \MSbard scheme and together this guarantees the cancellation 
 of the $\mu_\mathrm{UV}$-scale between the Wilson coefficients and the matrix elements.
 The scheme independence of the operators $O_V$, $O_T$ underlies 
or partly explains the small changes in the 
FFs when going from the pole- to the \MSbard scheme (cf.~Tab.~\ref{tab:scheme}).

\subsection{Remarks on fixing the Borel parameter}
\label{app:borel}

A sum rule for a FF $F(q^2)$ of a process $B \to P,V$-transition may be written as
%\footnote{The relation between the density in Eq.~\eqref{eq:SR} and that in Eq.~\eqref{eq:Fu} is
%$\rho_F(s,q^2)  = \frac{m_b}{m_B^2 f_B}  \bar  \rho_F(u(s),q^2) \frac{(m_b^2-q^2)}{(s-q^2)^2}$ through the change of variable 
% $u(s) = (m_b^2-q^2)/(s-q^2)$.}
\begin{equation}
\label{eq:SR}
F(q^2)_{M^2} = \int_{m_b^2}^{s_0}  \rho_F(s,q^2) e^{\frac{m_B^2-s}{M^2}}  ds \;,
\end{equation}
where $M^2$ is the Borel parameter. 
 The goal of this section is to show that two seemingly 
different methods for fixing $M^2$ are equivalent.  For this purpose we introduce the following 
notation
\begin{equation}
\vev{x(s)}_{q^2,M^2} \equiv  \int_{m_b^2}^{s_0} x(s)  \rho_F(s,q^2) e^{\frac{m_B^2-s}{M^2}}  ds \;.
\end{equation}
We note that $F(q^2)_{M^2} = \vev{1}_{q^2,M^2}$.
The two methods are:
\begin{itemize}
\item \emph{extremising the Borel parameter:} 
If one were to succeed in computing the sum rule exactly, which would imply\footnote{For the sake 
of illustration we employ the
narrow  width approximation which is justified for the $B$-meson.} 
\begin{equation}
\label{eq:exactSR}
 \rho_F(s,q^2) = \delta(s-m_B^2) F(q^2) + \Theta(s - s_c) \sigma_F(s) \;,
\end{equation}
then Eq.~\eqref{eq:SR} would remain valid for any Borel parameter.  
 In practice the partonic evaluation through the OPE is optimised by using a large Borel parameter,
with just the opposite being true  for the projection on the lowest hadronic state. 
Hence a compromise value has to be found, ideally in a region where $F(q^2)_{M^2}$ shows an extremum in $M^2$. 
This is imposed by
\begin{equation}
\label{eq:rule1}
0 = \frac{d}{d(1/ M^{2})} \ln F(q^2)_{M^2} =  \frac{m_B^2 \vev{1}_{q^2,M^2} - \vev{s}_{q^2,M^2} }{\vev{1}_{q^2,M^2}} \;.
\end{equation}

\item \emph{daughter sum rule in $m_B^2$:} 
One may write a daughter sum rule for $m_B^2$ as follows
\begin{equation}
\label{eq:rule2}
(m_B^2)_{M^2} = \frac{\vev{s}_{q^2,M^2}}{\vev{1}_{q^2,M^2}} \;.
\end{equation}
Note that using Eq.~\eqref{eq:exactSR} satisfies \eqref{eq:rule2} exactly as it should.

\end{itemize}
It is readily seen that Eqs.~(\ref{eq:rule1},\ref{eq:rule2}) are the same and hence the two methods are 
equivalent.

\section{Light-cone distribution amplitudes}
\label{app:LCDA}

\subsection{Distribution amplitudes including \texorpdfstring{$\mathbb{A}_\parallel$}{Aparallel} and the
new twist-5 \texorpdfstring{$\mathbb{G}^{v,a}_\perp$}{Gv,aperp}  DAs}
\label{app:DAnew}
 
 A general review on the subject of LCDAs can be found in \cite{Chernyak:1983ej} which is by now over thirty years old. 
 The main concepts for the vector DAs are explained, in some details, in the more modern write-up 
 \cite{Ball:1998sk}.

 Light-cone physics is conventionally discussed by introducing two light-like vectors say $z$ and $\hat{p}$ 
 (i.e. $z^2 = \hat{p}^2 =0 $).\footnote{The latter are often denoted by $n_{\pm}$ or [$n,\bar{n}$]  
with the two remaining directions being labelled by $\perp$.}  
The close to light-like separation $x$ and the meson momentum $p$ ($p^2 = m_V^2$) can be expressed as linear combinations 
of $z$ and $p$ 
\begin{equation}
\label{eq:zp}
 z_\mu =  x_\mu - p_\mu \frac{1}{m_V^2} ( x  p - \sqrt{ (x  p)^2 - x^2m_V^2}) \;, \quad
 \hat{p}_\mu =  p_\mu - \frac{1}{2} \frac{m_V^2}{\hat{p}  z} z_\mu \;,
\end{equation}
whereas any vector such as the vector meson polarisation $\eta(p)$ is decomposed 
\begin{equation}
\label{eq:n}
 \eta_\mu =   \frac{\eta z}{  \hat{p}  z} \hat{p}_\mu  +   \frac{\eta  \hat{p}}{  \hat{p} z} z_\mu + \eta^\perp_\mu \;,
\end{equation}
into $\eta^\perp$ and the $\hat{p}$ and $z$ direction e.g. \cite{Ball:1998sk}. Above $\hat{p}z \equiv  \hat{p} \cdot z $ etc is understood.

 The rigorous definition of the LCDAs is given for quark bilinears with light-like separation  (e.g. \cite{Ball:1998sk}).  
 Applying the decomposition in \eqref{eq:n} to the vector Dirac structure leads to the following parameterisation 
  \begin{equation}
 \label{eq:rigour}
 \matel{V(p)} {\bar{q}_1(0) \ga_\mu \, q_2(z) }{0} = 
  f_V^\parallel m_V
\int\!\! du \, e^{i\bar u \hat{p}z} \left\{ \frac{\eta z}{\hat{p} z}   \hat{p}_\mu\, \phi_\parallel(u)     + \eta^\perp_\mu g^v_\perp(u) -   \frac{\eta z \, m_V^2 }{2 (\hat{p}z)^2} z_\mu \, g_{\parallel,3}(u)  \right\} \;,
 \end{equation}
for a vector meson $V[q_1\bar{q}_2]$ coupling to a light-like 
separated vector quark-bilinear.\footnote{Above the Wilson line between $0$ and $z$, rendering the matrix element gauge invariant, is omitted for brevity.}
Considering all Dirac structures this amounts to a total of
 eight DAs $\phi_{\parallel,\perp}, g_\perp^{v,a}, h_\parallel^{(s,t)}, g_{\parallel,3}$  and $h_{\perp,3}$ which loosely   follow the nomenclature of the nucleon parton distributions functions. 
 It is readily verified, using $\eta p =0$, 
 $\int_0^1 f(u) = 1$ for $f = \phi_\parallel, g_\perp^v, g_{\parallel,3}$ and \eqref{eq:n}, that in the limit 
 $z \to 0$ the left hand side of  \eqref{eq:rigour} 
 reduces to $\eta_\mu f^\parallel_V m_V$ as required.
 Using \eqref{eq:zp} and \eqref{eq:n}
 this can be written in 
 terms of the actual momentum of the vector meson $p$ and the near light-like distance $x$ as 
 follows\footnote{Note that $ \langle V(p) | \bar{q}_1(0) \gamma_5 q_2(x) | 0\rangle = 0$ by parity conservation of  QCD.} \footnote{With due apologies we follow the notation in \cite{Ball:1998sk} and not the newer and more systematic notation introduced in \cite{Ball:2007zt} because of reasons of familiarity.  A dictionary between 
 the two notations is shown in Tab.~\ref{tab:trans}.} \footnote{\label{foot:29} The definition  of 
 $\mathbb{G}^v_\perp$ is adapted such that $\mathbb{A}_\parallel$ remains as in \cite{Ball:1998ff} 
 and ($\phi_\parallel$, $g_\perp^v$) and ($\mathbb{A}_\parallel$, $\mathbb{G}_\perp^v$) take on an analoguous role.}
 \begin{alignat}{2}
\label{eq:DAvec}
& \langle V(p) | \bar{q}_1(0)\gamma_\mu q_2(x)|0\rangle  \! & =\; & f_V^\parallel m_V
\int\!\! du \, e^{i\bar u px} \left\{ p_\mu\,\frac{\eta x}{px} \left(
\phi_\parallel(u) - g^v_\perp(u) + \frac{1}{16}\, m_V^2 x^2
(\mathbb{A}_\parallel(u) -\mathbb{G}^v_\perp(u) )   \right)\right.\nonumber\\[0.1cm]
&  & +&   \left.  \eta_\mu \left(
g^v_\perp(u)  +   \frac{m_V^2 x^2 }{16} \mathbb{G}^v_\perp(u)  \right) -  \frac{\eta x \, m_V^2}{2 (px)^2}\, x_\mu  
\left( g_{\parallel,3}(u)-2g^v_\perp(u) + \phi_\parallel(u)\right)\right\},\nonumber \\
& \langle V(p) | \bar{q}_1(0)\sigma_{\mu\nu} q_2(x)|0\rangle & = &
-if_V^\perp  \int\!\! du \, e^{i\bar u px} \left\{ \left(\eta_\mu p_\nu
- \eta_\nu p_\mu\right) (\phi_\perp(u) +
\frac{1}{16}\, m_V^2 x^2 \mathbb{A}_\perp(u))\right.\nonumber\\
& & & {} + \left( p_\mu
x_\nu - p_\nu x_\mu\right) \frac{\eta x}{(px)^2} \, m_V^2 \left(
h_\parallel^{(t)}(u)-\frac{1}{2}\, \phi_\perp(u) - \frac{1}{2}\,
h_{\perp,3}(u)\right)\nonumber\\
& & & \left. +
\frac{m_V^2}{2px} \left(\eta_\mu x_\nu - \eta_\nu x_\mu\right)
\left( h_{\perp,3}(u) - \phi_\perp(u)\right)\right\},\nonumber\\
& \langle V(p) | \bar{q}_1(0) \gamma_\mu\gamma_5q_2(x) | 0\rangle & = & \frac{1}{4}\,
f_V^\parallel      m_V \,\epsilon_{\mu\nu\rho\sigma} \eta^{*\nu}p^\rho
x^\sigma \int_0^1\!\! du\, e^{i\bar u px} \, \left( \tilde{g}^a_\perp(u)  +   \frac{m_V^2 x^2}{16} \tilde{\mathbb{G}}^a_\perp(u)  \right)  \;,\nonumber\\
& \langle V(p) | \bar{q}_1(0) q_2(x) | 0\rangle & = & \frac{i}{2}\,
f_V^\perp   \,(\eta x) m_V^2
\int_0^1\!\! du\, e^{i\bar u px} \, \tilde{h}_\parallel^{(s)} (u) \;,
\end{alignat}
where the  notation $g_{\parallel,3} = g_3$ and $h_{\perp,3} = h_3$ has been introduced (wrt to Ref.~\cite{Ball:1998sk}) in order to declare the polarisation of the DAs. Additionally 
\begin{equation}
\label{eq:tilde}
 \tilde{h}_\parallel^{(s)} =   ( 1   -   \delta_+ ) h_\parallel^{(s)} \;,  \quad  \tilde{g}^a_\perp(u) = 
 ( 1   -  \tilde \delta_+ ) g^a_\perp(u) \;,  \quad \tilde{\mathbb{G}}^a_\perp   = 
 ( 1   -  \tilde \delta_+ )   \mathbb{G}^a_\perp \;,
\end{equation}
 take into account valence quark mass corrections  
 \begin{equation}
 \delta_{\pm} \equiv \frac{\fL (m_{q_2}   \pm  m_{q_1})}{( \fT m_V)}  \;, \quad 
\tilde \delta_{\pm} \equiv \frac{\fT (m_{q_2}   \pm  m_{q_1})}{( \fL m_V)}   \;,
 \end{equation} 
 consistent with the normalisation 
 \begin{equation}
 I_1[\phi](1) = 1 \;,  \quad \phi = \{ \phi_{\parallel,\perp} ,g_\perp^{v,a}, h_\parallel^{(s,t)}, g_{\parallel,3},h_{\perp,3}\} \;,
 \quad I_1[ \phi ](u) \ \equiv  \ \int_0^u\!\! dv\, \phi(v),
 \end{equation}
  and the \eom of the LCDAs.
  The twist-4 meson mass corrections $\mathbb{A}_{\parallel,\perp}$ (wrt $\phi_{\parallel,\perp}$ DAs)
have been introduced in \cite{Ball:1998ff}.
The twist-5 meson mass corrections  $\mathbb{G}^{v,a}_\perp$ (wrt $g_{v,a}^\perp$ DAs)
are introduced in this work for the first time. The DAs $\mathbb{G}_\perp^{a,v}$ and $\mathbb{A}_{\parallel,\perp}$ are not subject to a particular normalisation whereas 
$I_1[  \mathbb{A}_\parallel - \mathbb{G}_\perp^v ] (1)  = 0$ is necessary and 
partly motivated the reinvestigation of the twist-4,5 DA in this work. More details follow just below.
We will see in section \ref{sec:determination}  that the four additional DAs  $\mathbb{A}_{\parallel,\perp}$ and $\mathbb{G}^{v,a}_\perp$ can be obtained from the eight basic ones (at least in the asymptotic limit).
 
A striking feature are the $(px)^{-1}$- and $(px)^{-2}$-terms which originate from the change of variables 
\eqref{eq:zp} and \eqref{eq:n}. From a conceptual viewpoint the limit $px \to 0$ ought to exist
and imply conditions on the DA which have to be obeyed automatically by the solutions.

More concretely the $1/px$ factors are removed by 
\begin{eqnarray}
\label{eq:I1}
\frac{1}{px}\int_0^1\!\! du\,e^{i\bar u px} \phi (u) \stackrel{I_1[\phi](1) =0 }{\longrightarrow} i\!\int_0^1\!\! du
\,e^{i\bar u px} I_1[ \phi](u)\;,
\end{eqnarray}
with the normalisation condition as indicated.
In Eq.~\eqref{eq:DAvec} this concerns the following  five functions combintations
\begin{alignat}{3}
\label{eq:IDA}
&   \frac{1}{i(px)} \left( \phi_\parallel -g^v_\perp  \right) &\;\to\;& \phi_\parallel^{(1)}(u) &\; =\; & I_1[ \phi_\parallel -g^v_\perp](u)     \;,  \nonumber\\
&  \frac{1}{i^2(px)^2}  \left(h_\parallel^{(t)}-\frac{1}{2}\,\phi_\perp-\frac{1}{2} \,h_{\perp,3} \right)   &\;\to\;& 
h_\parallel^{(t,2)}(u) &\; =\; & 
I_2  [ h_\parallel^{(t)}-\frac{1}{2}\,\phi_\perp-\frac{1}{2} \,h_{\perp,3} ](u) \;,
\nonumber\\ 
&  \frac{1}{i(px)}  \left( h_{\perp,3} -\phi_\perp \right) &\;\to\;&    h_{\perp,3}^{(1)}(u) &\; =\; &  I_1[ h_{\perp,3} -\phi_\perp](u)    \;,
\nonumber\\ 
& \frac{1}{i^2(px)^2}  \left(g_{\parallel,3}-2 g^v_\perp+
\phi_\parallel] \right)   &\;\to\;&   g_{\parallel,3}^{(2)}(u) &\; =\; &I_2 [ g_{\parallel,3}-2 g^v_\perp+
\phi_\parallel](u)     \;,  \nonumber\\
&    \frac{1}{i(px)} \left(  \mathbb{A}_\parallel  - \mathbb{G}_\perp^v \right)   &\;\to\;&   \mathbb{A}_\parallel^{(1)}(u) &\; =\; & I_1[ \mathbb{A}_\parallel- \mathbb{G}_\perp^v](u)   \;,
\end{alignat}
where $\xi = 2u -1$ and $I_2 = I_1 \circ I_1$ is a double application of \eqref{eq:I1}. 
The asymptotic form of the DAs has been indicated and the dots stand for non-asymptotic corrections. 
As stated earlier all DAs are such that unwanted boundary terms disappear which is guaranteed provided that
\begin{equation}
I_1[\phi](1) = 0 \;, \quad I_2[\phi](1) = 0 \;,
\end{equation}
with the first and both conditions applying to the case where the DA in \eqref{eq:IDA} is written in terms of 
$I_1$ and $I_2$ respectively. 
The integrated DAs are those
that appear in the explicit results quoted in section \ref{app:T-exresults}.

\begin{table}[ht]
$$
\begin{array}{l || ll | llll | llll | ll      }
\text{DA} & \phi_{\perp}   & \phi_{\parallel} & g_\perp^v  & \tilde{g}_\perp^a  &  \tilde{h}^{(s)}_\parallel  & h^{(t)}_\parallel  &  g^{(\parallel)}_3  & h^{(\perp)}_3  & 
 \mathbb{A}_\perp &  \mathbb{A}_\parallel  &    \mathbb{G}^v_\perp &     \mathbb{G}^a_\perp     \\[0.2cm]  \hline 
\text{NN} &  \phi_{2,V}^\perp & \phi_{2,V}^\parallel & \phi_{3,V}^\perp & \psi_{3,V}^\perp &  \psi_{3,V}^\parallel &  
\phi_{3,V}^\parallel &  \psi_{4,V}^\parallel & \psi_{4,V}^\perp  &  \phi_{4,V}^\perp & \phi_{4,V}^\parallel &  \phi_{5,V}^\perp &  \psi_{5,V}^\perp 
  \\[0.2cm]   \hline
\text{form} & - & - & 3.9  & 3.9   & 3.7    & 3.7   & 3.24  & 4.21  & 4.22  & - & - & - \\[0.2cm]  
\text{expl} & \eqref{eq:t2} & \eqref{eq:t2}  &  3.18    & 3.17    & 3.16  & 3.15    & 3.27  & 4.24 & 4.25  & (\ref{eq:trick}) & (\ref{eq:trick}) &(\ref{eq:trick}) \\[0.2cm]   \hline
\text{Dirac} & \sigma_{\mu \nu} & \ga_\mu  & \ga_\mu & \ga_\mu \ga_5 & \mathbf{1} & \sigma_{\mu \nu}  & \ga_\mu 
 & \sigma_{\mu \nu} & \sigma_{\mu \nu}  & \ga_\mu  & \ga_\mu  & \ga_\mu \ga_5   \\[0.2cm]   \hline
 \text{twist} & 2 & 2 & 3 & 3 & 3 & 3 & 4 & 4 & 4 & 4 & 5 & 5 
\end{array}
$$
\caption{\small Translation table between old \cite{Ball:1998sk,Ball:1998ff} and new notation (NN) \cite{Ball:2007rt,Ball:2007zt}. 
The third and fourth line indicate the reference to the formal and explicit solution of the DAs. 
The twist-3,4 DAs refer to Refs.~\cite{Ball:2007rt,Ball:2007zt} respectively.
The first two DAs are of leading twist $2$ and are not referenced since they have been known for a long time 
e.g. \cite{Chernyak:1983ej}.
The last three DAs are obtained in this work. 
The second last line denotes the Dirac structure of the DA with $\mathbf{1},\sigma_{\mu \nu}$ and $\ga_\mu, \ga_\mu \ga_5$ being chiral odd and chiral even respectively.}
\label{tab:trans}
\end{table}

\subsection{Determination of  \texorpdfstring{$m_V^2$}{mV²}-LCDA in asymptotic limit}
\label{sec:determination}

Introducing more LCDA means that more information is needed to solve for the DAs. 
We did not systematically aim to do this but present an alternative and possibly new way to determine 
the asymptotic form of the DA directly from \eqref{eq:rigour}. 
We expand  all quantities systematically 
to first order in  $m_V^2$, using Eq. \eqref{eq:zp}, including in particular the $\hat p z$ in exponential factor 
in \eqref{eq:rigour}, 
\begin{equation}
\hat{p} z = px \left( 1 - \frac{x^2 m_V^2}{(px)^2} \right)^{1/2} = px - \frac{1}{2} \frac{ x^2  m_V^2}{px} + {\cal O}(m_V^4)  \;.
\end{equation}
By matching the first power in $m_V^2$ this leads to the following 
identifications at the level of \emph{asymptotic DAs} (cf. footnote \ref{foot:29}) 
\begin{alignat}{4}
\label{eq:trick}
& \mathbb{A}_\parallel(u) &\;=\;&  - 4 I_1[\xi ( \phi_\parallel )](u)  \;  &\;+\;&  \; 4 I_2[ 4 g_\perp^v - g_{\parallel,3} - 3 \phi_\parallel](u)   &\;=\;&  24 u^2 \bar u^2     +\dots \;, \nonumber
  \\
& \mathbb{A}_\perp(u) &\;=\;& - 4 I_1[\xi \phi_\perp](u) \; &\;+\;&  \; 4 I_2[ h_{\perp,3} - \phi_\perp](u) &\;=\;&  
 24  u^2 \bar u^2 +\dots \;, \nonumber  \\
& \mathbb{G}_\perp^v(u) &\;=\;&  - 4 I_1[\xi g^v_\perp](u)  \;, & &  &\;=\;&  6 u \bar u (1-  u \bar u)  +\dots \;, \nonumber   \\
& \mathbb{G}_\perp^a(u) &\;=\;&  - 4 I_1[\xi g^a_\perp](u) \;, & &  &\;=\;& 12 u^2 \bar u^2 + \dots  \;,
\end{alignat}
where the dots stand for non-asymptotic corrections. This method, convenient as it is, can  not 
determine non-asymptotic corrections since one would need to expand in the $z$-coordinate of the quarks 
as well as the Wilson line. 
This leads to higher dimensional local operators and 3-particle DAs which are both non-asymptotic.  
Hence if only the asymptotic DAs are required then we do not need to do this expansion.
We might turn to a more systematic study of this method in future work.
Our confidence in this alternative method, for determining the asymptotic DAs,  is borne out of several consistency checks.
 \begin{itemize}
\item The DA $\mathbb{A}_\parallel $ is such $I_1[ \mathbb{A}_\parallel  -\mathbb{G}_\perp^v ](1) = 0$ which guarantees the, previously discussed, finite limit $ p \cdot x \to 0$. The latter allows for the substitution \eqref{eq:IDA}, i.e. the $1/px$-pole is removed by partial integration.\footnote{We remind the reader that 
$I_1[ \mathbb{A}_\parallel|_\text{BBL} ](1) \neq 0$ was our motivation to investigate the DAs. It should be mentioned that it is possible that the \eom can be satisfied to the given order in $m_V$ by using 
the ultrarelatvistic approximatoin $\eta(p) \to p/m$  for $\mathbb{A}_\parallel$. For consistency one should use the same approximation to determine the light-cone DAs.}
  
  \item The asymptotic form $\mathbb{A}_\perp$ from the literature is reproduced.

 \item We adapt the moment equation \cite{Ball:1998sk} [Eq.~4.7]
for $g_\perp^{a,v}$ to $\mathbb{G}_\perp^{a,v}$ which can be done by replacing $(n+2) \to (n+4)$ 
where the extra additive factor of $2$ originates from the extra power of $x^2$ in the LCDA expansion. 
The adapted moment equation in the limit of vanishing quark mass and no 3-particle DA becomes 
\begin{equation}
\frac{1}{2}(n+4) M_n^{(\mathbb{G}^a)} = M_n^{(\mathbb{G}^v)} \;, \quad M_n^{(\mathbb{G}^x)} \equiv \int_0^1 du \, \xi^n \,\mathbb{G}_\perp^x(u)  \;, \quad ( \xi \equiv 2u-1) \;.
\end{equation}
It is readily verified that the asymptotic DAs given in \eqref{eq:trick} satisfy the moment equation exactly.
\item Last but not least  the obtained asymptotic DAs do verify the  \eom \eqref{eq:eom_op} at the level 
of the correlation function and therefore FFs. This was our original motivation to look into 
this matter.
\end{itemize}

\subsection{Explicit DAs used for this work}
\label{sec:explicit}

In this section we provide the actual DAs used in this paper to the given approximation.
For more complete solutions for the lower DAs we refer to the references in Tab.~\ref{tab:trans}.
The conventional approach for twist-2 and twist-3 DAs is the expansion in conformal spin (e.g. Gegenbauer moments) analogous to the partial wave expansion of $SO(3)$. For the twist $4$ there is the conformal spin 
expansion as well as a renormalon model e.g. \cite{Ball:2007zt}. 
In this work we only solve for the asymptotic DA for twist $4$ which is the lowest order in the conformal expansion. We estimate the effect of this to be at the $1\%$-level which is well beyond the uncertainty.

For the twist-2 DA we expand up to second order in the Gegenbauer polynomials $C_{n}^{3/2}(\xi = 2u-1)$
\begin{equation}
\label{eq:t2}
\phi_{\perp,\parallel}(u) = 6 u (1-u) \left( 1 + 
  a^{\perp,\parallel}_{1} C_{1}^{3/2}(\xi) +   a^{\perp,\parallel}_{2} C_{2}^{3/2}(\xi)  \right),
\end{equation}
which is a standard approximation in view of the lack of reliable knowledge on higher moments.
The twist-3 DAs are obtained from the twist-2 and the leading twist-3 3-particle DAs \cite{Ball:1998sk}
\begin{alignat}{3}
& \tilde{g}^a_\perp &\;=\;&   \bar u \int_0^u\,dv\,\frac{\Psi_g(v)}{\bar v} + u
\int_u^1\,dv\,\frac{\Psi_g(v)}{v} &\;+\;& 6 u \bar u \left(\frac{10}{9} \zeta^\parallel_3  + \frac{5}{12} (\omega_3^\parallel -  
\tilde{\omega}_3^\parallel/2) \right)C_{2}^{3/2}(\xi)  \nonumber \\[0.1cm]
& g^v_\perp &\;=\;& \frac{1}{4}\left[
\int_0^u\,dv\,\frac{\Psi_g(v)}{\bar v} + 
\int_u^1\,dv\,\frac{\Psi_g(v)}{v}\right]&\;+\;&
5\zeta^\parallel_3   ( 3 \xi^2-1) + \frac{15}{32} (\omega_3^\parallel -  
\tilde{\omega}_3^\parallel/2)   \left( 3 - 30 \xi^2  + 35
\xi^4 \right) \;,\nonumber \\
& h_\parallel^{(t)} &\;= \;&  \frac{1}{2}\left[
\int_0^u\,dv\,\frac{\Psi_h(v)}{\bar v} + 
\int_u^1\,dv\,\frac{\Psi_h(v)}{v}\right]   &\;+\;&  \frac{15}{8 } \omega_3^\perp (3-30\xi^2+35\xi^4)   
    \;, \nonumber \\
& \tilde{h}_\parallel^{(s)} &\;=\;&  \bar u \int_0^u\,dv\,\frac{\Psi_h(v)}{\bar v} + u
\int_u^1\,dv\,\frac{\Psi_h(v)}{v} &\;+\;&
6 u \bar u  \left(  \frac{5}{18}\, \omega_3^\perp  C_{2}^{3/2}(\xi)  \right)   \;.
  \label{eq:t3}
\end{alignat}
with 
$\Psi_g(u) \equiv 2 \phi_\parallel(u) + \tilde{\de}_+ \xi \phi'_\perp(u) + \tilde{\de}_- \phi'_\perp(u)$ 
and $\Psi_h(u) \equiv  2 \phi_\perp(u) - \de_+(\phi'_\perp(u) -  \xi/2 \phi'_\perp(u)) +1/2 \de_- \phi'_\perp(u)$.
The contributions of the 2-particle DAs are given implicitly whereas the 3-particle DA-contributions have 
been given explicitly. The four parameters $\zeta_3, \omega_3^\perp, \omega_3^\parallel$ and $\tilde{\omega}_3^\parallel$ are $G$-parity even parameters of the three twist-3 3-particle DAs as given in [Eq.~3.11] in 
\cite{Ball:2007rt}.  $SU(3)$-breaking parameters can be neglected at the current level of precision.

As stated above for the twist-4 and twist-5 DAs we employ the asymptotic form which means that we set the Gegenbauer moments 
$a_{1,2}$, the 3-particle DA parameters and the quark masses to zero with respect to the more general solution. 
%The effect of this approximation ought to be roughly $1\%$ and is therefore justified from the pragmatic viewpoint.
 The asymptotic twist-4 DAs are given by 
\begin{alignat}{2}
\label{eq:t4}
 h_{\perp,3}  = 6 u \bar u \;, \quad  g_{\parallel,3}  = 6 u \bar u  \;, \quad  \mathbb{A}_\perp  =  24  u^2 \bar u^2  \;,  \quad   \mathbb{A}_\parallel =  24  u^2 \bar u ^2  
  \end{alignat}
and the newly introduced twist-5 DAs 
\begin{equation}
\label{eq:t5}
 \mathbb{G}^v_\perp  =   6 u \bar u (1-  u \bar u)   \;,  \quad  \mathbb{G}^a_\perp = 12 u^2 \bar u^2 \;.
\end{equation}
The determination of $\mathbb{A}_\parallel $ and   $\mathbb{G}^{v,a}_\perp $ are new and discussed 
in  the previous section.
The numerical input is given in Tab.~\ref{tab:input} except for the values for 
$\omega_3^\perp, \omega_3^\parallel$ and $\tilde{\omega}_3^\parallel$ which are taken from table one in
\cite{Ball:2007rt} and are related to the parameters previously used (e.g. \cite{Ball:2004rg}) as follows
$3/2 \zeta_3^\parallel \omega_3^V = \omega^\parallel_3$, $3/2 \zeta_3^\parallel \omega_3^T = \omega^\parallel_\perp$ and $1/2 \zeta_3^\parallel \omega_3^A = \tilde{\omega}^\parallel_3$. 
For the sake of completeness and clarity we give the form of the integrated asymptotic DAs given 
in \eqref{eq:IDA}:
\begin{alignat}{5}
&  \phi_\parallel^{(1)}(u) &\;=\;&  \frac{3}{2} \bar u u \xi  \;, \quad  & &  h_\parallel^{(t,2)}(u) &\;=\;&    \frac{3}{2} u^2 \bar u^2 \;,\quad 
 h_{\perp,3}^{(1)}(u)  =  0 \;,  \nonumber  \\[0.1cm]
 & g_{\parallel,3}^{(2)}(u)  &\;=\;&  - \frac{3}{2} u^2 \bar u^2  \;, \quad 
& &   \mathbb{A}_\parallel^{(1)}(u) &\;=\;&  3 \xi u^2 \bar u ^2  \;. 
\end{alignat}

\section{Decay constants from experiment}
\label{app:decay}

\subsection{The Neutral Decay Constants
  \texorpdfstring{$f_{\rho^0,\omega,\phi}$}{frho,omega,phi} from \texorpdfstring{$V^0 \to e^+e^-$}{V0->e+e-}}

We improve the discussion on the extraction of the decay 
constants of the $\rho^0$, $\omega$ and $\phi$ from $V^0 \to e^+ e^-$ with respect 
to the earlier work \cite{Ball:2006eu}.  The effects on the decay constants due to mixing are taken into account at the level of matrix elements. Previously 
the mixing was abstracted from the state mixing. The relation between the two is commented 
on  in section \ref{app:comment}.

The three vector mesons $\rho^0,\omega$ and $\phi$ are 
flavour neutral and can therefore  mix into each other through QCD and QED.  
The mixing of $\phi$--$\omega$ is driven by QCD, $\rho$--$\omega$ is due to QED 
and $m_{u,d}$-quark mass difference
 whereas $\phi$--$\rho$ requires both forces and can therefore be neglected.  
 In $V^0 \to e^+e^-$ the meson couple to the electromagnetic current as follows
 \begin{equation}
 \label{eq:jem}
 j_\mu^\text{em} = Q_s V_\mu^{\phi_I}  + \frac{Q_u+Q_d}{\sqrt{2}}  V_\mu^{\omega_I} +  \frac{Q_u-Q_d}{\sqrt{2}}  V_\mu^{\rho_I}
 \end{equation}
 with $Q_{u,d,s}$ being the charges of the quarks and the quark currents  are defined by 
 \begin{equation}
 \label{eq:ideal}
 V_\mu^{\omega_I,\rho^0_I} = 
 \frac{1}{\sqrt{2}} ( \bar u \ga_\mu u \pm \bar d \ga_\mu d) \;, \quad V_\mu^{\phi_I} = \bar s \ga_\mu s \;.
\end{equation}
The label $I$ stands for isospin as well as ideal mixing (i.e. $\phi$ being a pure $s\bar s$-state). 
The currents   $ V_\mu^{\omega_I,\phi_I} $ and   $ V_\mu^{\rho^0_I} $ are of isospin 
  $I = 0$ and $I=1$ for respectively.   

It is our goal to extract the following decay constants
 \begin{equation}
 \label{eq:Ziel}
 \matel{\phi}{ V_\mu^{\phi_I} }{0} =  \eta_\mu m_{\phi}  f^{\phi_I}_{\phi}\,, \quad  
 \hat{c}_{\omega} ^q \matel{\omega}{ \bar q \ga_\mu q }{0} =  \eta_\mu m_{\omega}  f^{(q)}_{\omega} \;, \quad 
 \hat{c}_{\rho^0}^q \matel{\rho^0}{ \bar q \ga_\mu q }{0}=  \eta_\mu m_{\rho^0}  f^{(q)}_{\rho^0} \;,
 \end{equation}
relevant for the description of flavour transition via the weak force.
Above $\eta$ denotes the polarisation vectors,  the $\parallel$ superscript on $\fL$ is omitted 
 and 
 $\hat{c}^{ u}_{\rho^0} = -\hat{c}^{ d}_{\rho^0}  =\hat{c}^{ u}_{\omega} = \hat{c}^{ d}_{\omega}=  \sqrt{2}$ are prefactors 
taking into account  the quark composition of the  wave functions.  
 The effect of the mixing is investigated in a two step procedure of 
 $\phi$--$\omega$ and $\rho$--$\omega$ mixing.

\subsubsection{Effective couplings to the electromagnetic current}

\paragraph{\texorpdfstring{$\phi$--$\omega$}{phi--omega} mixing:}
In order to asses the effect of the $\phi$--$\omega$ mixing  the following matrix elements  are needed
\begin{alignat}{2}
\label{eq:SVZ-elements}
&  \matel{\phi}{ V_\mu^{\omega_I} }{0}    &\;=\;& \eta_\mu m_\phi f^{\omega_I}_{\phi} \equiv  \epsilon(m_\phi) \eta_\mu m_\phi f^{\phi_I}_{\phi} \;, \nonumber \\ 
 &  \matel{\omega}{ V_\mu^{\phi_I} }{0}   &\; =\;& \eta_\mu m_\omega f^{\phi_I}_{\omega} \equiv  - \epsilon(m_\omega) \eta_\mu m_\omega f^{\omega_I}_{\omega} \;.
\end{alignat}
They have been computed   to be  
$\epsilon \equiv \epsilon(m_\phi) \simeq \epsilon(m_\omega) = 0.05(2)$ in the 
pioneering papers of QCD sum rules \cite{Shifman:1978by}.  
Note, the effect 
is driven by contributions of four quark condensates, estimated in the vacuum saturation approximation, 
and we have therefore assigned a conservative $40\%$ error to $\epsilon$. 

\paragraph{$\rho$--$\omega$ mixing:} 

The analogous  $\rho$, $\omega$  decay constants  have been computed in reference \cite{Maltman:1998xn} 
by using finite energy sum rules.   Effects are due to different QED and  $m_{u,d}$-quark mass 
differences.\footnote{Note computing the QED corrections to the local matrix element, i.e. which we call 
decay constant, is not the same as computing the QED corrections to the corresponding leptonic decays themselves. The latter are more complex,  requiring the computation of virtual and real corrections taken into 
account in the experimental analysis.}
Their results, neglecting the $\rho$--$\phi$ mixing,
 translates into the notation analogous to \eqref{eq:SVZ-elements} as follows:  $f_\rho^{\omega_I} = \sqrt{6}F_\rho^8 \simeq 5.9(12) \MeV$ 
and $f_\omega^{\rho_I} = \sqrt{2} F_\omega^3 = -4.8(10) \MeV$. We have enlarged the uncertainty 
in view of possible duality violation of finite energy sum rules \cite{Maltman}.

\subsubsection{Scaling factors due to mixing}

We parameterise the mixing effects in terms of correction  factors, denoted by $\kappa$, to the 
matrix element to the electromagnetic current,
\begin{alignat}{2}
&  \matel{\omega}{ j_\mu^\text{em}}{0} &=&  \eta_\mu m_\omega  f^{\omega_I}_{\omega}  \frac{Q_u+Q_d}{\sqrt{2}}  \kappa^{\omega[\phi]}\kappa^{\omega[\rho]} \;, \quad 
 \matel{\rho^0}{ j_\mu^\text{em}}{0} =  \eta_\mu m_{\rho^0}  f^{\rho^0_I}_{\rho^0}  \frac{Q_u-Q_d}{\sqrt{2}}  
\kappa^{\rho[\omega]} \;, \nonumber  \\[0.1cm]
& \matel{\phi}{ j_\mu^\text{em}}{0} &=&  \eta_\mu m_\phi  f^{\phi_I}_{\phi} Q_s \kappa^{\phi[\omega]} \;.
 \end{alignat}
Using the numerical input of the previous section we get 
\begin{alignat}{3}
\label{eq:kappa}
&  \kappa^{\phi[\omega]}   &\;\equiv\;& 
  \left( 1 +  \frac{f_\phi^{\omega_I} }{f_\phi^{\phi_I}}  \frac{Q_u+Q_d}{\sqrt{2}Q_s}  \right) \simeq 
   \left( 1 + \epsilon \frac{Q_u+Q_d}{\sqrt{2}Q_s}  \right) 
     &\;\simeq\;&  1/(1.037(16))   \;, 
    \nonumber \\
 &  \kappa^{\omega[\phi]}    &\;\equiv\;&   \left( 1 + \frac{f_\omega^{\phi_I} }{f_\omega^{\omega_I}}  \frac{\sqrt{2}Q_s}{Q_u+Q_d}  \right) \simeq
  \left( 1 - \epsilon \frac{\sqrt{2}Q_s}{Q_u+Q_d}  \right)   &\;\simeq\;& 1/(0.933(28)) \;,  \nonumber \\
&   \kappa^{\omega[\rho]}   &\;\equiv\;&  \left( 1 + \frac{f_{\omega}^{\rho^0_I}}{f_\omega^{\omega_I}} \frac{Q_u-Q_d}{Q_u+Q_d}  \right) &\;\simeq\;& 1/(1.08(2))  \;,  \nonumber \\
  &   \kappa^{\rho[\omega]}    &\;\equiv\;&
    \left( 1 + \frac{f_{\rho^0}^{\omega_I}}{f_{\rho^0}^{\rho^0_I}} \frac{Q_u+Q_d}{Q_u-Q_d}  \right) &\;\simeq\;& 1/(0.990(2))   \;.
\end{alignat}
The impact of the mixing on the extraction of the decay constants is heavily affected 
by the charge ratios $\sqrt{2} {Q_s}/({Q_u+Q_d} ) = -\sqrt{2}$ and $(Q_u-Q_d)/(Q_u+Q_d) = 3$.
The reader is reminded that  $\rho$--$\phi$ mixing is neglected since it requires the strong as well as 
the electromagnetic force which is expected to be a small effect.

The experimental branching ratios are \cite{Agashe:2014kda}
\begin{alignat}{2}
\label{eq:Bee}
&  \text{BR}(\rho^0 \to e^+ e^- ) &=& (4.72 \pm 0.05) \times 10^{-5}\,, \quad
\text{BR}(\omega  \to e^+ e^-) = (7.28 \pm 0.14) \times 10^{-5}\,,  
\nonumber \\
&  \text{BR}(\phi \to e^+ e^- ) &=& (2.95 \pm 0.30) \times 10^{-5}\,.
\end{alignat}
The theoretical expression for the decay rate is given by
\begin{equation}
\label{eq:Vee}
\Gamma(V^0 \to e^+ e^-) = \frac{4 \pi}{3} \frac{\alpha^2}{m_V} f_V^2
c_V  + O\left(\alpha,\frac{m_V^2}{M_W^2}\right)\,,
\end{equation}
where the coefficients $c_V$ in the limit of no mixing (i.e. $\kappa \to 0$) can be read-off
from 
\eqref{eq:jem}
$c_{\rho_I^0} = (Q_u-Q_d)^2/2 = 1/2$,  $c_{\omega_I} = (Q_u+Q_d)^2/2=
1/18$ and $c_{\phi_I} = Q_s^2 = 1/9$. The effect of mixing  leads to the following replacements 
$$
c_{\rho_I^0} \to c_{\rho_I^0} (\kappa^{\rho[\omega]})^2 \;, \quad c_{\omega_I} \to c_{\omega_I} 
(\kappa^{\omega[\rho]} \kappa^{\omega[\phi]} )^2 \;, \quad c_{\phi_I} \to c_{\phi_I}  (\kappa^{\phi[\omega]})^2 \;,
$$
and results in a shift of $\rho$ , $\omega$ and  $\phi$ decay constant of roughly $-1\%$, $0.5\%$ and $4\%$ respectively. 
It is noticed that the individual effects of the  $\omega$--$\rho$ and the  $\omega$--$\phi$ 
mixing are around $\pm 8\%$ but
do almost cancel each other out.

 Including the mixing effects we
 get the following decay constants for the currents \eqref{eq:ideal}
\begin{alignat}{7}
\label{eq:fdecayI}
& f^{\rho_I}_{\rho^0} &\;=\;&  (215.6 &\;\pm\;&  2_\text{Br} &\;\pm\;& 1_{\Gamma_{\rho}} &\;\pm\;& 1_{\omega\!-\!\rho  }   &\;\pm\;&  0\phantom{_{\omega\!-\!\phi }} )\, \text{MeV}   &\;=\;& 216(3)\MeV \,,\nonumber \\ 
& f^{\omega_I}_{\omega}   &\;=\;&  (196.5 &\;\pm\;&  2_\text{Br} &\;\pm\;& 1_{\Gamma_\omega}  
 &\;\pm\;& 4_{\omega\!-\!\rho  } 
&\;\pm\;&  6_{\omega\!-\!\phi } 
 ) \, \text{MeV}  &\;=\;&   197(8) \MeV \,, \nonumber \\
& f^{\phi_I}_{\phi}  &\;=\;&  (233.0 &\;\pm\;&  2_\text{Br} &\;\pm\;& 1_{\Gamma_\phi}  &\;\pm\;& 3_{\phi\!-\!\omega } &\;\pm\;&  0\phantom{_{\omega\!-\!\phi }} 
) \, \text{MeV}  &\;=\;& 233(4)\MeV\,,
\end{alignat}
where  the uncertainties in the other input parameters are irrelevant. 
Errors are added in quadrature in the final result.
To clarify the notation we quote the example  $f^{\rho_I}_{\rho^0} = f_{\rho^0} / \kappa^{\rho[\omega]} \simeq 0.99 f_{\rho^0}$ with 
$f_{\rho^0}$ from \eqref{eq:Vee}.
 
 Finally we get for the $\rho$ and $\omega$ decay constants coupling directly to $u$ and $d$ quark currents
 \begin{alignat}{3}
 \label{eq:fflavour}
 &  f^{(u)}_{\rho^0}  &=&  f^{\rho^0_I}_{\rho^0}  +  f^{\omega_I}_{\rho^0} &=&  (215.6 + 5.9)\MeV = 221.5 (3) \MeV 
 \;,  \nonumber \\
&  f^{(d)}_{\rho^0}  &=&  f^{\rho^0_I}_{\rho^0}  -  f^{\omega_I}_{\rho^0} &=&  (215.6 - 5.9)\MeV = 209.7 (3) \MeV 
 \;,  \nonumber \\
&  f^{(u)}_{\omega}  &=&  f^{\omega_I}_{\omega}  +  f^{\rho^0_I}_{\omega} &=&  (196.5 - 4.8)\MeV = 191.7 (8)    \MeV \;,  \nonumber \\
&  f^{(d)}_{\omega}  &=&  f^{\omega_I}_{\omega}  -  f^{\rho^0_I}_{\omega} &=&  (196.5 + 4.8)\MeV = 201.3 (8)    \MeV   \;,
 \end{alignat}
 where we have taken the same uncertainties as in \eqref{eq:fdecayI}.
 
% and the corrections to \eqref{eq:Vee} are expected to be 
%smaller than the total uncertainty.
%$\rho$--$\omega$ mixing has a negligible effect on $f_{\rho^0}$, 
%but raises $f_\omega$ by 2 to 3$\,\text{MeV}$. $\omega$--$\phi$ mixing
%is much more relevant and lowers $f_\omega$ by about 
%$10 \, \text{MeV}$ and $f_\phi$ by about $13 \, \text{MeV}$ . 

\subsubsection{Comment on state mixing versus decay constant mixing}
\label{app:comment}

The mixing of   states and  decay constants are related but can be quantitatively different.\footnote{This is particularly enhanced for $\eta$--$\eta'$ system because of the 
effect of the chiral anomaly.}
The former is one of the effects contributing to the latter. Below we present evidence that in 
reality the mixing of states dominates the mixing of  the $\omega-\phi$ decay  constants.

For example if one assumes that
$f_{\omega}^{\omega_I} \simeq f_{\phi}^{\phi_I}$ ($SU(3)_F$-symmetry for which there is empirical 
evidence), $|f_{\omega_I}^{\phi_I}| \ll |f_{\omega_I}^{\phi}|$ and takes into account the $\phi$--$\omega$
state mixing 
\begin{equation}
\state{\omega} \sim \state{\omega_I} - \epsilon_{\omega\phi} 
\state{\phi_I}\,,
\qquad
\state{\phi} \sim \state{\phi_I} + \epsilon_{\omega\phi} \state{\omega_I}\,,
\end{equation}
one arrives at
\begin{equation}
\label{eq:hmeps}
\epsilon_{\omega\phi}  \frac{m_\omega}{m_\phi}   \simeq  \epsilon(m_\phi)   \;,
\quad  
\epsilon_{\omega\phi}  \frac{m_\phi}{m_\omega}    \simeq \epsilon(m_\omega) \;,
\end{equation} 
which is reasonably well satisfied. 
A recent determination of the mixing angle by  the KLOE 
collaboration is given by  $\epsilon_{\omega\phi} = 3.32(9)^{\circ} \simeq 0.58(2)$.
Using, as previously \cite{Shifman:1978by},
 $\epsilon \equiv \epsilon(m_\phi) \simeq \epsilon(m_\omega) = 0.05(2)$, 
Eq.~\eqref{eq:hmeps} is equivalent to $0.45 \simeq 0.5(2)$ and $0.7 \simeq 0.5(2)$ which 
is satisfied within errors. It is to be concluded  that the effect of $\phi$--$\omega$
 decay constant mixing  is driven by the state-mixing. 

One could put forward the same procedure for the $\rho$--$\omega$ system but there 
are complications. The $\rho$--$\omega$ system is more delicate since the closeness 
of the two states means that the mixing angle is effectively a complex number because 
diagonal and off-diagonal self energies are complex. 
The off-diagonal self energy acquires an imaginary part through the isospin violating $\omega \to \pi\pi \to \rho$ transition; a circumstance which has been neglected in the literature for a long time! 
The off-diagonal self energy has been determined to be  $\Pi_{\rho \omega} (m_{\rho}^2) \simeq ( - 4620 \pm 220_\text{model} \pm 170_\text{data}) + (- 6100 \pm 1800_\text{model} \pm 1110_\text{data})i  \MeV^2$  \cite{Wolfe:2010gf}
 by using a recent BaBar analysis \cite{Aubert:2009ad}. 
 The value of the mixing angle through  
$\epsilon_{\rho \omega} = \Pi_{\rho \omega}  (m_\rho^2)/ \big( (m_\omega-
i \Gamma_\omega/2)^2 - (m_\rho-i \Gamma_\rho/2)^2\big)$ then comes with 
a large error; especially on the real part which is decisive. 
The small error on previous determinations turned out to be an artefact of neglecting 
the imaginary part of the off-diagonal self energy \cite{Maltman:1998xn}.  
In view of this situation we chose  to directly use the computations on the mixing 
of the decay constants and abandon the mixing of state picture.

\subsection{Charged decay constants from \texorpdfstring{$\tau^+ \to V^+ \nu$}{tau->V nu} decays}

The same standard procedure is applied as in \cite{Ball:2006eu} including in addition the sizeable 
leading electroweak corrections (due to  a $\ln (m_Z/m_\tau)$-term) \cite{Braaten:1990ef,Grossmann:2015lea}. 
Implementing this amounts  to making the replacement $\text{BR}(\tau^+ \to K^{*+} \nu)_\text{here} =  
1.015 \cdot \text{BR}(\tau^+ \to K^{*+} \nu)_{\mbox{\cite{Ball:2006eu}}}$.
Using the input parameters 
  $|V_\text{us}| \simeq 0.225$, $|V_\text{ud}| \simeq 0.974$, 
 $ \text{BR}(\tau^+ \to \rho^+ \nu) = 25.22 \pm 0.33$ and 
  $\text{BR}(\tau^+ \to K^{*+} \nu)=
1.20 \pm 0.07$, $\alpha(m_V) \simeq  1/135.4$  and the $\tau$ lifetime  $\tau_\tau =  (290.3 \pm 0.5 ) \cdot  10^{-15} s$ 
  \cite{Agashe:2014kda} we get
\begin{equation}
\label{eq:fdecay-charged}
f_{K^*} = 204(7)\MeV \;, \quad f_{\rho^+} = 210(4) \MeV \;.
\end{equation}

\subsection{Final results summarised}

In view of the many details discussed and numbers quoted we summarise our results 
for the reader's convenience. The difference in the charged and neutral $\rho$ decay constants is $6 \MeV$ and we therefore choose to average them  $\bar f^{\rho_{I}}_{\rho} = 213(5)\MeV$ slightly enhancing 
the uncertainty. The final results for the decay constants 
coupling to the currents \eqref{eq:ideal} are then taken from \eqref{eq:fdecayI}, \eqref{eq:fdecay-charged} and the above mentioned average
\begin{alignat}{4}
\label{eq:ffinal}
& \bar f^{\rho^{I}}_{\rho} &=& 213(5)\MeV \;, \quad  & & f^{\omega_I}_{\omega}    &= &   197(8) \MeV \,,  \nonumber \\
& f^{\phi_I}_{\phi}&=& 233(4)\MeV \;, \quad & & f_{K^*} &=& 204(7)\MeV \;.
\end{alignat}
The decay constant for the  pure flavour currents to the 
$\rho^0$- and $\omega$-meson are \eqref{eq:fflavour}
\begin{alignat}{4}
\label{eq:ffinal2}
 &  f^{(u)}_{\rho^0}  &=&   221.5 (3) \MeV  \;, \quad 
& &    f^{(d)}_{\rho^0}   &=& 209.7 (3) \MeV \nonumber \\
&  f^{(u)}_{\omega}  &=&  191.7 (8)  \MeV 
& & f^{(d)}_{\omega}  &=&   201.3 (8) \MeV   \;.
 \end{alignat}

In our tables and computation we will use the decay constants \eqref{eq:ffinal} omitting 
the additional labels. 
The results of $f_{K^*}$ is consistent with \cite{Grossmann:2015lea}, 
$\phi$--$\omega$ is treated similarly to \cite{Koenig:2015pha}, 
whereas our discussion on $\rho$--$\omega$ mixing is more detailed in terms of explicit results.
 We would like to add a comment concerning QED corrections. 
 The experimental analyses are performed using photon showers (e.g.  Photos~\cite{Golonka:2005pn}) and subtracting the large part of the final state photons. A  fully consistent treatment of QED corrections might be carried out
in the future for which we may expect a global shift (i.e. multiplicative factor in front of all decay constants) 
 at or below the $1\%$-level.

As for the flavour specific decay constants 
we leave it to the reader to scale the $B \to \rho^0, \omega$ FFs appropriately.  For example for  
\begin{alignat}{3}
\label{eq:scale-F}
& F^{B \to \rho^0}_{b \to u} &=&\;  k_{(\rho^0,u)} F^{B \to \rho^0}  &\simeq& \; 1.040\, F^{B \to \rho^0} \;,  \nonumber \\ 
& F^{B \to \omega}_{b \to u} &=&\;  k_{(\omega,u)} F^{B \to \omega} 
&\simeq&\;  0.973 \,F^{B \to \omega} \;,
\end{alignat}
where $F$ stands for any of the seven FFs.
The scale factors $k$ are $ k_{(\rho^0,q)} =f^{(q)}_{\rho^0}/\bar f^{\rho^{I}}_{\rho} $ and $ k_{(\rho^0,q)} =f^{(q)}_{\omega}/  f^{\omega^{I}}_{\omega} $ which upon using 
\eqref{eq:ffinal}  and \eqref{eq:ffinal2}  amount to
 \begin{equation}
 k_{(\rho^0,u)} = 1.040 \;, \quad k_{(\rho^0,d)} = 0.985 \;, \quad k_{(\omega,u)} = 0.973 \;, \quad k_{(\omega,d)} = 1.022  \;.
\end{equation}
Scaling the FF as in \eqref{eq:scale-F} is a reasonable procedure since, in practical 
 computations,  all 
input parameters of the DA are made dependent on the normalisation of the longitudinal decay constant.

\section{Conversion between form factor bases}
\label{app:FF-hel}

\subsection{Helicity basis}

In this appendix we give the projection of the FFs onto the helicity basis which 
is convenient for the computation of angular observables.
Using the Jacob Wick polarisation tensors\footnote{Cf.~appendix  A \cite{Hiller:2013cza}  where the polarisation tensors 
$\eta$ and $\epsilon$ correspond to  $\ga$ and  $\beta$ respectively.} we define:
\begin{equation}
X^{(\rho)} =  \epsilon_\mu(\rho) \matel{K^*(p,\eta(m(\rho)))}{\bar s \Gamma_X^\mu b}{\bar B(p_B)}  \;,   \quad m(t)=m(0) = 0 \;, \; m(\pm ) = \pm 
\end{equation}
where $\rho =0,\pm,t $ is the polarisation index which is not summed over  and
$\Gamma_T^\mu =  i q_\nu \sigma^{\mu\nu} (1 \pm \ga_5)$,
$\Gamma_\V^\mu = \ga^\mu(1\mp \ga_5)$ and  $\Gamma_{\cal D}^\mu =  (2 i \!\stackrel{\leftarrow}{D})^{\mu}(1 \!\pm\!\gamma_5)$ correspond to tensor, vector and derivative FFs. 
 We get
\begin{eqnarray}
\label{eq:XX}
X^{(\perp)} &=&  \frac{1}{\sqrt{2}}( X^{(+)} - X^{(-)})  =   i \sqrt{2}\sqrt{ \la(q^2)} X_1 \;, \nonumber  \\
X^{(\parallel)} &=&  \frac{1}{\sqrt{2}}( X^{(+)} + X^{(-)})  =  \pm  i \sqrt{2}(m_B^2-m_{K^*}^2) X_2 \;, \nonumber   \\
X^{(0)}  &=&   \mp i \frac{\sqrt{q^2}(m_B^2+3 m_V^2-q^2) }{2 m_{K^*} } X_0 \;,  \nonumber   \\
X^{(t)}  &=&     \mp i \frac{\sqrt{\la(q^2)}}{2} X_P  \;,
\end{eqnarray}
where $ X_0 \equiv X_2 - c_{23}(q^2) X_3 $ with
\begin{alignat}{2}
\label{eq:c23la}
& c_{23}(q^2) &\;\equiv\;&  \frac{\la(q^2)}{(m_B^2-m_{K^*}^2)  (m_B^2+3 m_{K^*}^2-q^2)} 
= 1 + {\cal O}(q^2/m_B^2,m_{K^*}^2/m_B^2)\;, \nonumber \\
&  \la(q^2)&\;\equiv\;& ((m_B+m_{K^*})^2-q^2)((m_B-m_{K^*})^2-q^2) \;,
\end{alignat}
and $\la$ being 
the  K\"all\'en-function. We infer that at the kinematic endpoint where $\la =0$, only the $X_2$ structure contributes in accordance with general findings on endpoint symmetries \cite{Hiller:2013cza}.

 For  
$X=T, \V, {\cal D}$, $X_i$ is given by $T_i$ (with $T_P \equiv 0$)  $\V_i$ and ${\cal D}_i$ 
in Eq.~\eqref{eq:VAs} and \eqref{eq:DD} respectively.
The relation of $T_0$ and $\V_0$ to $T_{23}$ and $A_{12}$ used in the literature (e.g.~\cite{Horgan:2013hoa}) is as follows:
\begin{equation}
\label{eq:convert}
T_0 = \frac{8 m_B m_{K^*}^2}{(m_B+m_{K^*}) (m_B^2+3 m_{K^*}^2-q^2)} T_{23} \;, \quad 
\V_0 = \frac{- 16 m_B m_{K^*}^2} {q^2 (m_B^2+3 m_{K^*}^2-q^2)}A_{12}
\end{equation}
where 
\begin{align}
\label{eq:hm}
A_{12}%&=%\frac{(-A_1/(m_B-m_V)) \left(m_B^2-m_V^2\right) \left(m_B^2+3 m_V^2-q^2\right)-\lambda ( 2 m_  V/q^2 A_3) }{8 m_B m_{K^*}^2 \left(m_B-m_{K^*}\right)}  \\
 &=  \frac{ \left(m_B+m_{K^*}\right){}^2 \left(m_B^2-m_{K^*}^2-q^2\right)A_1 -\la(q^2) A_2  }{16 m_B m_{K^*}^2 \left(m_B+m_{K^*}\right)}
\nonumber \\ 
&=  \frac{q^2/2  (m_B^2 + 3 m_{K^*}^2 - q^2) A_1+ 
    \la(q^2) m_{K^*}/(m_B + m_{K^*}) A_3} {8 m_B m_{K^*}^2 (m_B - m_{K^*})} \nonumber  \\
T_{23}&=\frac{ \left(m_B^2-m_{K^*}^2\right) \left(m_B^2+3 m_{K^*}^2-q^2\right)T_2-\la(q^2)  T_3}{8 m_B m_{K^*}^2 \left(m_B-m_{K^*}\right)}.
\end{align}
We further notice that
\begin{equation}
\label{eq:A12-A0}
A_{12}(0)  =  \frac{m_B^2 - m_{K^*}^2}{8 m_B m_{K^*}} A_3(0) =   \frac{m_B^2 - m_{K^*}^2}{8 m_B m_{K^*}} A_0(0)  \;,
\end{equation}
which we implement, besides $T_1(0) = T_2(0)$, into the fit as a constraint.

\begin{table}[h]
$$
\begin{array}{    llll   | l    }
 \sigma_{\mu\nu}q^\nu & \ga_\mu & \ga_\mu \ga_5 & \ga_5 &   \text{type} \\[0.2cm] \hline
 T_{1,2,3} & V & A_{1,3,0} & A_0 &  \text{traditional} \\[0.2cm]
 T_{1,2,3} & \V_1 & \V_{2,3,P} & \V_P  & \text{\eom} \mbox{\eqref{eq:eom1}-\eqref{eq:eom4}} \\[0.2cm]
 T_{\perp,\parallel} \sim T_{1,2} \;, T_0 \sim T_{23} & \V_\perp \sim \V_1 & V_{\parallel} \sim \V_2  \;,   
 \V_0 \sim A_{12} \;, \V_P &  \V_P &  \text{helicity}
\end{array}
$$
\caption{\small The conversion factors between the traditional and the \eom FFs is given in 
\eqref{eq:VAs}. The $0$-helicity FF are given by $T[\V]_0(q^2) = T[\V]_2(q^2) - c_{23}(q^2)  T[\V]_3(q^2)$ 
with the kinematic function given as in  \eqref{eq:c23la} and   
the $q^2$ dependence of the factor relating $T[\V]_{\perp,\parallel} \sim T[\V]_{1,2}$ can be read off from 
\eqref{eq:XX}. The $0$ helicity FFs $T_{23}$ and $A_{12}$ whose notation is inspired by 
\eqref{eq:hm} are related to $T_0$ and $\V_0$ as given in \eqref{eq:convert}.}
\label{tab:overview}
\end{table}

\subsection{Overview of form factor notation}

Not including the derivative FFs there are seven independent FFs of which all  others
are linear combinations. 
The basis $T_{1,2,3}$, $V$ and $A_{1,3,0}$ is the traditional basis (e.g. \cite{Ball:2004rg}; note: 
$A_2$ is linearly dependent on $A_{0,3}$ cf.~\eqref{eq:VAs}). 
The basis $T_{1,2,3}$, and $\V_{1,2,3,P}$ is suited for the \eom and the conversion between the two is given 
in \eqref{eq:VAs}. The helicity basis $T[\V]_{\perp,\parallel,0}$ and $\V_P$ is  suited for phenomenology with $T[\V]_{\perp,\parallel} \sim T[\V]_{1,2}$ and 
$T[\V]_0 = T[\V]_2(q^2) - c_{23}(q^2)  T[\V]_3(q^2)$. The $0$-helicity FFs $A_{12}$ and $T_{23}$ have been introduced in \cite{Horgan:2015vla} and their relation to the traditional basis is given in \eqref{eq:hm}. 
An overview is given in Tab.~\ref{tab:overview}.

%{Hiller:2013cza

%\begin{align}
%\mathcal{B}_{V,0}&=A_{12}\frac{8  m_B m_V}{\sqrt{\lambda}}\\
%\mathcal{B}_{V,1}&=V\frac{\sqrt{2} \sqrt{q^2} }{m_B+m_V}\\
%\mathcal{B}_{V,2}&=A_1\frac{\sqrt{2}  \sqrt{q^2} \left(m_B+m_V\right)}{\sqrt{\lambda}}\\
%\mathcal{B}_{V,t}&=A_0\\
%\mathcal{B}_{T,0}&=T_{23}\frac{4 q^2  m_B m_V}{\sqrt{q^2 \lambda} \left(m_B+m_V\right)}\\
%\mathcal{B}_{T,1}&=T_1\sqrt{2} \\
%\mathcal{B}_{T,2}&=T_2\frac{\sqrt{2}  \left(m_B^2-m_V^2\right)}{\sqrt{\lambda}}\\
%\end{align}

\section{Plots of form factors as a function of \texorpdfstring{$z$}{z}}
\label{app:z}

\begin{figure}
\includegraphics[width=\textwidth]{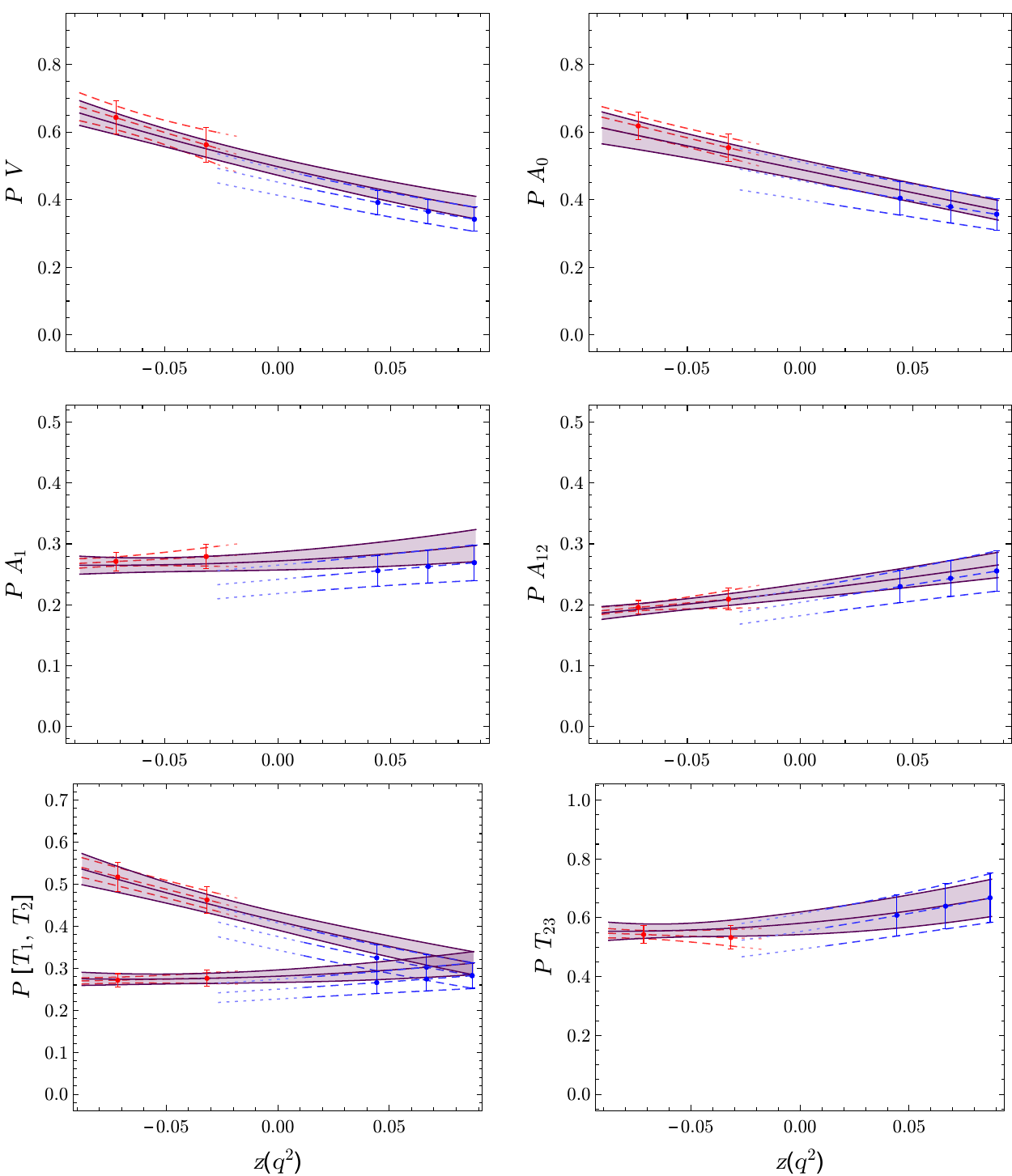}
\caption{\small Combined LCSR  and lattice fit to $B\to K^*$ FFs, where lattice data points are indicated in red, LCSR points in blue, the gray solid band shows the combined 3-parameter fit and the red dashed band the 2-parameter lattice fit from Ref.~\cite{Horgan:2015vla}.}
\label{fig:lcsr-latt-bks}
\end{figure}

\begin{figure}
\includegraphics[width=\textwidth]{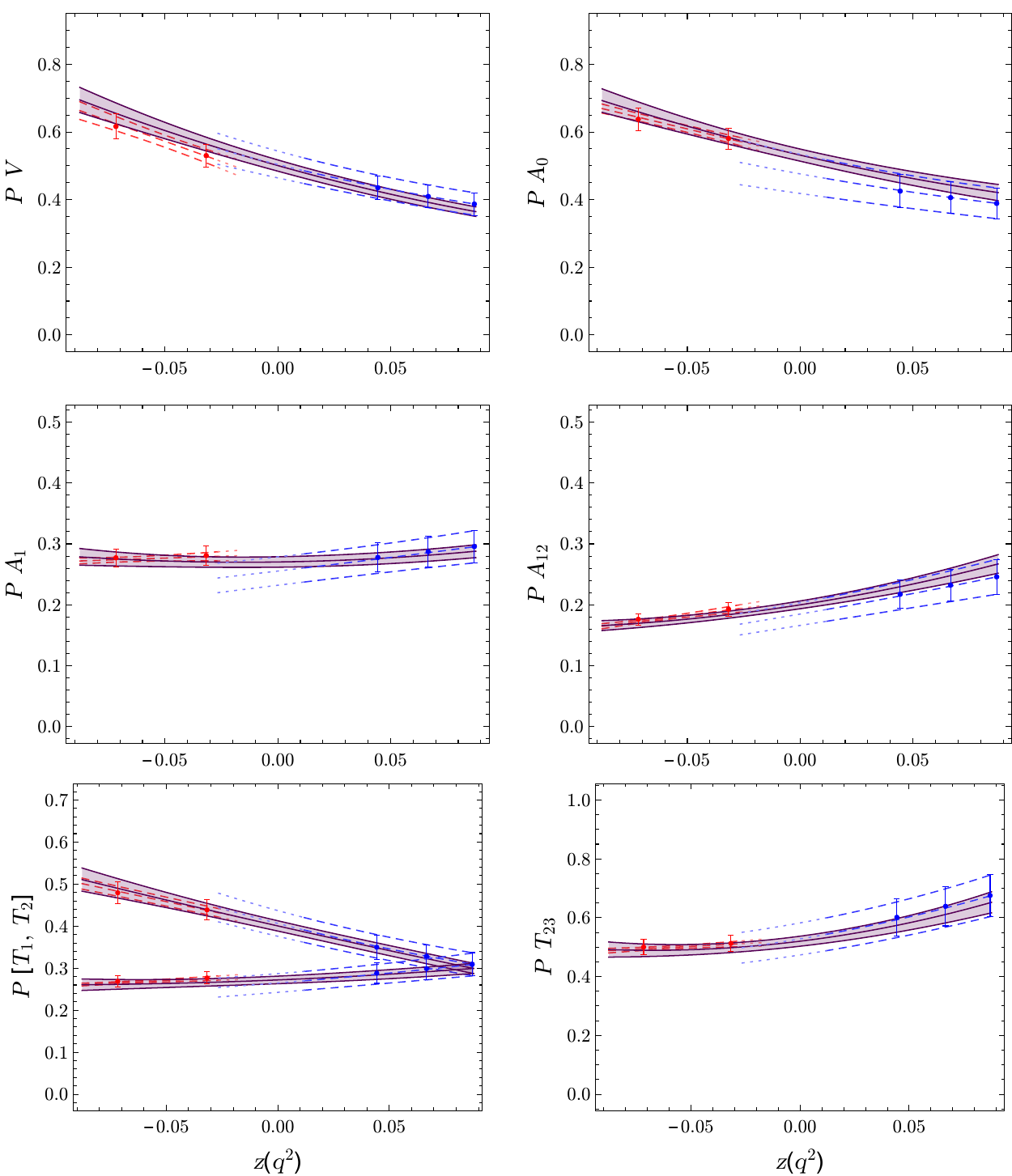}
\caption{\small Combined LCSR  and lattice fit to $B_s\to \phi$ FFs, where lattice data points are indicated in red, LCSR points in blue, the gray solid band shows the combined 3-parameter fit and the red dashed band the 2-parameter lattice fit from Ref.~\cite{Horgan:2015vla}.}
\label{fig:lcsr-latt-bsphi}
\end{figure}

\begin{figure}
\includegraphics[width=\textwidth]{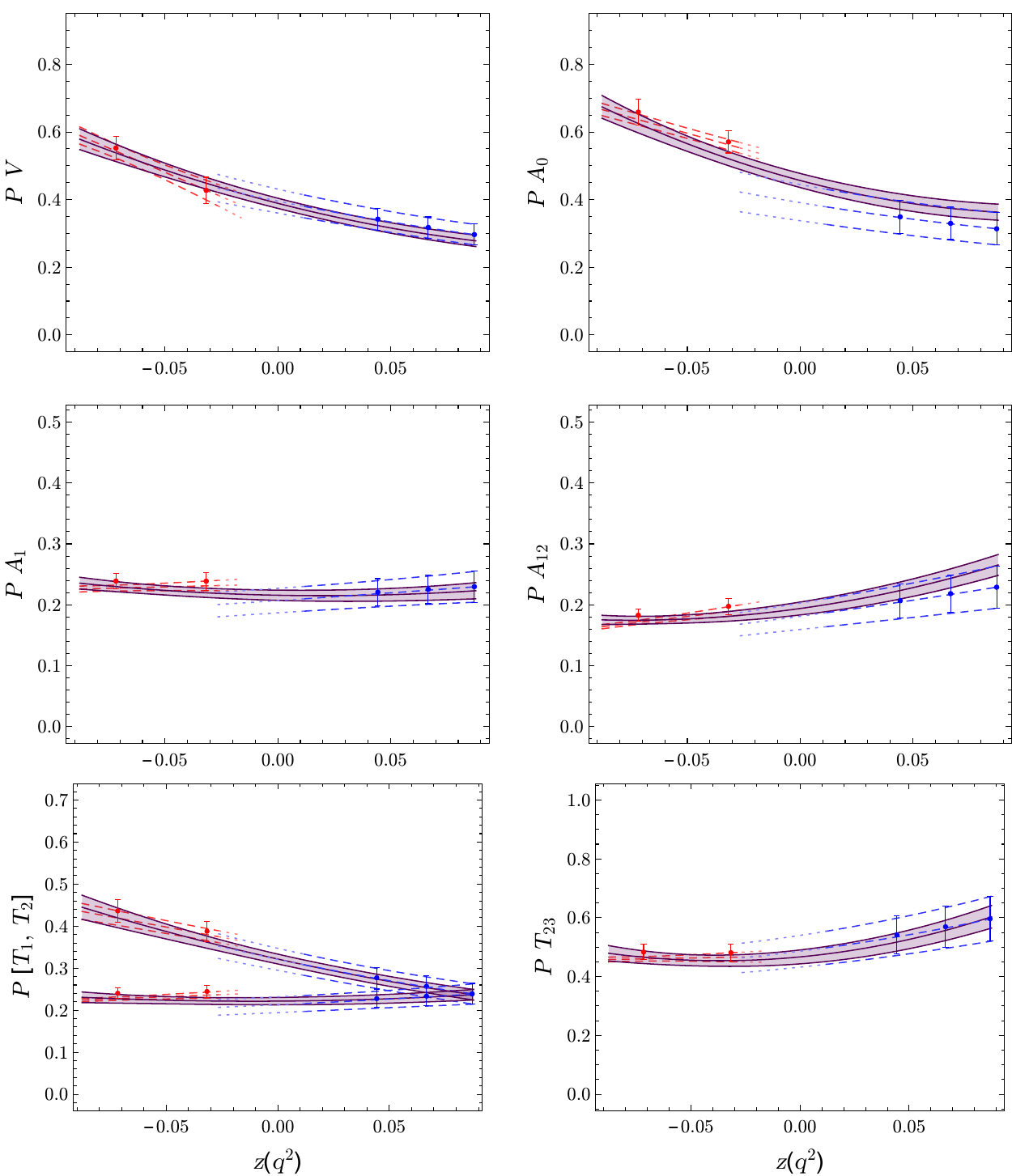}
\caption{\small Combined LCSR  and lattice fit to $B_s\to \bar K^*$ FFs, where lattice data points are indicated in red, LCSR points in blue, the gray solid band shows the combined 3-parameter fit and the red dashed band the 2-parameter lattice fit from Ref.~\cite{Horgan:2015vla}.}
\label{fig:lcsr-latt-bsks}
\end{figure}

The plots of the FFs in the $z$-variable can be found in Figs.~\ref{fig:lcsr-latt-bks},\ref{fig:lcsr-latt-bsphi}, and \ref{fig:lcsr-latt-bsks} for the modes $B \to K^*$, $B_s \to \phi$ and $B_s \to \bar K^*$, respectively. 

\section{SSE coefficients}
\label{sec:coeff}

In this appendix we list the central values and uncertainties of the SSE expansion coefficients of the $B\to K^*$, $B\to\rho$, $B\to\omega$, $B_s\to\phi$ and $B_s \to  \bar K^*$ FFs from LCSR (Tab.~\ref{tab:fit}) as well as the combined fits to LCSR and lattice data for the $B\to K^*$ and $B_s\to\phi$ FFs (Tab.~\ref{tab:fit-combined}).
Note that of the 21 parameters for each transition, two are in fact redundant due to the exact relations \eqref{eq:coeffrel}.\footnote{Due to this redundancy, the $21\times21$ covariance matrices do not have full rank.
Invertible covariance matrices can be obtained by removing the two redundant rows and columns.}

In addition to these central values and uncertainties, we also provide the full correlation and covariance matrices as ancillary files downloadable from the arXiv preprint page. The data are contained in 5 JSON files named \texttt{[Process]\char`_[Fit].json}, where \texttt{[Process]} is \texttt{BKstar} for $B\to K^*$, \texttt{Brho} for $B\to\rho$, \texttt{Bomega} for $B\to\omega$, \texttt{Bsphi} for $B_s\to\phi$ and \texttt{BsKstar} for $B_s\to\bar K^*$ FFs; \texttt{[Fit]} is \texttt{LCSR} for the fit to LCSR only (valid at low $q^2$) and \texttt{LCSR-Lattice} for the combined fit valid in the full $q^2$ range.

The JSON format can be easily used in \texttt{Mathematica}. For example, reading in the file for the $B\to K^*$ LCSR FFs,
\begin{verbatim}
  data = Import["BKstar_LCSR.json"]
\end{verbatim}
the central value of $\alpha_0^{T_1}$ can be accessed simply via
\begin{verbatim}
  OptionValue[data, "central" -> "T1" -> "a0"]
\end{verbatim}
and the correlation between $\alpha_1^{A_0}$ and $\alpha_2^{V}$ as
\begin{verbatim}
  OptionValue[data, "correlation" -> "A0V" -> "a1a2"]
\end{verbatim}
and similarly for the objects \verb+"uncertainty"+ and \verb+"covariance"+.
In Python, the corresponding commands would read
\begin{verbatim}
  import json
  with open('BKstar_LCSR.json') as file:
    data = json.load(file)
\end{verbatim}
and the parameters can be accessed via
\begin{verbatim}
  data['central']['T1']['a0']
  data['correlation']['A0V']['a1a2']
\end{verbatim}
etc.

\begin{table}[ht]
\renewcommand{\arraystretch}{1.3}
\centering
\begin{tabular}{lccccc}
\hline
\input{Coeffs}
\end{tabular}
\caption{\small Fit results for the SSE expansion coefficients in the fit to the LCSR computation only. These numbers are provided (to higher accuracy) in electronic form along with the full correlation matrices as arXiv ancillary files. 
The LCSR FFs are usually taken to be valid in the range from $0$ to $14 \GeV^2$.}
\label{tab:fit}
\end{table}

\begin{table}[ht]
\renewcommand{\arraystretch}{1.3}
\centering
\begin{tabular}{lcccc}
\hline
\input{Coeffs-combined}
\end{tabular}
\caption{\small Fit results for the SSE expansion coefficients in the combined LCSR + lattice fit. These numbers are provided (to higher accuracy) in electronic form along with the full correlation matrices as arXiv ancillary files.}
\label{tab:fit-combined}
\end{table}

\section{Lifetime effect in \texorpdfstring{$B_s\to\phi\mu^+\mu^-$}{Bs->phi mu+mu-}}\label{app:ADeltaGamma}

To compare the experimental measurement of the $B_s\to\phi\mu\mu$ branching ratio and angular observables from an untagged data sample to the theoretical predictions, the difference in finite width $\Delta \Gamma_s$ between the $B_s$ mass eigenstates of widths $\Gamma_L$ and $\Gamma_H$ has to be taken into account \cite{DeBruyn:2012wj,Descotes-Genon:2015hea}.
This leads to a difference between experimentally accessible time-integrated CP-averaged observables $\mathcal O_\text{exp}$
and the theoretical definition of CP-averaged observables $\mathcal O_\text{theo}$ in the flavour eigenstate basis.
The former and the latter are defined as
\begin{align}
\label{eq:exth}
\mathcal O_\text{exp} &= \int_0^\infty \frac{dt}{\tau_{B_s}}\,\mathcal O(t)\,,
&
\mathcal O_\text{theo} &= \mathcal O (t=0)\,,
\end{align}
where $\tau_{B_s}$ is the lifetime of the $B_s$
\begin{equation}
\mathcal  O (t)(B_s\to\phi\mu^+\mu^-) = \frac{1}{2} \left[
\mathcal O(B_s(t)\to\phi\mu^+\mu^-) + \mathcal O(\bar B_s(t)\to\phi\mu^+\mu^-) 
\right]  \;.
\end{equation}
In the case where only vector operators 
 are present (i.e. $H_\mathrm{eff} \sim  \bar b \gamma_\mu (\ga_5) s \bar \ell  \ga^\mu(\ga_5) \ell$),
the time-dependent CP-averaged observables $\mathcal  O (t)$ can be written as functions $F_{\mathcal O}$ of bilinears 
of time-dependent transversity amplitudes $\mathcal J_{b X,a Y}(t)$
\begin{align}
\mathcal  O (t) &= F_{\mathcal O}(\mathcal J_{b X,a Y}(t)) \;,
\\[0.1cm]
\mathcal J_{b X,a Y}(t)
&=
A_b^X(t)
A_a^Y(t)^*
+
\bar A_b^X(t)
\bar A_a^Y(t)^* \;,
\end{align}
where $a,b = 0,\parallel,\perp$ are the vector meson polarisation indices and $X,Y = L,R$ denote the chirality structure of the lepton production.  The  CP-conjugated amplitude is
\begin{equation}
\bar A_a^{L,R} = \eta_a A_a^{L,R} (\phi_w\to -\phi_w) \;,
\end{equation}
where $\eta_{\parallel,0}=+1$ and $\eta_\perp=-1$ are the CP-eigenvalues of the amplitudes and $(\phi_w\to -\phi_w)$ refers to the conjugation of all weak (CP-odd) phases.
Defining the quantity
\begin{align}
\label{eq:rxi}
\xi_a^\lambda &= -e^{-i\phi_s} \frac{A_a^\lambda}{\bar{A}_a^\lambda}  \;,
\end{align}
where $\phi_s$ is the $B_s$ mixing phase, one can write
\begin{align}
\mathcal J_{b X,a Y}(t)&= \mathcal J_{b X,a Y}(0)
\frac{1}{2}\left[
(e^{-\Gamma_Lt}+e^{-\Gamma_Ht})
- \mathcal A_{\Delta\Gamma}^{b X,a Y}
(e^{-\Gamma_Lt}-e^{-\Gamma_Ht})
\right] \;,
\end{align}
with  
\begin{align}
\mathcal A_{\Delta\Gamma}^{b X,a Y} &\equiv  -\frac{ \xi_b^{X*} +\xi_a^Y}{1+ \xi_b^{X*} \xi_a^Y}\,.
\end{align}
In summary  for the experimental and theoretical expression in \eqref{eq:exth}, we obtain
\begin{eqnarray}
J_{b X,a Y}|_\mathrm{exp} &=&  \int_0^\infty \frac{dt}{\tau_{B_s}}\,\mathcal J_{b X,a Y}(t)
=
\frac{1+y_s\mathcal A_{\Delta\Gamma}^{b X,a Y}}{1-y_s^2}
\mathcal J_{b X,a Y}(0)  \;, \nonumber \\ [0.1cm]
J_{b X,a Y}|_\mathrm{theo} &=&  \mathcal J_{b X,a Y}(0)  \;,
\end{eqnarray}
where 
\begin{equation}
y_s \equiv \frac{\Gamma_L - \Gamma_H}{\Gamma_L + \Gamma_H} =\frac{\Delta \Gamma_s}{2\Gamma_s}\;, \quad\Gamma_s= \frac{1}{\tau_{B_s}} = \frac{1}{2}( \Gamma_L + \Gamma_H) \;.
\end{equation}

As a simple example, we consider the differential branching ratio at low $q^2$ in the SM, within naive factorisation and the heavy quark and massless lepton limit, where the transversity amplitudes read
\begin{align}
A_\perp^{L,R}
&= \sqrt{2} N m_{B_s} (1-\hat s) \left(C_9^\text{eff} \mp C_{10} + \frac{2m_bm_B}{q^2} C_7^\text{eff}\right)\xi_\perp
\;, \nonumber \\
A_\parallel^{L,R} &= - A_\perp^{L,R}
\;, \nonumber \\
A^{L,R}_0 &= - \frac{N m_{B_s}^2(1-\hat s)^2}{2 m_\phi \sqrt{\hat s}}
\left(C_9^\text{eff} \mp C_{10} + \frac{2m_b}{m_B} C_7^\text{eff}\right) \xi_\parallel \;,
\end{align}
with $\hat s=q^2/m_{B_s}^2$ and $N$ being a normalisation factor including the CKM elements $V_{tb}V_{ts}^*$. Note that the soft FFs $\xi_{\parallel,\perp}$ are not to be confused with the ratio of amplitudes  
in \eqref{eq:rxi}.
One finds
\begin{align}
\xi_{\perp}^{L,R} &= -1
\;, &
\xi_{\parallel}^{L,R} &= +1
\;, &
\xi_{0}^{L,R} &= +1
\;.
\end{align}
The theoretical and experimental CP-averaged differential branching ratios in the assumed limit read
\begin{align}
\left. \frac{d\overline{\text{BR}}}{dq^2} \right|_\mathrm{theo} &= \tau_{B_s} \left(
|A_\perp^{L}|^2+
|A_\perp^{R}|^2+
|A_\parallel^{L}|^2+
|A_\parallel^{R}|^2+
|A_0^{L}|^2+
|A_0^{R}|^2
\right)
\nonumber   \\
&= \tau_{B_s} \left[
2\left(|A_\perp^{L}|^2+
|A_\perp^{R}|^2\right)+
|A_0^{L}|^2+
|A_0^{R}|^2
\right]
\end{align}
and 
\begin{align}
\left. \frac{d\overline{\text{BR}}}{dq^2} \right|_\mathrm{exp} &= \tau_{B_s} \sum_{a=\perp,\parallel,0}
\left(
\frac{1+y_s\mathcal A_{\Delta\Gamma}^{a}}{1-y_s^2}
\right)\left(
|A_a^{L}|^2+
|A_a^{R}|^2\right)
\nonumber   \\
&= \tau_{B_s} \left[
\frac{2}{1-y_s^2}\left(|A_\perp^{L}|^2+
|A_\perp^{R}|^2\right)+
\frac{1}{1+y_s}
\left(
|A_0^{L}|^2+
|A_0^{R}|^2
\right)
\right] \;,
\label{eq:BRlifetime}
\end{align}
where
$$
\mathcal A_{\Delta\Gamma}^{a}= \mathcal A_{\Delta\Gamma}^{aL,aL}= \mathcal A_{\Delta\Gamma}^{aR,aR} = -\eta_a \;.
$$
At low $q^2$, the sizeable  longitudinal polarization fraction of the $\phi$-meson  signals that
the last term in \eqref{eq:BRlifetime} dominates, so the time-integrated branching
ratio at low $q^2$ is \textit{suppressed} by ${\cal O}(y_s)$,
where $y_s=0.62(5)$ \cite{Amhis:2014hma},
with respect to the prompt one. This is in agreement with the findings in
\cite{Descotes-Genon:2015hea}.

\bibliographystyle{JHEP}
\bibliography{input}

\end{document}

%% file: FF-at-zero.tex
& $B\to K^*$ & $B\to\rho$ & $B\to\omega$ & $B_s\to\phi$ & $B_s\to K^*$ \\
\hline
$A_0(0)$ & $0.356 \pm 0.046$ & $0.356 \pm 0.042$ & $0.328 \pm 0.048$ & $0.389 \pm 0.045$ & $0.314 \pm 0.048$ \\
$A_1(0)$ & $0.269 \pm 0.029$ & $0.262 \pm 0.026$ & $0.243 \pm 0.031$ & $0.296 \pm 0.027$ & $0.230 \pm 0.025$ \\
$A_{12}(0)$ & $0.256 \pm 0.033$ & $0.297 \pm 0.035$ & $0.270 \pm 0.040$ & $0.246 \pm 0.029$ & $0.229 \pm 0.035$ \\
$V(0)$ & $0.341 \pm 0.036$ & $0.327 \pm 0.031$ & $0.304 \pm 0.038$ & $0.387 \pm 0.033$ & $0.296 \pm 0.030$ \\
$T_1(0)$ & $0.282 \pm 0.031$ & $0.272 \pm 0.026$ & $0.251 \pm 0.031$ & $0.309 \pm 0.027$ & $0.239 \pm 0.024$ \\
$T_2(0)$ & $0.282 \pm 0.031$ & $0.272 \pm 0.026$ & $0.251 \pm 0.031$ & $0.309 \pm 0.027$ & $0.239 \pm 0.024$ \\
$T_{23}(0)$ & $0.668 \pm 0.083$ & $0.747 \pm 0.076$ & $0.683 \pm 0.090$ & $0.676 \pm 0.071$ & $0.597 \pm 0.076$ \\
\hline

%% file: Obs-BKs.tex
\multirow{10}{1cm}{$10^{7}~\frac{d\text{BR}}{dq^2}$}
 & $[0.1,1]$ & $0.897 \pm 0.035 \pm 0.147 \pm 0.050$ \\
 & $[1,2]$ & $0.436 \pm 0.017 \pm 0.094 \pm 0.014$ \\
 & $[2,3]$ &  $ 0.400  \pm  0.015  \pm  0.091  \pm  0.010 $ \\
 & $[3,4]$ & $ 0.409  \pm  0.016  \pm  0.091  \pm  0.008 $ \\
 & $[4,5]$ &  $ 0.432  \pm  0.016  \pm  0.091  \pm  0.010 $ \\
 & $[5,6]$ & $ 0.461  \pm  0.018  \pm  0.093  \pm  0.012 $ \\
 & $[1.1,2.5]$ & $ 0.420  \pm  0.016  \pm  0.093  \pm  0.013 $ \\
 & $[2.5,4]$ &  $	0.406 \pm 0.088 \pm 0.087 \pm 0.094$ \\
 & $[4,6]$ & $ 0.447  \pm  0.017  \pm  0.092  \pm  0.011 $ \\
 & $[1.1,6]$ &  $ 0.426  \pm  0.016  \pm  0.091  \pm  0.009 $ \\
\hline
\multirow{10}{1cm}{$A_\text{FB}$}
 & $[0.1,1]$ & $ -0.093  \pm  0.000  \pm  0.012  \pm  0.001 $ \\
 & $[1,2]$ & $ -0.140  \pm  0.002  \pm  0.036  \pm  0.010 $ \\
 & $[2,3]$ &  $ -0.072  \pm  0.002  \pm  0.021  \pm  0.020 $ \\
 & $[3,4]$ & $ 0.010  \pm  0.002  \pm  0.011  \pm  0.026 $ \\
 & $[4,5]$ &  $ 0.085  \pm  0.002  \pm  0.023  \pm  0.030 $ \\
 & $[5,6]$ & $ 0.152  \pm  0.002  \pm  0.034  \pm  0.031 $ \\
 & $[1.1,2.5]$ &  $ -0.122  \pm  0.002  \pm  0.033  \pm  0.013 $ \\
 & $[2.5,4]$ & $-0.010 \pm 0.011 \pm 0.011 \pm 0.010$ \\
 & $[4,6]$ &  $ 0.120  \pm  0.002  \pm  0.029  \pm  0.031 $ \\
 & $[1.1,6]$ &  $ 0.014  \pm  0.002  \pm  0.011  \pm  0.025 $ \\
\hline
\multirow{10}{1cm}{$F_L$}
 & $[0.1,1]$ &  $ 0.330  \pm  0.004  \pm  0.064  \pm  0.018 $ \\
 & $[1,2]$ & $ 0.749  \pm  0.004  \pm  0.053  \pm  0.019 $ \\
 & $[2,3]$ &  $ 0.825  \pm  0.001  \pm  0.041  \pm  0.009 $ \\
 & $[3,4]$ & $ 0.805  \pm  0.000  \pm  0.046  \pm  0.005 $ \\
 & $[4,5]$ &  $ 0.757  \pm  0.000  \pm  0.055  \pm  0.011 $ \\
 & $[5,6]$ & $ 0.702  \pm  0.001  \pm  0.062  \pm  0.016 $ \\
 & $[1.1,2.5]$ & $ 0.782  \pm  0.003  \pm  0.048  \pm  0.016 $ \\
 & $[2.5,4]$ &  $0.812 \pm 0.044 \pm 0.047 \pm 0.046$ \\
 & $[4,6]$ &  $ 0.728  \pm  0.001  \pm  0.059  \pm  0.013 $ \\
 & $[1.1,6]$ & $ 0.768  \pm  0.001  \pm  0.051  \pm  0.006 $ \\

%% file: Obs-PBKs.tex
\multirow{10}{1cm}{$S_4$}
 & $[0.1,1]$ & $ 0.093  \pm  0.000  \pm  0.005  \pm  0.003 $ \\
 & $[1,2]$ & $ 0.005  \pm  0.001  \pm  0.009  \pm  0.010 $ \\
 & $[2,3]$ &  $ -0.096  \pm  0.001  \pm  0.015  \pm  0.013 $ \\
 & $[3,4]$ & $ -0.163  \pm  0.001  \pm  0.019  \pm  0.013 $ \\
 & $[4,5]$ &  $ -0.206  \pm  0.001  \pm  0.019  \pm  0.011 $ \\
 & $[5,6]$ & $ -0.233  \pm  0.000  \pm  0.017  \pm  0.009 $ \\
 & $[1.1,2.5]$ & $ -0.027  \pm  0.001  \pm  0.010  \pm  0.011 $ \\
 & $[2.5,4]$ & $-0.148 \pm 0.018 \pm 0.019 \pm 0.018$ \\
 & $[4,6]$ & $ -0.220  \pm  0.001  \pm  0.018  \pm  0.010 $ \\
 & $[1.1,6]$ & $ -0.145  \pm  0.001  \pm  0.016  \pm  0.012 $ \\
\hline
\multirow{10}{1cm}{$S_5$}
 & $[0.1,1]$ & $ 0.254  \pm  0.000  \pm  0.009  \pm  0.004 $ \\
 & $[1,2]$ & $ 0.110  \pm  0.004  \pm  0.017  \pm  0.020 $ \\
 & $[2,3]$ &$ -0.090  \pm  0.004  \pm  0.017  \pm  0.027 $ \\
 & $[3,4]$ & $ -0.222  \pm  0.003  \pm  0.024  \pm  0.028 $ \\
 & $[4,5]$ & $ -0.306  \pm  0.002  \pm  0.025  \pm  0.025 $ \\
 & $[5,6]$ & $ -0.360  \pm  0.002  \pm  0.022  \pm  0.022 $ \\
 & $[1.1,2.5]$ & $ 0.048  \pm  0.004  \pm  0.016  \pm  0.023 $ \\
 & $[2.5,4]$ & $-0.192 \pm 0.023 \pm 0.023 \pm 0.023$ \\
 & $[4,6]$ &  $ -0.334  \pm  0.002  \pm  0.023  \pm  0.024 $ \\
 & $[1.1,6]$ & $ -0.185  \pm  0.003  \pm  0.019  \pm  0.026 $ \\
\hline
\multirow{10}{1cm}{$P_4'$}
 & $[0.1,1]$ & $ 0.240  \pm  0.001  \pm  0.006  \pm  0.007 $ \\
 & $[1,2]$ & $ 0.014  \pm  0.003  \pm  0.022  \pm  0.025 $ \\
 & $[2,3]$ &  $ -0.273  \pm  0.005  \pm  0.029  \pm  0.042 $ \\
 & $[3,4]$ & $ -0.430  \pm  0.003  \pm  0.021  \pm  0.031 $ \\
 & $[4,5]$ & $ -0.491  \pm  0.001  \pm  0.016  \pm  0.020 $ \\
 & $[5,6]$ & $ -0.518  \pm  0.001  \pm  0.014  \pm  0.013 $ \\
 & $[1.1,2.5]$ & $ -0.070  \pm  0.004  \pm  0.026  \pm  0.032 $ \\
 & $[2.5,4]$ & $-0.398 \pm 0.022 \pm 0.022 \pm 0.022 $ \\
 & $[4,6]$ & $ -0.504  \pm  0.001  \pm  0.015  \pm  0.016 $ \\
 & $[1.1,6]$ & $ -0.358  \pm  0.003  \pm  0.022  \pm  0.029 $ \\
\hline
\multirow{10}{1cm}{$P_5'$}
 & $[0.1,1]$ &  $ 0.653  \pm  0.002  \pm  0.009  \pm  0.012 $ \\
 & $[1,2]$ & $ 0.280  \pm  0.008  \pm  0.031  \pm  0.043 $ \\
 & $[2,3]$ & $ -0.254  \pm  0.011  \pm  0.044  \pm  0.082 $ \\
 & $[3,4]$ & $ -0.585  \pm  0.008  \pm  0.035  \pm  0.070 $ \\
 & $[4,5]$ & $ -0.732  \pm  0.005  \pm  0.029  \pm  0.051 $ \\
 & $[5,6]$ & $ -0.799  \pm  0.003  \pm  0.028  \pm  0.039 $ \\
 & $[1.1,2.5]$ &  $ 0.126  \pm  0.009  \pm  0.038  \pm  0.057 $ \\
 & $[2.5,4]$ &  $-0.517 \pm 0.043 \pm 0.040 \pm 0.039$ \\
 & $[4,6]$ & $ -0.765  \pm  0.004  \pm  0.028  \pm  0.044 $ \\
 & $[1.1,6]$ & $ -0.459  \pm  0.008  \pm  0.034  \pm  0.064 $ \\

%% file: Obs-BRBpKs.tex
\multirow{10}{1cm}{$10^{7}~\frac{d\text{BR}}{dq^2}$}
 & $[0.1,1]$ & $ 0.923  \pm  0.036  \pm  0.155  \pm  0.052 $ \\
 & $[1,2]$ & $ 0.474  \pm  0.018  \pm  0.102  \pm  0.015 $ \\
 & $[2,3]$ & $ 0.438  \pm  0.017  \pm  0.099  \pm  0.010 $ \\
 & $[3,4]$ & $ 0.448  \pm  0.017  \pm  0.098  \pm  0.009 $ \\
 & $[4,5]$ & $ 0.472  \pm  0.018  \pm  0.099  \pm  0.011 $ \\
 & $[5,6]$ & $ 0.502  \pm  0.019  \pm  0.100  \pm  0.014 $ \\
 & $[1.1,2.5]$ & $ 0.458  \pm  0.017  \pm  0.101  \pm  0.013 $ \\
 & $[2.5,4]$ & $0.445 \pm 0.095 \pm 0.094 \pm 0.102$ \\
 & $[4,6]$ & $ 0.487  \pm  0.018  \pm  0.099  \pm  0.012 $ \\
 & $[1.1,6]$ & $ 0.466  \pm  0.018  \pm  0.099  \pm  0.009 $ \\

%% file: Obs-Bsphi.tex
\multirow{10}{1cm}{$10^{7}~\frac{d\text{BR}}{dq^2}$}
 & $[0.1,1]$ & $ 1.067  \pm  0.042  \pm  0.155  \pm  0.058 $ \\
 & $[1,2]$ & $ 0.497  \pm  0.019  \pm  0.099  \pm  0.017 $ \\
 & $[2,3]$ & $ 0.450  \pm  0.017  \pm  0.096  \pm  0.011 $ \\
 & $[3,4]$ & $ 0.459  \pm  0.017  \pm  0.096  \pm  0.009 $ \\
 & $[4,5]$ & $ 0.484  \pm  0.018  \pm  0.097  \pm  0.011 $ \\
 & $[5,6]$ & $ 0.516  \pm  0.019  \pm  0.099  \pm  0.015 $ \\
 & $[1.1,2.5]$ & $ 0.476  \pm  0.018  \pm  0.098  \pm  0.014 $ \\
 & $[2.5,4]$ & $ 0.455  \pm  0.017  \pm  0.096  \pm  0.009 $ \\
 & $[4,6]$ & $ 0.500  \pm  0.019  \pm  0.098  \pm  0.013 $ \\
 & $[1.1,6]$ & $ 0.479  \pm  0.018  \pm  0.097  \pm  0.010 $ \\
\hline
\multirow{10}{1cm}{$F_L$}
 & $[0.1,1]$ & $ 0.311  \pm  0.004  \pm  0.057  \pm  0.017 $ \\
 & $[1,2]$ & $ 0.732  \pm  0.003  \pm  0.051  \pm  0.020 $ \\
 & $[2,3]$ & $ 0.813  \pm  0.001  \pm  0.039  \pm  0.010 $ \\
 & $[3,4]$ & $ 0.791  \pm  0.001  \pm  0.045  \pm  0.006 $ \\
 & $[4,5]$ & $ 0.739  \pm  0.001  \pm  0.054  \pm  0.011 $ \\
 & $[5,6]$ & $ 0.682  \pm  0.001  \pm  0.060  \pm  0.016 $ \\
 & $[1.1,2.5]$ & $ 0.767  \pm  0.003  \pm  0.046  \pm  0.017 $ \\
 & $[2.5,4]$ & $ 0.799  \pm  0.001  \pm  0.043  \pm  0.006 $ \\
 & $[4,6]$ & $ 0.710  \pm  0.001  \pm  0.057  \pm  0.014 $ \\
 & $[1.1,6]$ & $ 0.752  \pm  0.001  \pm  0.050  \pm  0.006 $ \\
\hline
\multirow{10}{1cm}{$S_4$}
 & $[0.1,1]$ & $ 0.088  \pm  0.000  \pm  0.005  \pm  0.002 $ \\
 & $[1,2]$ &  $ 0.003  \pm  0.001  \pm  0.009  \pm  0.010 $ \\
 & $[2,3]$ & $ -0.099  \pm  0.001  \pm  0.016  \pm  0.013 $ \\
 & $[3,4]$ & $ -0.166  \pm  0.001  \pm  0.019  \pm  0.012 $ \\
 & $[4,5]$ & $ -0.208  \pm  0.001  \pm  0.018  \pm  0.010 $ \\
 & $[5,6]$ & $ -0.234  \pm  0.000  \pm  0.016  \pm  0.008 $ \\
 & $[1.1,2.5]$ & $ -0.029  \pm  0.001  \pm  0.011  \pm  0.011 $ \\
 & $[2.5,4]$ & $ -0.151  \pm  0.001  \pm  0.018  \pm  0.013 $ \\
 & $[4,6]$ & $ -0.221  \pm  0.000  \pm  0.017  \pm  0.009 $ \\
 & $[1.1,6]$ & $ -0.146  \pm  0.001  \pm  0.016  \pm  0.012 $ \\

%% file: Coeffs.tex
& $B\to K^*$ & $B\to\rho$ & $B\to\omega$ & $B_s\to\phi$ & $B_s\to K^*$ \\
\hline
$\alpha_0^{A_0}$ & $0.36 \pm 0.05$ & $0.36 \pm 0.04$ & $0.33 \pm 0.05$ & $0.39 \pm 0.05$ & $0.31 \pm 0.05$ \\
$\alpha_1^{A_0}$ & $-1.04 \pm 0.27$ & $-0.83 \pm 0.20$ & $-0.83 \pm 0.30$ & $-0.78 \pm 0.26$ & $-0.66 \pm 0.23$ \\
$\alpha_2^{A_0}$ & $1.12 \pm 1.35$ & $1.33 \pm 1.05$ & $1.42 \pm 1.25$ & $2.41 \pm 1.48$ & $2.57 \pm 1.44$ \\
\hline
$\alpha_0^{A_1}$ & $0.27 \pm 0.03$ & $0.26 \pm 0.03$ & $0.24 \pm 0.03$ & $0.30 \pm 0.03$ & $0.23 \pm 0.03$ \\
$\alpha_1^{A_1}$ & $0.30 \pm 0.19$ & $0.39 \pm 0.14$ & $0.34 \pm 0.24$ & $0.48 \pm 0.19$ & $0.27 \pm 0.19$ \\
$\alpha_2^{A_1}$ & $-0.11 \pm 0.48$ & $0.16 \pm 0.41$ & $0.09 \pm 0.57$ & $0.29 \pm 0.65$ & $0.13 \pm 0.56$ \\
\hline
$\alpha_0^{A_{12}}$ & $0.26 \pm 0.03$ & $0.30 \pm 0.03$ & $0.27 \pm 0.04$ & $0.25 \pm 0.03$ & $0.23 \pm 0.03$ \\
$\alpha_1^{A_{12}}$ & $0.60 \pm 0.20$ & $0.76 \pm 0.20$ & $0.66 \pm 0.26$ & $0.76 \pm 0.20$ & $0.60 \pm 0.21$ \\
$\alpha_2^{A_{12}}$ & $0.12 \pm 0.84$ & $0.46 \pm 0.76$ & $0.28 \pm 0.98$ & $0.71 \pm 0.96$ & $0.54 \pm 1.12$ \\
\hline
$\alpha_0^{V}$ & $0.34 \pm 0.04$ & $0.33 \pm 0.03$ & $0.30 \pm 0.04$ & $0.39 \pm 0.03$ & $0.30 \pm 0.03$ \\
$\alpha_1^{V}$ & $-1.05 \pm 0.24$ & $-0.86 \pm 0.18$ & $-0.83 \pm 0.29$ & $-1.03 \pm 0.25$ & $-0.90 \pm 0.27$ \\
$\alpha_2^{V}$ & $2.37 \pm 1.39$ & $1.80 \pm 0.97$ & $1.72 \pm 1.24$ & $3.50 \pm 1.55$ & $2.65 \pm 1.33$ \\
\hline
$\alpha_0^{T_1}$ & $0.28 \pm 0.03$ & $0.27 \pm 0.03$ & $0.25 \pm 0.03$ & $0.31 \pm 0.03$ & $0.24 \pm 0.02$ \\
$\alpha_1^{T_1}$ & $-0.89 \pm 0.19$ & $-0.74 \pm 0.14$ & $-0.72 \pm 0.22$ & $-0.87 \pm 0.19$ & $-0.76 \pm 0.20$ \\
$\alpha_2^{T_1}$ & $1.95 \pm 1.10$ & $1.45 \pm 0.77$ & $1.41 \pm 1.01$ & $2.75 \pm 1.19$ & $2.08 \pm 1.00$ \\
\hline
$\alpha_0^{T_2}$ & $0.28 \pm 0.03$ & $0.27 \pm 0.03$ & $0.25 \pm 0.03$ & $0.31 \pm 0.03$ & $0.24 \pm 0.02$ \\
$\alpha_1^{T_2}$ & $0.40 \pm 0.18$ & $0.47 \pm 0.13$ & $0.41 \pm 0.23$ & $0.58 \pm 0.19$ & $0.34 \pm 0.19$ \\
$\alpha_2^{T_2}$ & $0.36 \pm 0.51$ & $0.58 \pm 0.46$ & $0.46 \pm 0.57$ & $0.89 \pm 0.71$ & $0.52 \pm 0.61$ \\
\hline
$\alpha_0^{T_{23}}$ & $0.67 \pm 0.08$ & $0.75 \pm 0.08$ & $0.68 \pm 0.09$ & $0.68 \pm 0.07$ & $0.60 \pm 0.08$ \\
$\alpha_1^{T_{23}}$ & $1.48 \pm 0.49$ & $1.90 \pm 0.43$ & $1.65 \pm 0.62$ & $2.11 \pm 0.46$ & $1.58 \pm 0.56$ \\
$\alpha_2^{T_{23}}$ & $1.92 \pm 1.96$ & $2.93 \pm 1.81$ & $2.47 \pm 2.19$ & $4.94 \pm 2.25$ & $3.65 \pm 3.27$ \\
\hline

%% file: Coeffs-combined.tex
& $B\to K^*$ & $B_s\to\phi$ & $B_s\to K^*$ \\
\hline
$a_0^{A_0}$ & $0.37 \pm 0.03$ & $0.42 \pm 0.02$ & $0.36 \pm 0.02$ \\
$a_1^{A_0}$ & $-1.37 \pm 0.26$ & $-0.98 \pm 0.24$ & $-0.36 \pm 0.20$ \\
$a_2^{A_0}$ & $0.13 \pm 1.63$ & $3.27 \pm 1.36$ & $8.03 \pm 2.07$ \\
\hline
$a_0^{A_1}$ & $0.30 \pm 0.03$ & $0.29 \pm 0.01$ & $0.22 \pm 0.01$ \\
$a_1^{A_1}$ & $0.39 \pm 0.19$ & $0.35 \pm 0.10$ & $0.24 \pm 0.16$ \\
$a_2^{A_1}$ & $1.19 \pm 1.03$ & $1.70 \pm 0.79$ & $1.77 \pm 0.85$ \\
\hline
$a_0^{A_{12}}$ & $0.27 \pm 0.02$ & $0.27 \pm 0.02$ & $0.27 \pm 0.02$ \\
$a_1^{A_{12}}$ & $0.53 \pm 0.13$ & $0.95 \pm 0.13$ & $1.12 \pm 0.11$ \\
$a_2^{A_{12}}$ & $0.48 \pm 0.66$ & $2.15 \pm 0.48$ & $3.43 \pm 0.78$ \\
\hline
$a_0^{V}$ & $0.38 \pm 0.03$ & $0.36 \pm 0.01$ & $0.28 \pm 0.02$ \\
$a_1^{V}$ & $-1.17 \pm 0.26$ & $-1.22 \pm 0.16$ & $-0.82 \pm 0.19$ \\
$a_2^{V}$ & $2.42 \pm 1.53$ & $3.74 \pm 1.73$ & $5.08 \pm 1.42$ \\
\hline
$a_0^{T_1}$ & $0.31 \pm 0.03$ & $0.30 \pm 0.01$ & $0.24 \pm 0.01$ \\
$a_1^{T_1}$ & $-1.01 \pm 0.19$ & $-1.10 \pm 0.08$ & $-0.75 \pm 0.15$ \\
$a_2^{T_1}$ & $1.53 \pm 1.64$ & $0.58 \pm 1.00$ & $2.49 \pm 1.37$ \\
\hline
$a_0^{T_2}$ & $0.31 \pm 0.03$ & $0.30 \pm 0.01$ & $0.24 \pm 0.01$ \\
$a_1^{T_2}$ & $0.50 \pm 0.17$ & $0.40 \pm 0.08$ & $0.31 \pm 0.15$ \\
$a_2^{T_2}$ & $1.61 \pm 0.80$ & $1.04 \pm 0.61$ & $1.58 \pm 0.93$ \\
\hline
$a_0^{T_{23}}$ & $0.67 \pm 0.06$ & $0.65 \pm 0.04$ & $0.60 \pm 0.04$ \\
$a_1^{T_{23}}$ & $1.32 \pm 0.22$ & $2.10 \pm 0.33$ & $2.40 \pm 0.27$ \\
$a_2^{T_{23}}$ & $3.82 \pm 2.20$ & $6.74 \pm 1.80$ & $9.64 \pm 2.03$ \\
\hline